\documentclass[11pt]{article}
\usepackage{amsfonts,amsmath,amssymb}
\usepackage{enumerate}
\usepackage{hyperref}
\usepackage{bbm}
\usepackage{nicefrac}
\usepackage[all]{xy}
\usepackage{graphicx}
\usepackage{bm}
\usepackage{makecell}
\usepackage[table]{xcolor}

\usepackage{upgreek}

\usepackage{cite} 

\usepackage{booktabs}

\newcommand{\be}{\begin{equation}}
\newcommand{\ee}{\end{equation}}

\newcommand{\bea}{\begin{eqnarray}}
\newcommand{\eea}{\end{eqnarray}}

\newcommand{\bes}{\begin{subequations}}
\newcommand{\ees}{\end{subequations}}

\newcommand{\cN}{{\cal N}}
\newcommand{\cA}{{\cal A}}

\def\sst#1{{\scriptscriptstyle #1}}
\def\oneone{\rlap 1\mkern4mu{\rm l}}

\def\0{{\sst{(0)}}}
\def\1{{\sst{(1)}}}
\def\2{{\sst{(2)}}}
\def\3{{\sst{(3)}}}
\def\4{{\sst{(4)}}}
\def\5{{\sst{(5)}}}
\def\6{{\sst{(6)}}}
\def\7{{\sst{(7)}}}
\def\8{{\sst{(8)}}}

\def\cA{{{\cal A}}}

\def\cH{{{\cal H}}}

\allowdisplaybreaks  

\usepackage{multirow}
\usepackage{rotating}

\newcommand{\ba}{\begin{align}}
\newcommand{\ea}{\end{align}}

\newcommand{\bse}{\begin{subequations}}
\newcommand{\ese}{\end{subequations}}

\allowdisplaybreaks

\newcommand{\tmu}{\tilde{\mu}}
\newcommand{\hD}{\hat{D}}
\newcommand{\hrho}{\hat{\rho}}

\begin{document}

\makeatletter
\renewcommand{\theequation}{\thesection.\arabic{equation}}
\@addtoreset{equation}{section}
\makeatother

\begin{titlepage}

\begin{flushright}
IFT-UAM/CSIC-19-111 \\
%
\end{flushright}

\vspace{5pt}

   \begin{center}
   \baselineskip=16pt

   \begin{Large}\textbf{
Minimal $D=4$ truncations of type IIA}
   \end{Large}

\vspace{25pt}

{\large  Oscar Varela }

\vspace{30pt}

	\begin{small}

   {\it Department of Physics, Utah State University, Logan, UT 84322, USA}

	\vspace{15pt}
          
   {\it Departamento de F\'\i sica Te\'orica and Instituto de F\'\i sica Te\'orica UAM/CSIC , \\
   Universidad Aut\'onoma de Madrid, Cantoblanco, 28049 Madrid, Spain} 
		
	\end{small}

\vskip 70pt

\end{center}

\begin{center}
\textbf{Abstract}
\end{center}

\begin{quote}

Consistent embeddings are found of the minimal $\mathcal{N} = 2$ and $\mathcal{N} = 3$ gauged supergravities in four dimensions into its maximally supersymmetric, $\mathcal{N} = 8$, counterpart with a dyonic ISO(7) gauging. These minimal truncations retain the metric along with relevant U(1) and SO(3) R-symmetry gauge fields selected from the ISO(7) ones. The remaining ISO(7) gauge fields are turned off, with subtleties introduced by the dyonic gauging, and the scalars are fixed to their expectation values at the $\mathcal{N} = 2$ and $\mathcal{N} = 3$ vacua of the $\mathcal{N} = 8$ theory. Using the truncation formulae for massive type IIA supergravity on the six-sphere to $D=4$ $\mathcal{N} = 8$ ISO(7) supergravity, the minimal $D=4$ $\mathcal{N} = 2$ and $\mathcal{N} = 3$ gauged supergravities are then uplifted consistently to ten dimensions.

\end{quote}

\vfill

\end{titlepage}

\tableofcontents



\section{Introduction}


Despite their limited field contents, minimal, or pure, gauged supergravitites prove very useful for holography when embedded in string or M-theory. These theories involve the gravity multiplet only,  stripped of additional matter multiplets. Thus, they capture holographically universal aspects of large classes of superconformal field theories that are governed exclusively by the R-symmetry, and that are independent of details involving particular matter couplings or flavour symmetries, see \cite{Liu:2009dm,DHoker:2009mmn,Gauntlett:2011mf,Martelli:2012sz,Davison:2013bxa,Martelli:2013aqa,Cassani:2014zwa,Benini:2015bwz,Genolini:2016ecx,Ammon:2017ded,Blazquez-Salcedo:2017cqm,Azzurli:2017kxo,Bobev:2017uzs,Blazquez-Salcedo:2017kig,Cabo-Bizet:2018ehj,BenettiGenolini:2019jdz,Gang:2019uay}. In this paper, I will present dimensional reductions of ten-dimensional massive type IIA supergravity \cite{Romans:1985tz} to the minimal $\cN=2$ \cite{Fradkin:1976xz,Freedman:1976aw} and $\cN=3$ \cite{Freedman:1976aw} gauged supergravities in four dimensions. These will be obtained by consistent truncation on the internal six-dimensional geometries corresponding to the $\cN=2$ and $\cN=3$ AdS$_4$ solutions of type IIA recently constructed in \cite{Guarino:2015jca} and \cite{Pang:2015vna,DeLuca:2018buk}. Such consistent truncations should exist, based on the general arguments of \cite{Gauntlett:2007ma,Cassani:2019vcl} (see also \cite{Duff:1985jd,Pope:1987ad}). I will show by direct construction that this is indeed the case.

A natural strategy to build consistent truncations of string or M-theory down to pure gauged supergravities relies on the existence of a $G$-structure description \cite{Gauntlett:2002sc} of the background geometry. A Kaluza-Klein truncation ansatz can be constructed involving the lower-dimensional fields together with forms on the internal geometry suitably selected among those defining the $G$-structure \cite{Gauntlett:2007ma}. Consistency is then shown by enforcing the higher-dimensional equations of motion on the lower-dimensional ones while making use of the torsion classes. This technique or variations thereof has been fruitfully applied in various contexts \cite{Pope:1985jg,Pope:1985bu,Tsikas:1986rx,Buchel:2006gb,Gauntlett:2006ai,Gauntlett:2007ma,Gauntlett:2007sm,Passias:2015gya,Malek:2017njj,Hong:2018amk,Malek:2018zcz,Liu:2019cea,Cassani:2019vcl,Larios:2019lxq}. $G$-structure analyses exist \cite{Petrini:2009ur,Lust:2009mb,Passias:2018zlm} for the $\cN=2$ AdS$_4$ type IIA solution of \cite{Guarino:2015jca} and generalisations thereof. The construction of the associated consistent truncation would thus be amenable in this case to the use of this technology. Unfortunately, for the $\cN=3$ AdS$_4$ solution, a workable $G$-structure description is not readily available --although it should follow from the Killing spinor analysis of \cite{DeLuca:2018buk}. 

A different strategy can nevertheless be employed for the cases at hand, which relies on the existence of a truncation to a larger (in fact, maximal) supergravity on the relevant geometries. More concretely, the internal space for both the $\cN=2$ \cite{Guarino:2015jca} and $\cN=3$ \cite{Pang:2015vna,DeLuca:2018buk} AdS$_4$ solutions corresponds to a topological six-sphere, $S^6$ (equipped with metrics that display isometry groups smaller than the largest possible one, SO(7)). Not unrelated to this fact is the existence of a truncation of massive IIA supergravity on $S^6$ \cite{Guarino:2015jca,Guarino:2015vca} (see also \cite{Ciceri:2016dmd,Cassani:2016ncu,Inverso:2016eet}) down to $D=4$ $\cN=8$ supergravity with a dyonically gauged \cite{Dall'Agata:2012bb,Dall'Agata:2014ita,Inverso:2015viq} $\textrm{ISO}(7) = \textrm{SO}(7) \ltimes \mathbb{R}^7$ gauge group \cite{Guarino:2015qaa}. In fact, the $\cN=2$ \cite{Guarino:2015jca} and $\cN=3$ \cite{Pang:2015vna,DeLuca:2018buk} AdS$_4$ solutions of massive IIA arise as the $S^6$ uplifts of critical points of ISO(7) supergravity that break the $\cN=8$ supersymmetry of the theory to $\cN=2$ \cite{Guarino:2015jca} and $\cN=3$ \cite{Gallerati:2014xra}. In this context, the argument of \cite{Gauntlett:2007ma} indirectly implies that truncations should exist, within four dimensions, of the $\cN=8$ ISO(7) theory to the minimal $\cN=2$ and $\cN=3$ supergravities around the corresponding critical points. I will show that this is indeed the case.

The desired type IIA truncations can thus be constructed by a two-step process. First of all, truncate $D=4$ $\cN=8$ ISO(7) supergravity to the minimal $\cN=2$ and $\cN=3$ gauged supergravities around the corresponding critical points. This is achieved by leaving the $D=4$ metric and suitable combinations of the massless vectors dynamical, while placing restrictions on the remaining fields. These restrictions amount to turning off the remaining massless vectors and the massive ones, as well as freezing the scalars to their vacuum expectation values (vevs). Some of the gauge fields that need to be truncated out are the dyonically gauged non-compact ones of ISO(7). Interestingly, this is done by writing them in terms of the surviving compact R-symmetry gauge fields and their Hodge duals, rather than by setting them to zero. Secondly, bring these field restrictions to the $D=10$ to $D=4$ consistent truncation formulae of \cite{Guarino:2015jca,Guarino:2015vca}, in order to find the embedding into massive type IIA. The consistency of the IIA truncation to the full $\cN=8$ supergravity \cite{Guarino:2015jca,Guarino:2015vca}, along with the consistency of the further truncations within $D=4$, translates into the consistency of the IIA truncation to the minimal $D=4$ $\cN=2$ and $\cN=3$ theories.

The minimal $\cN=2$ and $\cN=3$ theories arise as different, non-overlapping subsectors of the parent $\cN=8$ supergravity. Indeed, the relevant U(1) and SO(3) R-symmetry gauge groups are embedded differently into the SO(7) compact subgroup of the parent ISO(7) gauge group of the $\cN=8$ theory. More precisely, this U(1) is not the Cartan subgroup of this SO(3). The $\cN=3$ gauged supergravity can nevertheless be truncated to $\cN=2$ by retaining the U(1) Cartan subgroup of SO(3). This yields an alternative embedding of $\cN=2$ supergravity into $\cN=8$ and type IIA. In any case, in order to construct the embeddings into $\cN=8$, it is thus convenient to start from a smaller sector of the $\cN=8$ supergravity that is more manageable than the full theory, yet is large enough to contain both minimal truncations. A suitable such sector is the $\cN=4$ one built in \cite{Guarino:2019jef}, which is reviewed in section \ref{sec:N=4subsec} for convenience. The minimal theories are then embedded into this $\cN=4$ sector and the full $\cN=8$ theory in section \ref{sec:minISO7truncs}, and then uplifted to type IIA in section \ref{sec:minIIAtruncs}. Section \ref{sec:discussion} concludes, and technical details are left for the appendices.


\section{An intermediate $\cN=4$ sector} \label{sec:N=4subsec}


With the aim of finding consistent truncations of the $D=4$ $\cN=8$ theory to minimal $D=4$ $\cN=2$ and $\cN=3$ supergravites, it is useful to start from a suitable subsector of the $\cN=8$ theory. The relevant subsector should be small enough so that an explicit parametrisation for its fields can be introduced, and large enough to contain both disjoint  pure $\cN=2$ and $\cN=3$ subsectors. The $\cN=4$ subsector constructed in \cite{Guarino:2019jef} suits that purpose. In this section, I will review the aspects of that $\cN=4$ model that are relevant for the present discussion, referring to that reference for further details\footnote{\label{fn:IndexConvs} I follow the notation of \cite{Guarino:2019jef} and \cite{DeLuca:2018buk} with minor changes. Indices $i=1,2,3$ here and in \cite{Guarino:2019jef,DeLuca:2018buk} label the fundamental representation of $\textrm{SO}(3)^\prime$ in (\ref{embedding2});  $a=0,1,2,3$ here ($\alpha=0,1,2,3$ in \cite{Guarino:2019jef} and $\hat{i} = 0,1,2,3$ in \cite{DeLuca:2018buk}) label the fundamental of $\textrm{SO}(4)^\prime$; and $\hat{\imath}=1,2,3$ here ($a=1,2,3$ in \cite{Guarino:2019jef}) label the fundamental of $\textrm{SO}(3)_{\textrm{L}}$. Since the fundamentals of $\textrm{SO}(3)^\prime$ and $\textrm{SO}(3)_{\textrm{L}}$ are irreducible under the diagonal $\textrm{SO}(3)_{\textrm{d}}$, tensors under $\textrm{SO}(3)_{\textrm{d}}$ can be effectively labelled by identifying $i$ and $\hat{\imath}$, formally removing the hat on the latter. This identification was implicitly used in \cite{DeLuca:2018buk}.}. This $\cN=4$ model arises as the subsector of the $\cN=8$ theory that retains all singlets under a certain SU(2) subgroup of the ISO(7) gauge group of the $\cN=8$ theory. This SU(2) is defined via the following embedding into $\textrm{SO}(7) \subset \textrm{ISO}(7)$,
\begin{equation} \label{embedding1}
\textrm{SO}(7) \supset \textrm{G}_2  \supset \textrm{SU} (3) \supset \textrm{SU} (2) \; ,
\end{equation}
with the triplet of SU(3) branching as $\mathbf{3} \rightarrow \mathbf{2} + \mathbf{1}$ under SU(2). Equivalently, this  $\textrm{SU}(2) \equiv \textrm{SO}(3)_\textrm{R}$ is also embedded in SO(7) through
\begin{equation} \label{embedding2}
\textrm{SO}(7) \, \supset \, 
  \textrm{SO}(3)^\prime \times \textrm{SO}(4)^\prime \; , 
\qquad \textrm{with}  \qquad
\textrm{SO}(4)^\prime \equiv \textrm{SO}(3)_{\textrm{L}} \times \textrm{SO}(3)_{\textrm{R}} \; . 
\end{equation}
The sector of $\cN=8$ ISO(7) supergravity invariant under the intermediate SU(3) in (\ref{embedding1}) is $\cN=2$, and was studied in section 3 of \cite{Guarino:2015qaa}. That sector contains, among others, the $\cN=2$, $\textrm{SU}(3) \times \textrm{U}(1)$--invariant critical point. The sector invariant under the diagonal subgroup, $\textrm{SO}(3)_\textrm{d}$, of $ \textrm{SO}(4) \equiv \textrm{SO}(3)^\prime \times \textrm{SO}(3)_{\textrm{L}} $ defined in (\ref{embedding2}) is $\cN=1$ and was studied in section 5 of the same reference. The latter sector contains the $\cN=3$, $\textrm{SO}(3)_\textrm{d} \times \textrm{SO}(3)_{\textrm{R}}$--invariant critical point. It is clear that the sector of $\cN=8$ supergravity that is invariant under $\textrm{SU}(2) \equiv \textrm{SO}(3)_\textrm{R}$ contains both subsectors and thus both critical points of interest. It also contains both minimal $\cN=2$ and $\cN=3$ subsectors.

The $\textrm{SO}(3)_\textrm{R}$--invariant sector of $\cN=8$ ISO(7) supergravity corresponds to an $\cN=4$ supergravity coupled to three vector multiplets. The scalar manifold is therefore
\begin{eqnarray} 
\label{ScalManN=4}
\frac{\textrm{SL}(2, \mathbb{R})}{\textrm{SO}(2)}  \times  \frac{\textrm{SO}(6,3)}{\textrm{SO}(6) \times \textrm{SO}(3)} \; .
\end{eqnarray}
The scalar fields of (\ref{ScalManN=4}) are collectively denoted as $q^u$, $u=1, \ldots , 20$. Specifically, $\varphi$, $\chi$ denote the gravity multiplet scalars which parametrise the first factor. The vector multiplet scalars, which parametrise the second factor, are $\phi_i$, $h^i{}_j$, $a_{ij}$, $b^{\hat \imath}{}_j$, with $i = 1,2,3$ and $\hat \imath = 1,2,3$, see footnote \ref{fn:IndexConvs}. The scalars $h^i{}_j$ are only defined for $i<j$, and $a_{ij} = -a_{ji}$. It is helpful to introduce the $3 \times 3$ matrices $\bm{a}$ and $\bm{b}$ defined to have components $a_{ij}$ and $b^{\hat \imath}{}_j$, respectively. It is also useful to note that the scalars $\phi_i$, $h^i{}_j$ parametrise a $\textrm{GL}(3,\mathbb{R})/\textrm{SO}(3)$ submanifold of the second factor in (\ref{ScalManN=4}) with coset representative ${\bm \nu}$ given in (2.9) of \cite{Guarino:2019jef} and scalar matrix ${\bm m} \equiv  {\bm \nu}^\textrm{T} {\bm \nu}$. The components of ${\bm m}$ will be denoted $m_{ij}$. 

The nine vectors of the model gauge (dyonically, in the frame inherited from the $\cN=8$ theory), an
\begin{equation} \label{eq:gaugegroup}
\textrm{ISO}(3)^\prime \times \textrm{SO}(3)_\textrm{L} \equiv  \left( \textrm{SO}(3)^\prime \ltimes \mathbb{R}^3 \right) \times \textrm{SO}(3)_\textrm{L}
\end{equation}
gauge group, where $\textrm{SO}(3)^\prime$ and $\textrm{SO}(3)_\textrm{L} $ were defined in (\ref{embedding2}), and $\mathbb{R}^3$ are the three translations of $\textrm{ISO}(7) \equiv \textrm{SO}(7) \ltimes \mathbb{R}^7$ that commute with $\textrm{SO}(4)^\prime$. The electric gauge fields associated to each factor on the r.h.s.~of (\ref{eq:gaugegroup}) are respectively denoted $A^{\prime i}$, $A^{(\textrm{t}) i}$, $A^{(\textrm{L}) \hat{\imath}}$, their field strengths  $H^{\prime i}_\2$, $H^{(\textrm{t}) i}_\2$, $H^{(\textrm{L}) \hat{\imath}}_\2$, and the corresponding magnetic duals $\tilde{A}^{\prime}_i$, $\tilde{A}^{(\textrm{t})}_i$, $\tilde{A}^{(\textrm{L})}_{\hat{\imath}}$ and $\tilde{H}^{\prime}_{\2 i} $, $\tilde{H}^{(\textrm{t})}_{\2 i} $, $\tilde{H}^{(\textrm{L})}_{\2 \hat{\imath}}$. Collectively, these are denoted as
\begin{equation} \label{gaugefields}
A^\Lambda = \big( A^{\prime i} \, , \,   A^{(\textrm{L}) \hat{\imath}} \, , \, A^{(\textrm{t}) i} \big) \; , \quad \Lambda =  1 , \ldots , 9 \; ,
\end{equation}
and similarly for the field strengths, $H^\Lambda_\2$, and the magnetic duals, $\tilde{A}_\Lambda$, $\tilde{H}_{\2 \Lambda}$. In the gauged theory, the electric field strengths explicitly read
\begin{eqnarray} \label{N=4electricFS}
&& H_\2^{\prime  i } = d A^{\prime  i } + \tfrac12 \, g \epsilon^i{}_{jk} \, A^{\prime  j } \wedge A^{\prime  k } \; , \nonumber \\[5pt]
&& H_\2^{(\textrm{L})  \hat{\imath} } = d A^{ (\textrm{L})  \hat{\imath}   } + \tfrac12  \, g \epsilon^{\hat{\imath}}{}_{  \hat{\jmath}   \hat{k}  } \, A^{ (\textrm{L})   \hat{\jmath}  } \wedge A^{(\textrm{L})   \hat{k}  } \; , \nonumber \\[5pt]
&& H_\2^{( \textrm{t})  i } = d A^{( \textrm{t})  i  } +  g \epsilon^i{}_{jk} \, A^{\prime  j } \wedge A^{( \textrm{t})  k } - \tfrac12  m \epsilon^i{}_j{}^k   \, A^{\prime  j } \wedge \tilde{A}^{( \textrm{t})}_k + m B^i \; ,
\end{eqnarray}
and their magnetic counterparts,
\begin{eqnarray} \label{N=4magneticFS}
&& \tilde{H}^{\prime}_{\2 i } = d \tilde{A}^{\prime}_{i} + \tfrac12 \, g   \epsilon_{ij}{}^k  \, A^{\prime  j } \wedge \tilde{A}^{\prime}_ {k } + \tfrac12 \, g   \epsilon_{ij}{}^k  \, A^{ (\textrm{t})  j } \wedge \tilde{A}^{(\textrm{t}) }_ {k } - \tfrac12 \, m   \epsilon_{i}{}^{jk}  \,  \tilde{A}^{(\textrm{t}) }_ {j} \wedge \tilde{A}^{(\textrm{t}) }_ {k }  + g  \epsilon_{ij}{}^k   \,  B_k{}^j  \; , 
\nonumber \\[5pt]
&& \tilde{H}^{(\textrm{L})  }_{\2  \hat{\imath}}   = d  \tilde{A}^{(\textrm{L})  }_{  \hat{\imath}}  + \tfrac12  \, g \epsilon_{\hat{\imath} \hat{\jmath} }{}^{ \hat{k}  } \, A^{ (\textrm{L})   \hat{\jmath}  } \wedge \tilde{A}^{(\textrm{L})}_ {\hat{k}  }  + g  B_{\hat{\imath}} \; , \nonumber \\[5pt]
&& \tilde{H}^{( \textrm{t})}_{ \2 i }  \equiv d \tilde{A}^{( \textrm{t})}_{ i }  + \tfrac12  g \epsilon_{ij}{}^k \, A^{\prime  j } \wedge \tilde{A}^{( \textrm{t})}_k  + g \delta_{ij} B^j  \; .
\end{eqnarray}
In (\ref{N=4electricFS}), (\ref{N=4magneticFS}), $g$ and $m$ are the electric and magnetic gauge couplings of the parent $\cN=8$ ISO(7) theory, and $B^i$, $B_i{}^j$, $B_{\hat{\imath}}$ are $\textrm{SO}(3)_\textrm{R}$--invariant \cite{Guarino:2019jef} two-form potentials in the restricted \cite{Guarino:2015qaa} tensor hierarchy \cite{deWit:2008ta}. Their explicit three-form field strengths will not be needed. The three-form potentials in the hierarchy will not be needed either, short of generically keeping track of an auxiliary three-form whose four-form field strength becomes the Freund-Rubin term upon uplift to type IIA, as in \cite{Guarino:2015vca,Varela:2015uca,DeLuca:2018buk}.

The bosonic Lagrangian of this $\cN=4$ subsector reads \cite{Guarino:2019jef}
{\setlength\arraycolsep{2pt}
\begin{eqnarray}
\label{N=4Lagrangian}
{\cal L} &=&  R  \, \textrm{vol}_4 + h_{uv} Dq^u \wedge * Dq^v  + \tfrac12 {\cal I}_{\Lambda \Sigma} \, H_\2^\Lambda  \wedge * H_\2^\Sigma + \tfrac12 {\cal R}_{\Lambda \Sigma} \, H_\2^\Lambda  \wedge  H_\2^\Sigma  - m B^i \wedge \tilde{H}^{(\textrm{t})}_{\2 i}      \\
&&   +\tfrac12 g m \delta_{ij} B^i \wedge B^j    +\tfrac14 m \epsilon^{ij}{}_k \, \tilde{A}^{(\textrm{t})}_i \wedge \tilde{A}^{(\textrm{t})}_j \wedge dA^{\prime k}  +\tfrac18 g  m \, \tilde{A}^{(\textrm{t})}_i \wedge \tilde{A}^{(\textrm{t})}_j \wedge A^{\prime i} \wedge A^{\prime j}   - V  \, \textrm{vol}_4  \; .   \nonumber 
\end{eqnarray}
In this paper, no significant role will be played by either the non-linear scalar selfcouplings, or the higher-rank tensor hierarchy fields as already indicated. For that reason, I simply refer to (2.15)--(2.17) of \cite{Guarino:2019jef} for the scalar kinetic terms $h_{uv} Dq^u \wedge * Dq^v$, to (2.25) of that reference for the scalar potential $V$, and to appendix B therein for the dualisation conditions for the two- and three-form potentials in the tensor hierarchy. The gauge field self-couplings via their field strengths (\ref{N=4electricFS}), (\ref{N=4magneticFS}) will be important, as will their couplings to the scalars. These couplings occur minimally, through the covariant derivatives $Dq^u$, explicitly given by
\begin{eqnarray} \label{CovDers}
&& D m _{ij} = d m _{ij} + 2 g \, \epsilon^k{}_{h(i} m_{j) k } \, A^{\prime h}  \; , \nonumber \\
&& D a_{ij} = d a_{ij} -2 g \,  \epsilon^k{}_{h [i} a_{j] k }  \,  A^{\prime h} + \epsilon_{ijk} \big( g A^{(\textrm{t}) k } - m \, \delta^{kh} \tilde{A}^{(\textrm{t}) }_h  \big) \; , \nonumber \\
&& D b^{\hat{\imath}}{}_j = d b^{\hat{\imath}}{}_j - g \, \epsilon_{jk}{}^h  A^{\prime k}    b^{\hat{\imath}}{}_h - g \epsilon^{\hat{\imath}}{}_{ \hat{\jmath} \hat{k} } \,  A^{(\textrm{L})   \hat{\jmath} } \, b^{ \hat{k} }{}_j  \; , 
\end{eqnarray}
and non-minimally through the gauge kinetic matrix $\cN_{\Lambda \Sigma} = {\cal R}_{\Lambda \Sigma} + i {\cal I}_{\Lambda \Sigma}$. In the basis (\ref{gaugefields}), this is
\begin{eqnarray}
\label{GaugeKinMat}
\cN = \cN^{\textrm{T}}  = 
\left(
\begin{array}{lll}
\cN_1 & \cN_2  & \cN_3 \\
\cN_2^{\textrm{T}} & \cN_4  & \cN_5   \\
\cN_3^{\textrm{T}} & \cN_5^{\textrm{T}}  & \cN_6
\end{array}
\right) \; . \qquad
\end{eqnarray} 
Defining the scalar-dependent $3 \times 3$ matrix, 
\begin{equation}
\bm{N} \equiv -i e^{-\varphi} (1+ e^{2\varphi} \chi^2 ) \bm{m}  -( - \chi +  i e^{-\varphi}  )  \, {\bm b}^\textrm{T} {\bm b} \; ,
\end{equation}
the blocks that compose (\ref{GaugeKinMat}) read
{\setlength\arraycolsep{2pt}
\begin{eqnarray} \label{GaugeKinMatBlocks}
\cN_1 &=& -i e^{\varphi}  \bm{m}  - \big( i e^{\varphi}  \chi \,  \bm{m} -\tfrac12  {\bm b}^\textrm{T} {\bm b}  - \bm{a} \big)   \bm{N}^{-1}  \big( i e^{\varphi}  \chi \,  \bm{m} -\tfrac12  {\bm b}^\textrm{T} {\bm b}  + \bm{a} \big)  \; , \nonumber \\[5pt]
\cN_2 &=&  \tfrac{1}{\sqrt{2}} \,  {\bm b}^\textrm{T} -  \tfrac{1}{\sqrt{2}} \,  ( - \chi +  i e^{-\varphi}  )  \big(  i e^{\varphi}  \chi \,  \bm{m} -\tfrac12  {\bm b}^\textrm{T} {\bm b}  - \bm{a} \big)   \bm{N}^{-1}  {\bm b}^\textrm{T}    \; , \nonumber \\[5pt]
\cN_3 &=& \big( i e^{\varphi}  \chi \,  \bm{m} -\tfrac12  {\bm b}^\textrm{T} {\bm b}  - \bm{a} \big)   \bm{N}^{-1}    \; , \nonumber \\[5pt]
\cN_4 &=& - \tfrac12 ( - \chi +  i e^{-\varphi}  ) \, \oneone_3 - \tfrac12( - \chi +  i e^{-\varphi}  )^2   \, {\bm b}  \bm{N}^{-1}  {\bm b}^\textrm{T}     \; , \nonumber \\[5pt]
\cN_5 &=&  \tfrac{1}{\sqrt{2}} \,  ( - \chi +  i e^{-\varphi}  )  \, {\bm b}  \bm{N}^{-1}     \; , \nonumber \\[5pt]
\cN_6 &=& - \bm{N}^{-1}     \; .
\end{eqnarray}
}Of the tensor dualisation conditions, the only ones that will play a role here are the vector-vector duality relations,
\begin{equation} \label{VectorDualityRelations}
\tilde{H}_{\2 \Lambda } = {\cal R}_{\Lambda \Sigma} \, H^\Sigma_\2 + {\cal I}_{\Lambda \Sigma} \, * H^\Sigma_\2 \; ,
\end{equation}
with ${\cal R}_{\Lambda \Sigma}$ and ${\cal I}_{\Lambda \Sigma}$ the scalar matrices just defined. 

For future reference, the equations of motion that derive from the Lagrangian (\ref{N=4Lagrangian}), upon variation of the scalars $q^u$, the electric vectors $A^\Lambda$, and the metric $g_{\mu\nu}$ are
\begin{eqnarray} \label{EomsMainText}
&& D \big( h_{uv} * D q^v \big)  -\tfrac12 ( \partial_u h_{vw} ) D q^v \wedge * Dq^w  +\tfrac12 \partial_u V \, \textrm{vol}_4 \nonumber \\
&& \qquad - \tfrac14 (  \partial_u {\cal I}_{\Lambda \Sigma} \big) \, H_\2^\Lambda  \wedge * H_\2^\Sigma - \tfrac14 (  \partial_u {\cal R}_{\Lambda \Sigma} \big) \, H_\2^\Lambda  \wedge  H_\2^\Sigma =0 \; , \nonumber \\[8pt]
&& D \tilde{H}_{\2 \Lambda}   + 2 h_{uv} \, k_\Lambda^u * Dq^v = 0  \; ,    \nonumber \\[8pt]
&& R_{\mu \nu} = h_{uv} D_\mu q^u D_\nu q^v + \tfrac12 V g_{\mu\nu} 
-\tfrac12 {\cal I}_{\Lambda \Sigma} \big( H^\Lambda_{\mu\lambda} H^\Sigma_{\nu}{}^\lambda - \tfrac14 g_{\mu\nu}  H^\Lambda_{\rho \sigma} H^{\Sigma \, \rho \sigma} \big) \; ,
\end{eqnarray}
with $k_\Lambda^u$ the Killing vectors on the second factor of the scalar manifold (\ref{ScalManN=4}) that can be read off from the covariant derivatives (\ref{CovDers}). In addition, the variation w.r.t.~the magnetic gauge field $\tilde{A}^{(\textrm{t})}_i$ yields the last three components, in the basis (\ref{gaugefields}), of the vector duality relations (\ref{VectorDualityRelations}). Finally, the variation w.r.t.~the two-form potential $B^i$ gives a dualisation condition for its field strength $H^{ i }_\3$,
\begin{equation} \label{3FormDualityRelations}
H^i_{\3 } = h_{uv} \, k^{u i} * Dq^v  \; ,
\end{equation}
with $k^{ui}$ three of the Killing vectors $k_\Lambda^u$.

The vector equations of motion in (\ref{EomsMainText}) are equivalent to the Bianchi identities for the magnetic field strengths (\ref{N=4magneticFS}), once the duality hierarchy \cite{Bergshoeff:2009ph} is employed. The Bianchi identities for their electric counterparts (\ref{N=4electricFS}) read
\begin{equation} \label{ElectricBianchisMainText}
D H^{\prime i}_\2 = 0 \; , \qquad 
D H^{\textrm{(L)} \hat{\imath}}_\2 = 0 \; , \qquad 
D H^{\textrm{(t)} i }_\2 = m \, H^{ i }_\3 \; ,
\end{equation}
with the covariant derivatives defined as
\begin{eqnarray} \label{covDersFieldStr}
&& D H^{\prime i}_\2 \equiv d H^{\prime i}_\2  + g \epsilon^i{}_{jk} \, A^{\prime  j } \wedge H^{\prime k}_\2  \; , \nonumber \\[5pt]
&& D H_\2^{(\textrm{L})  \hat{\imath} } \equiv d H_\2^{(\textrm{L})  \hat{\imath} } + g \epsilon^{\hat{\imath}}{}_{  \hat{\jmath}   \hat{k}  } \, A^{ (\textrm{L})   \hat{\jmath}  } \wedge H_\2^{(\textrm{L})  \hat{k} } \; , \nonumber \\[5pt]
&& D H_\2^{( \textrm{t})  i } = d  H_\2^{( \textrm{t})  i }  +  g \epsilon^i{}_{jk} \, A^{\prime  j } \wedge H_\2^{( \textrm{t})  k }   + \epsilon_{ijk} \big( g A^{(\textrm{t}) j } - m \, \delta^{jh} \tilde{A}^{(\textrm{t}) }_h  \big) \wedge  H_\2^{( \textrm{t})  k }  \; .
\end{eqnarray}
Finally, the covariant derivatives of the magnetic field strengths (\ref{N=4magneticFS}), which feature in the vector equations of motion in (\ref{EomsMainText}), are
\begin{eqnarray} \label{covDersFieldStrMagn}
&& D \tilde{H}^{\prime}_{\2 i } \equiv d \tilde{H}^{\prime}_{\2 i }   + g \epsilon_{ij}{}^k \, A^{\prime  j } \wedge \tilde{H}^{\prime}_{\2 k }  + \epsilon_{ij}{}^k \big( g A^{(\textrm{t}) j } - m \, \delta^{jh} \tilde{A}^{(\textrm{t}) }_h  \big) \wedge   \tilde{H}^{( \textrm{t})}_{\2 k } \; , \nonumber \\[5pt]
&& D  \tilde{H}^{(\textrm{L})  }_{\2  \hat{\imath}}    \equiv d  \tilde{H}^{(\textrm{L})  }_{\2  \hat{\imath}} + g \epsilon_{\hat{\imath}  \hat{\jmath} }{}^{ \hat{k}  } \, A^{ (\textrm{L})   \hat{\jmath}  } \wedge \tilde{H}^{(\textrm{L})  }_{\2  \hat{k}}\; , \nonumber \\[5pt]
&& D \tilde{H}^{( \textrm{t})}_{ \2 i }  \equiv d  \tilde{H}^{( \textrm{t})}_{ \2 i }   +  g \epsilon_{ij}{}^k \, A^{\prime  j } \wedge  \tilde{H}^{( \textrm{t})}_{ \2 k }  \; .
\end{eqnarray}

As argued briefly above and in more detail in \cite{Guarino:2019jef}, the present $\cN=4$ model contains the SU$(3)$-- and $\textrm{SO}(4)$--invariant sectors of $\cN=8$ ISO(7) supergravity constructed in \cite{Guarino:2015qaa}. In that reference, the SU(3)--invariant scalars were denoted by $\varphi$, $\chi$, $\phi$, $a$, $\zeta$, $\tilde \zeta$, the electric and magnetic gauge fields by $A^0$, $A^1$, $\tilde{A}_0$, $\tilde{A}_1$ and their field strengths by $H^0_\2$, $H^1_\2$, $\tilde{H}_{\2 0}$, $\tilde{H}_{\2 1}$.  This SU(3)--invariant field content is recovered from the $\cN=4$ model by identifying $\varphi$, $\chi$ here and there, and further identifying the scalars as
\begin{eqnarray} \label{N=4toSU3inv}
& \phi_1 = \phi_2 \equiv \sqrt{2} \, \phi \; , \; \phi_3 = \sqrt{2} \,  \varphi \; , \; h^i{}_j = 0 \; , \; a_{12} = -a \; , \; a_{13}= a_{23} = 0 \; , \nonumber \\
& b^{\hat{3}}{}_3 = -\sqrt{2} \,  \chi \; , \; b^{\hat{1}}{}_1 = b^{\hat{2}}{}_2 \equiv -\tfrac{1}{\sqrt{2}} \, \zeta \; , \; b^{\hat{1}}{}_2 = -b^{\hat{2}}{}_1  \equiv \tfrac{1}{\sqrt{2}} \, \tilde{\zeta} \; , \nonumber \\
& b^{\hat{1}}{}_3 = b^{\hat{2}}{}_3 = b^{\hat{3}}{}_1 = b^{\hat{3}}{}_2 = 0 \; ,
\end{eqnarray}
the electric vectors as
\begin{equation} \label{N=4toSU3invVecs}
A^0 \equiv  -A^{(\textrm{t}) 3 }  \; , \; 
A^1 \equiv A^{\prime 3} = -\tfrac12 A^{(\textrm{L}) \hat 3 } \; , \; 
A^{\prime 1} =A^{\prime 2} =  A^{(\textrm{L}) \hat 1} = A^{(\textrm{L}) \hat 2} = A^{(\textrm{t}) 1} =   A^{(\textrm{t}) 2}  = 0 \; ,
\end{equation}
and their magnetic duals as
\begin{equation} \label{N=4toSU3invVecsMagnetic}
\tilde{A}_0 \equiv  -\tilde{A}^{(\textrm{t})}_3   \; , \; 
\tilde{A}_1 \equiv  3 \tilde{A}^{\prime}_3 = -3 \tilde{A}^{(\textrm{L})}_{\hat 3 } \; , \; 
\tilde{A}^{\prime}_{ 1} =\tilde{A}^{\prime}_{ 2} =  \tilde{A}^{(\textrm{L})}_{ \hat 1} = \tilde{A}^{(\textrm{L})}_{ \hat 2} = \tilde{A}^{(\textrm{t})}_{ 1} =   \tilde{A}^{(\textrm{t})}_{ 2}  = 0 \; .
\end{equation}
With these identifications, the present $\cN=4$ model reduces to the SU(3)--invariant sector as given in section 3 of \cite{Guarino:2015qaa}. The SO(4)--invariant sector, in turn, contains four scalars, $\varphi^\prime$, $\chi^\prime$, $\phi^\prime$, $\rho^\prime$ (denoted in \cite{Guarino:2015qaa} with no primes), and no vectors. Setting to zero the vectors of the $\cN=4$ model and identifying the scalars as 
\begin{eqnarray} \label{N=4toSO4inv}
& \varphi \equiv \phi^\prime \; , \;  \chi \equiv \rho^\prime \; , \;   \phi_1 = \phi_2 = \phi_3 \equiv \sqrt{2} \, \varphi^\prime \; , \; h^i{}_j = 0 \; , \; a_{ij} = 0 \; , \;  b^{\hat{\imath}}{}_j \equiv  - \sqrt{2} \, \chi^\prime \,\delta^{\hat{\imath}}_j \; ,
\end{eqnarray}
the SO(4)--invariant sector is recovered, as given in section 5 of \cite{Guarino:2015qaa}. 

The $\cN=4$ model contains all previously known AdS vacua of $\cN=8$ dyonic ISO(7) supergravity (see table 1 of \cite{Guarino:2015qaa} for a summary) and, as shown in \cite{Guarino:2019jef}, some new non-supersymmetric vacua. In particular, the $\cN=2$, $\textrm{SU}(3) \times \textrm{U}(1)$--invariant AdS vacuum is located within the $\cN=4$ theory at
\begin{eqnarray} \label{N=2vevs}
& e^{6\varphi} = e^{3\sqrt{2} \, \phi_3} =  \tfrac{64}{27} \left( \tfrac{g}{m} \right)^2 \, , \quad 
e^{3\sqrt{2} \, \phi_1} = e^{3\sqrt{2} \, \phi_2} =   8 \left( \tfrac{g}{m} \right)^2 \, , \quad
b^{\hat{3}}{}_3 = -\sqrt{2} \,  \chi = \tfrac{1}{\sqrt{2}} \left(\tfrac{m}{g} \right)^\frac13  \; ,  \nonumber \\
& a_{ij} = 0 \; , \qquad h^i{}_j = 0 \; , \qquad   b^{\hat{1}}{}_1 = b^{\hat{1}}{}_2 = b^{\hat{1}}{}_3 = b^{\hat{2}}{}_1 =b^{\hat{2}}{}_2 = b^{\hat{2}}{}_3 = b^{\hat{3}}{}_1 = b^{\hat{3}}{}_2 = 0 \; ,
\end{eqnarray}
while the $\cN=3$, $\textrm{SO}(4)$--invariant critical point occurs at
\begin{eqnarray} \label{N=3vevs}
& e^{6\varphi} =  \tfrac{4}{27} \left( \tfrac{g}{m} \right)^2 \, , \qquad
\chi = -2^{-\frac13} \left(\tfrac{m}{g} \right)^\frac13 , \qquad
e^{3\sqrt{2} \, \phi_1} = e^{3\sqrt{2} \, \phi_2} = e^{3\sqrt{2} \, \phi_3} =    \tfrac{256}{27}  \left( \tfrac{g}{m} \right)^2 \, , \;  \nonumber \\
& a_{ij} = 0 \; , \qquad  h^i{}_j = 0 \; , \qquad  
 b^{\hat{\imath}}{}_j =   -2^{-\frac56} \left(\tfrac{m}{g} \right)^\frac13  \, \delta^{\hat{\imath}}_j \; .
\end{eqnarray}
Unlike in the SU(3) and $\textrm{SO}(4)$--invariant sectors which respectively contain them, these vacua preserve their full $\cN=2$ and $\cN=3$ supersymmetries within the $\cN=4$, $\textrm{SO}(3)_\textrm{R}$--invariant sector \cite{Guarino:2019jef}. See that reference for the allocation of the spectra within the $\cN=4$ theory in OSp$(4|2)$ and OSp$(4|3)$ multiplets, and \cite{Pang:2017omp,Gallerati:2014xra} for the supersymmetric spectra within the full $\cN=8$ ISO(7) supergravity.


\section{Minimal truncations of ISO(7) supergravity} \label{sec:minISO7truncs}


I will now move on to show that the $\cN=4$ subsector of $\cN=8$ dyonic ISO(7) supergravity discussed in \cite{Guarino:2019jef} and reviewed in the previous section can be further truncated around the vacua (\ref{N=2vevs}), (\ref{N=3vevs}) to the minimal $\cN=2$ and $\cN=3$ gauged supergravities (\ref{minimalN=2}), (\ref{minimalN=3}). The conventions for these theories are specified in appendix \ref{sec:N=2N=3Conv}.

\subsection{General strategy} \label{sec:minISO7truncsGenStrat}

The goal is to show that the Bianchi identities (\ref{ElectricBianchisMainText}) and the equations of motion (\ref{EomsMainText}) of the $\cN=4$ model are satisfied on the field equations (\ref{minimalN=2eoms}), (\ref{minimalN=3eoms}) of the minimal theories. In order to do this, the metric must be left dynamical; the scalars must be frozen to the respective vevs (\ref{N=2vevs}), (\ref{N=3vevs}); the two-form potentials set to zero, $B^i = B_i{}^j = B_{\hat{\imath}} =0$; the three-form potentials similarly turned off (except for the generic three-form argued above); and the $\cN=2$ and $\cN=3$ dynamical vectors must be conveniently selected among those of the $\cN=4$ theory that gauge the residual R-symmetry. Thus, for the $\cN=2$ and the $\cN=3$ minimal truncations, the retained vectors must respectively gauge a U(1) and the diagonal SO(3)$_\textrm{d}$ of the compact subgroup $\textrm{SO}(3)^\prime \times \textrm{SO}(3)_\textrm{L}$ of the $\cN=4$ gauge group (\ref{eq:gaugegroup}). The vectors that gauge the remaining compact generators must be truncated out, even if they remain massless at the relevant vacuum. The vectors $A^{(\textrm{t}) i }$ that gauge the non-compact $\mathbb{R}^3$ factor of the gauge group (\ref{eq:gaugegroup}) become massive at both, and any other, vacua. The gauge fields $A^{(\textrm{t}) i }$ must  therefore be truncated out too, albeit not by naively setting them to zero but, as will be argued below, by relating their field strengths to the R-symmetry vector field strengths and their Hodge duals.

To see this, note that the requirement that the scalars $q^u$ be frozen to their vevs can be phrased in a covariant way by requiring that they are covariantly constant, $D q^ u = 0$. This must happen even if some of the vectors $A^\Lambda$, $\tilde{A}_\Lambda$ are left dynamical. Bringing the vevs (\ref{N=2vevs}) and (\ref{N=3vevs}) to the definition (\ref{CovDers}) of the covariant derivatives, this requirement translates into the algebraic relation
\begin{equation} \label{ConstraintAt}
g A^{(\textrm{t}) i } - m \, \delta^{ij} \tilde{A}^{(\textrm{t}) }_j = 0 \; ,
\end{equation}
for the non-compact gauge fields at both the $\cN=2$ and the $\cN=3$ vacua. Of course, $A^{(\textrm{t}) i } = 0$, $\tilde{A}^{(\textrm{t}) }_i = 0$ is a valid solution to the constraint (\ref{ConstraintAt}). However, for the cases of interest, it can be checked that the equations of motion then imply that all vectors must vanish $A^\Lambda = 0$, $\tilde{A}_\Lambda = 0$. This choice thus leads to the $\cN=2$ and $\cN=3$ AdS vacuum solutions. One is thus led to enquire whether there are more general solutions with $A^{(\textrm{t}) i } $, $\tilde{A}^{(\textrm{t}) }_i $ non-vanishing, along with some of the compact vectors. It turns out that there are. Using the definitions (\ref{N=4electricFS}), (\ref{N=4magneticFS}) of the vector field strengths, (\ref{ConstraintAt}) can be seen to imply the constraint
\begin{equation} \label{ConstraintHt}
g H^{(\textrm{t}) i }_\2 - m \, \delta^{ij} \tilde{H}^{(\textrm{t}) }_{\2 j } = 0 \; ,
\end{equation}
at the level of the field strengths. Equation (\ref{ConstraintHt}) can be rewritten with the help of the duality relations (\ref{VectorDualityRelations}), by trading the magnetic field strengths $\tilde{H}^{(\textrm{t}) }_{\2 i }$ for $q^u$--dependent combinations of the electric field strengths $H_\2^\Lambda$ and their Hodge duals. Taking the Hodge dual of the resulting equation, further algebraically independent constraints are generated that allow one to solve for $H^{(\textrm{t}) i }_\2$ and $\tilde{H}^{(\textrm{t}) }_{\2 i }$ in terms of the dynamical R-symmetry gauge fields.

Under these assumptions, the $\cN=4$ field equations must then be shown to reduce to those, (\ref{minimalN=2eoms}), (\ref{minimalN=3eoms}), of the minimal $\cN=2$ and $\cN=3$ theories. With $q^u$ set to their vevs and $Dq^u = 0$, the $\cN=4$ equations of motion (\ref{EomsMainText}) and Bianchi identities (\ref{ElectricBianchisMainText}) give rise to the following algebraic constraints on the gauge field strengths
\begin{equation} \label{ConstraintHFromScalarEoms}
(  \partial_u {\cal I}_{\Lambda \Sigma} \big) \, H_\2^\Lambda  \wedge * H_\2^\Sigma + (  \partial_u {\cal R}_{\Lambda \Sigma} \big) \, H_\2^\Lambda  \wedge  H_\2^\Sigma =0 \; ,
\end{equation}
and  to the equations
\begin{equation} \label{eqsSimplified}
D H^\Lambda_\2 = 0 \; , \quad 
D \tilde{H}_{\2 \Lambda } = 0 \; , \quad 
R_{\mu \nu} =  \tfrac12 V g_{\mu\nu} 
-\tfrac12 {\cal I}_{\Lambda \Sigma} \big( H^\Lambda_{\mu\lambda} H^\Sigma_{\nu}{}^\lambda - \tfrac14 g_{\mu\nu}  H^\Lambda_{\rho \sigma} H^{\Sigma \, \rho \sigma} \big) \; .
\end{equation}
The gauge covariant derivatives here are given in (\ref{covDersFieldStr}) and (\ref{covDersFieldStrMagn}). In (\ref{ConstraintHFromScalarEoms}), the derivatives $\partial_u {\cal I}_{\Lambda \Sigma}$ and $\partial_u {\cal R}_{\Lambda \Sigma}$ w.r.t.~the scalar fields $q^u$ are computed from the explicit expression (\ref{GaugeKinMat})--(\ref{GaugeKinMatBlocks}) of the gauge kinetic matrix, and then evaluated at the scalar vevs. In (\ref{eqsSimplified}), the potential $V$ becomes the cosmological constant at each critical point (see {\it e.g.}~tables 3 and 4 of \cite{Guarino:2015qaa}) and the gauge kinetic matrix ${\cal I}_{\Lambda \Sigma}$ is also evaluated at the constant scalar vevs. At this stage, the constraint (\ref{ConstraintHFromScalarEoms}) on the vector field strengths must be identically satisfied, and  (\ref{eqsSimplified}) must reduce to the field equations (\ref{minimalN=2eoms}), (\ref{minimalN=3eoms}) of the $\cN=2$ and $\cN=3$ theories upon suitable rescalings of the non-vanishing gauge fields and the metric. In particular, the $\cN=4$ (and $\cN=8$) metric $g_{\mu\nu}$ and the metrics $\bar{g}_{\mu\nu}$ of the $\cN=2$ and $\cN=3$ theories are related by the constant rescaling
\begin{equation} \label{metricrescale}
\bar{g}_{\mu\nu} =  -\tfrac16 \, g^{-2} \, V \,  g_{\mu\nu} \; ,
\end{equation}
in terms of the relevant cosmological constants $V$. These rescalings involve the $\cN=4$ (and $\cN=8$) non-vanishing gauge couplings $g$ and $m$ through $V$. The magnetic gauge coupling $m$ turns out to be eventually rescaled away, while the electric gauge coupling survives as the coupling $g$ of the minimal $\cN=2$ and $\cN=3$ theories.

The restricted tensor hierarchy fields can be safely disregarded in all this process. The three-form field strengths are set to zero. Their dualisation conditions are equivalent to projections of the scalar equations of motion, and these are identically satisfied for $Dq^u = 0$ and vanishing three-form field strengths (see (\ref{3FormDualityRelations}) and, more generally, appendix B of \cite{Guarino:2019jef}). A non-vanishing three-form potential related to the type IIA Freund-Rubin term plays an auxiliary role.

\subsection{Truncation to minimal $\cN=2$ supergravity} \label{sec:N=2intoN=8}

Let us make the above discussion more explicit for the truncation to minimal $\cN=2$ supergravity around the $\cN=2$ vacuum. Similar operations were performed in \cite{Azzurli:2017kxo} starting from the $\cN=2$ SU(3)--invariant sector \cite{Guarino:2015qaa} of $\cN=8$ ISO(7) supergravity, to show that a particular black hole was a solution of the $\cN=8$ theory. The present analysis turns out to be a simple extension of \cite{Azzurli:2017kxo} to general dynamical metric and graviphoton. 

First of all, the scalars need to be frozen to their vevs (\ref{N=2vevs}). Since the dynamics must be SU(3)--invariant, the $\cN=4$ electric gauge fields must in turn be truncated as in (\ref{N=4toSU3invVecs}),
\begin{equation} \label{N=4toSU3invVecsResc1}
A^{\prime 1} =A^{\prime 2} =  A^{(\textrm{L}) \hat 1} = A^{(\textrm{L}) \hat 2} = A^{(\textrm{t}) 1} =   A^{(\textrm{t}) 2}  = 0 \; ,
\end{equation}
with similar relations for the magnetic duals, and
\begin{equation} \label{N=4toSU3invVecsResc2}
A \equiv 3  A^{\prime 3} = -\tfrac32 A^{(\textrm{L}) \hat 3 } \; .
\end{equation}
Bringing (\ref{N=2vevs}), (\ref{N=4toSU3invVecsResc1}), (\ref{N=4toSU3invVecsResc2}) to the scalar covariant derivatives (\ref{CovDers}), all of them become zero, even for $A \neq 0$, except $D a_{ij}$ which produces the constraint (\ref{ConstraintAt}) with $A^{(\textrm{t}) 1} =   A^{(\textrm{t}) 2}  = 0$, $\tilde{A}^{(\textrm{t}) }_{1 } =  \tilde{A}^{(\textrm{t}) }_{ 2 }  = 0$. One may insist in setting $A^{(\textrm{t}) 3} = 0 $, $ \tilde{A}^{(\textrm{t}) }_{3 }  = 0$ as well, but then the equations of motion enforce $A =0$ and the $\cN=2$ AdS vacuum is recovered. 

Instead, one can proceed with $A^{(\textrm{t}) 3} \neq 0 $, $ \tilde{A}^{(\textrm{t}) }_{3 }  \neq 0$. Setting to zero the two-form potentials and using (\ref{N=4toSU3invVecsResc1}), (\ref{N=4toSU3invVecsResc2}) the electric vector field strengths (\ref{N=4electricFS}) become
\begin{equation} \label{N=4toSU3invVecsResc1FieldStr}
H^{\prime 1}_\2 =H^{\prime 2}_\2 =  H^{(\textrm{L}) \hat 1}_\2 = H^{(\textrm{L}) \hat 2}_\2 = H^{(\textrm{t}) 1}_\2 =   H^{(\textrm{t}) 2}_\2  = 0 \; ,
\end{equation}
and give the following non-vanishing Abelian field strength:
\begin{equation} \label{N=4toSU3invVecsResc2FieldStr}
F \equiv dA = 3  H^{\prime 3}_\2 = -\tfrac32 H^{(\textrm{L}) \hat 3 }_\2 \; .
\end{equation}
The compact gauge field $A^{\prime 3} = -\tfrac12 A^{(\textrm{L}) \hat 3 } $ gauges the residual U(1) R-symmetry. In (\ref{N=4toSU3invVecsResc2}) and (\ref{N=4toSU3invVecsResc2FieldStr}), this gauge field has been identified with the graviphoton $A$ of the $\cN=2$ theory (\ref{minimalN=2}) and its field strength $F$, up to a suitably chosen normalisation. On  (\ref{N=2vevs}), (\ref{N=4toSU3invVecsResc1FieldStr}), the vector duality relations (\ref{VectorDualityRelations}) simplify to
\begin{equation} \label{VecRel0}
\tilde{H}^\prime_{\2 1} = \tilde{H}^\prime_{\2 2} = \tilde{H}^{\textrm{(L)}}_{\2 \hat{1}} = \tilde{H}^{\textrm{(L)}}_{\2 \hat{2}} =  \tilde{H}^{(\textrm{t}) }_{\2 1 } =  \tilde{H}^{(\textrm{t}) }_{\2 2 }  = 0 \; ,
\end{equation}
and
{\setlength\arraycolsep{2pt}
\begin{eqnarray} \label{VecRel2}
\tilde{H}^\prime_{\2 3} &=& - \tilde{H}^{\textrm{(L)}}_{\2 \hat{3}} =  \tfrac{5}{14} g^{\frac13} m^{-\frac13} H^{(\textrm{t}) 3 }_\2  - \tfrac{\sqrt{3}}{14} g^{\frac13} m^{-\frac13} * H^{(\textrm{t}) 3 }_\2 - \tfrac{4}{21} g^{-\frac13} m^{\frac13}  F - \tfrac{2}{7\sqrt{3}} g^{-\frac13} m^{\frac13} * F \; ,  \nonumber  \\[5pt]
 \tilde{H}^{(\textrm{t}) }_{\2 3 } &=& \tfrac{1}{7} g m^{-1} H^{(\textrm{t}) 3 }_\2 - \tfrac{3\sqrt{3}}{7} g m^{-1} *H^{(\textrm{t}) 3 }_\2  +\tfrac{5}{14} g^{\frac13} m^{-\frac13} F - \tfrac{\sqrt{3}}{14} g^{\frac13} m^{-\frac13} * F   \; , 
\end{eqnarray} 
}and the constraints (\ref{ConstraintAt}), (\ref{ConstraintHt}) reduce to
\begin{eqnarray} \label{VecRel1}
g A^{(\textrm{t}) 3 } - m \, \tilde{A}^{(\textrm{t}) }_3 = 0 \; , \qquad 
g H^{(\textrm{t}) 3 }_\2 - m \, \tilde{H}^{(\textrm{t}) }_{\2 3 } = 0 \; . 
\end{eqnarray}
Combining (\ref{VecRel2}) with the second relation in (\ref{VecRel1}) and further taking Hodge duals, a set of equations is obtained that allows one to solve for $H^{(\textrm{t}) 3 }_\2$, $\tilde{H}^{(\textrm{t}) }_{\2 3 }$ and $\tilde{H}^\prime_{\2 3} = - \tilde{H}^{\textrm{(L)}}_{\2 \hat{3}} $ in terms of $F$ and $*F$ as
{\setlength\arraycolsep{0pt}
\begin{eqnarray} \label{twoformsembeddingN=2} 
& H^{(\textrm{t}) 3 }_\2  =   \tfrac16 \, g^{-\frac23} \, m^{\frac23} \big( F -\sqrt{3} * F \big)  \;  , \qquad
& 
 \\[5pt]
& \tilde{H}^{(\textrm{t}) }_{\2 3 }  =   \tfrac16 \, g^{\frac13} \, m^{-\frac13} \big( F -\sqrt{3} * F \big)  \;  , \qquad
&\tilde{H}^\prime_{\2 3} = - \tilde{H}^{\textrm{(L)}}_{\2 \hat{3}} =  - \tfrac16 \, g^{-\frac13} \, m^{\frac13} \big( F +\sqrt{3} * F \big) \; .  \nonumber
\end{eqnarray}
}
Finally, the metric is rescaled as in (\ref{metricrescale}),
\begin{equation} \label{N=2andN=8metrics}
\bar{g}_{\mu\nu} = 2\sqrt{3} \, g^{\frac13} m^{-\frac13} \, g_{\mu\nu} \; .
\end{equation}

At this point, all of the vectors of the $\cN=4$ model have either been set to zero or written in terms of the $\cN=2$ graviphoton and its Hodge dual. It only remains to verify that, with these definitions, the constraints (\ref{ConstraintHFromScalarEoms}) are satisfied and the equations (\ref{eqsSimplified}) give rise to the $\cN=2$ field equations (\ref{minimalN=2eoms}). Bringing (\ref{N=4toSU3invVecsResc1FieldStr}), (\ref{N=4toSU3invVecsResc2FieldStr}) and $ H^{(\textrm{t}) 3 }_\2$ given in (\ref{twoformsembeddingN=2}) to (\ref{ConstraintHFromScalarEoms}), and using the explicit expression (\ref{GaugeKinMat})--(\ref{GaugeKinMatBlocks}) for the gauge kinetic matrix, some calculation reveals that these constraints are indeed satisfied identically for any $F$. Next, the Bianchi identities and vector equations of motion in (\ref{eqsSimplified}) reduce to
\begin{equation}
d F -\sqrt{3} \, d * F = 0 \; , \qquad 
d F + \sqrt{3} \, d * F = 0 \; . 
\end{equation}
These are straightforwardly satisfied on the $\cN=2$ Bianchi identity and Maxwell equation in (\ref{minimalN=2eoms}). Finally, the $\cN=4$ Einstein equation in (\ref{eqsSimplified}) becomes, after some calculation,
\begin{equation} \label{N=2EinsteinfromN=8}
\bar{R}_{\mu\nu} = -3 g^2 \bar{g}_{\mu\nu} + \tfrac12 \big( F_{\mu \sigma} F_\nu{}^\sigma -\tfrac14 \bar{g}_{\mu\nu} \, F_{\rho \sigma} F^{\rho \sigma}  \big) +\tfrac{6 \sqrt{3}}{7} \left( \tfrac{g}{m} \right)^{\frac23} \big( F_{\lambda ( \mu} \epsilon_{\nu )}{}^{\lambda \sigma \tau } F_{ \sigma \tau} + \tfrac14 \, \bar{g}_{\mu\nu} \epsilon_{\lambda \rho \sigma \tau} F^{ \lambda \rho}  F^{ \sigma \tau}  \big)  .
\end{equation}
The last parenthesis is identically zero, as can be checked by giving concrete values to the indices in tangent space. Thus, (\ref{N=2EinsteinfromN=8}) reduces to the $\cN=2$ Einstein equation in (\ref{minimalN=2eoms}). 

To summarise, the identifications (\ref{N=2vevs}), (\ref{N=4toSU3invVecsResc1})--(\ref{VecRel0}), (\ref{twoformsembeddingN=2}), (\ref{N=2andN=8metrics}) have been shown to define a consistent truncation of the $D=4$ $\cN=4$ model \cite{Guarino:2019jef} of section \ref{sec:N=4subsec}, to minimal $D=4$ $\cN=2$ gauged supergravity (\ref{minimalN=2}). In retrospect, it was critical to get rid of the magnetically gauged non-compact vector $A^{(\textrm{t}) 3 }_\2$ with field strength $H^{(\textrm{t}) 3 }_\2$, not by naively setting it to zero, but by writing it in terms of the surviving graviphoton $F$ and its Hodge dual through the relations (\ref{twoformsembeddingN=2}). Setting $H^{(\textrm{t}) 3 }_\2 = 0$ also leads to $F=0$.

\subsection{Truncation to minimal $\cN=3$ supergravity} \label{sec:N=3intoN=8}

The truncation to minimal $\cN=3$ supergravity proceeds similarly. First of all, the scalars are fixed to their $\cN=3$ vevs (\ref{N=3vevs}). On this vacuum, only the vectors\footnote{Here, $\delta^i_{\hat{\jmath}}$ is the invariant tensor of $\textrm{SO}(3)_\textrm{d}$. This symbol can be removed by notationally identifying the indices $i$ and $\hat{\imath}$ as suggested in footnote \ref{fn:IndexConvs}. I will leave both types of indices explicit in this section, but will identify them in section \ref{TrunctoN=3}.} $\frac12 \big( A^{\prime  i } + \delta^i_{\hat{\jmath}} \, A^{ (\textrm{L})  \hat{\jmath}   } \big) $ that gauge the diagonal $\textrm{SO}(3)_\textrm{d}$ R-symmetry remain massless, while the anti-diagonal combinations  $ \frac12 \big( A^{\prime  i } - \delta^i_{\hat{\jmath}} \, A^{ (\textrm{L})  \hat{\jmath}   } \big) $ become massive. Thus, the latter combinations must be truncated out by identifying the SO(3) graviphoton $A^i$ of the $\cN=3$ theory (\ref{minimalN=3}) with
\begin{equation} \label{N=4toN=3Vecs}
A^i \equiv A^{\prime i} = \delta^i_{\hat{\jmath}} \, A^{ (\textrm{L})  \hat{\jmath}   }  \; .
\end{equation}
Together with (\ref{N=4toN=3Vecs}) we also have, from (\ref{N=3FielStrengths}) and (\ref{N=4electricFS}), the following identifications at the level of the electric field strengths:
\begin{equation} \label{N=4toN=3VecsFieldStr}
F^i \equiv H_\2^{\prime  i } = \delta^i_{\hat{\jmath}} \, H_\2^{(\textrm{L})  \hat{\jmath} } \; .
\end{equation}
Satisfactorily, the covariant derivatives of  $H_\2^{\prime  i }$ and $ H_\2^{(\textrm{L})  \hat{\imath} } $ in (\ref{covDersFieldStr}) then coincide with themselves and with that of $F^i$ defined in (\ref{N=3FielStrengthsBianchis}). Bringing the vevs (\ref{N=3vevs}) and the vector identifications (\ref{N=4toN=3Vecs}) to (\ref{CovDers}), all the covariant derivatives are seen to vanish, except the ones of the St\"uckelberg scalars $a_{ij}$, which produce the relation (\ref{ConstraintAt}). The latter can be solved by letting $A^{(\textrm{t}) i} = 0 $, $ \tilde{A}^{(\textrm{t}) }_{i}  = 0$, but then the equations of motion also set $F^i =0$. This leads to the $\cN=3$ AdS vacuum solution. 

Alternatively, $A^{(\textrm{t}) i} $ and $ \tilde{A}^{(\textrm{t}) }_{i}  $ can be left non-vanishing but still subject to the constraint (\ref{ConstraintAt}), with their field strengths $H^{(\textrm{t}) i }_\2$, $\tilde{H}^\prime_{\2 i}$ subject to the constraint (\ref{ConstraintHt}). On (\ref{N=3vevs}), (\ref{N=4toN=3VecsFieldStr}), the vector duality relations (\ref{VectorDualityRelations}) give
{\setlength\arraycolsep{2pt}
\begin{eqnarray} \label{VecRelN=3}
\delta^{ij} \tilde{H}^\prime_{\2 j} = \tfrac12  \delta^{i \hat{\jmath}}  \tilde{H}^{\textrm{(L)}}_{\2 \hat{\jmath}} & = &  \tfrac{5}{14} \, 2^{-\frac23} \, g^{\frac13}   m^{-\frac13} H^{(\textrm{t}) i }_\2  - \tfrac{\sqrt{3}}{14} \, 2^{-\frac23} \, g^{\frac13} m^{-\frac13} * H^{(\textrm{t}) i }_\2  \nonumber \\
&& - \tfrac{1}{7} \, 2^{\frac23} \, g^{-\frac13} m^{\frac13}  F^i - \tfrac{\sqrt{3}}{7}  \, 2^{-\frac13} \, g^{-\frac13} m^{\frac13} * F^i \; ,  
\end{eqnarray} 
}
\vspace{-10pt}
\begin{equation}
\delta^{ij}  \tilde{H}^{(\textrm{t}) }_{\2 j } =  \tfrac{1}{7} g m^{-1}   H^{(\textrm{t}) i }_\2  - \tfrac{3\sqrt{3}}{7}  g m^{-1} * H^{(\textrm{t}) i }_\2   + \tfrac{15}{14} \, 2^{-\frac23} \, g^{\frac13}   m^{-\frac13}  F^i - \tfrac{3\sqrt{3}}{14}  \, 2^{-\frac23} \, g^{\frac13} m^{-\frac13} * F^i \; . \nonumber 
\end{equation}
Combining the equation for  $\tilde{H}^{(\textrm{t}) }_{\2 i }$ in (\ref{VecRelN=3}) with the constraint (\ref{ConstraintHt}), and further taking Hodge duals, a set of equations is obtained that allows one to solve for the electric field strength $H^{(\textrm{t}) i }_\2$ and the magnetic $\tilde{H}^{(\textrm{t}) }_{\2 i }$, $\tilde{H}^\prime_{\2 i} =  \tfrac12  \delta_i^{\hat{\jmath}} \tilde{H}^{\textrm{(L)}}_{\2 \hat{\jmath}} $ in terms of $F^i$ and $*F^i$ as
{\setlength\arraycolsep{0pt}
\begin{eqnarray} \label{twoformsembeddingN=3} 
&  H^{(\textrm{t}) i }_\2  = 2^{-\frac53} \,  g^{-\frac23} \, m^{\frac23} \big( F^i -\sqrt{3} * F^i \big)  \;  , \qquad
%
%
%
 \delta^{ij} \tilde{H}^{(\textrm{t}) }_{\2 j }  =  2^{-\frac53} \,  g^{\frac13} \, m^{-\frac13} \big( F^i -\sqrt{3} * F^i \big)  \;  , \nonumber  \\[5pt]
&\delta^{ij} \tilde{H}^\prime_{\2 j} = \tfrac12  \delta^{i \hat{\jmath}}  \tilde{H}^{\textrm{(L)}}_{\2 \hat{\jmath}}  =  -2^{-\frac73} \, g^{-\frac13} \, m^{\frac13} \big( F^i +\sqrt{3} * F^i \big) \; .  
\end{eqnarray}
}
Finally, the metric must be rescaled as in (\ref{metricrescale}):
\begin{equation} \label{N=3andN=8metrics}
\bar{g}_{\mu\nu} = \tfrac{16}{3\sqrt{3}} \, 2^\frac13 \, g^{\frac13} m^{-\frac13} \, g_{\mu\nu}  \; .
\end{equation}

Now, it only remains to verify that, with these definitions, the constraints (\ref{ConstraintHFromScalarEoms}) are satisfied and the equations (\ref{eqsSimplified}) give rise to the $\cN=3$ field equations (\ref{minimalN=3eoms}). Bringing (\ref{N=4toN=3VecsFieldStr}) and $ H^{(\textrm{t}) i }_\2$ given in (\ref{twoformsembeddingN=3}) to (\ref{ConstraintHFromScalarEoms}), some calculation shows that these constraints indeed check out identically for arbitrary $F^i$. The Bianchi identities and equations of motion for  $H_\2^{\prime  i }$ and $ H_\2^{(\textrm{L})  \hat{\imath} } $ immediately reduce to their $\cN=3$ counterparts in (\ref{minimalN=3eoms}), while those for $ H_\2^{(\textrm{t}) i } $ give
\begin{equation} \label{N=3gaugeEqsfromN=4}
D F^i -\sqrt{3} \, D * F^i = 0 \; , \qquad 
D F^i + \sqrt{3} \, D * F^i = 0 \; ,
\end{equation}
with the covariant derivatives defined as in (\ref{N=3FielStrengthsBianchis}). Equations (\ref{N=3gaugeEqsfromN=4}) are equivalent to the $\cN=3$ Bianchi identity and Maxwell equation in (\ref{minimalN=3eoms}). Finally, the $\cN=4$ Einstein equation in (\ref{eqsSimplified}) becomes, after some calculation,
\begin{eqnarray} \label{N=3EinsteinfromN=8}
\bar{R}_{\mu\nu} & =& -3 g^2 \bar{g}_{\mu\nu} + 2 \big( F^i_{\mu \sigma} F_{i \, \nu}{}^\sigma -\tfrac14 \bar{g}_{\mu\nu} \, F^i_{\rho \sigma} F_i^{\rho \sigma}  \big) \nonumber \\
&& 
 + \tfrac{512}{63  \sqrt{3}} \, 2^{\frac23} \left( \tfrac{g}{m} \right)^{\frac23}  \big( F^i_{\lambda ( \mu} \epsilon_{\nu )}{}^{\lambda \sigma \tau } F_{i \, \sigma \tau} + \tfrac14 \, \bar{g}_{\mu\nu} \epsilon_{\lambda \rho \sigma \tau} F^{ i \, \lambda \rho}  F_i^{ \sigma \tau}  \big) \; .
\end{eqnarray}
The last parenthesis is identically zero, as can be easily checked in tangent space by giving concrete values to the indices, and (\ref{N=3EinsteinfromN=8}) reduces to the $\cN=3$ Einstein equation in (\ref{minimalN=3eoms}). 

In summary, I have shown, at the level of the bosonic field equations, that (\ref{N=3vevs}), (\ref{N=4toN=3Vecs}), (\ref{N=4toN=3VecsFieldStr}), (\ref{twoformsembeddingN=3}), (\ref{N=3andN=8metrics}) define a consistent truncation of the $D=4$ $\cN=4$ subsector \cite{Guarino:2019jef} of $\cN=8$ ISO(7) supergravity reviewed in section \ref{sec:N=4subsec}, to minimal $D=4$ $\cN=3$ gauged supergravity (\ref{minimalN=3}). In turn, minimal $D=4$ $\cN=3$ supergravity (\ref{minimalN=3}) can be further truncated consistently to the $\cN=2$ minimal theory (\ref{minimalN=2}). This is achieved by turning off two of the SO(3) Yang-Mills fields and by selecting the $\cN=2$ graviphoton to lie in the $\textrm{U}(1)$ Cartan subgroup of $\textrm{SO}(3)$, as in (\ref{N=3toN=2}). As discussed in appendix \ref{sec:GaugeGroupEmbedding}, this U(1) Cartan subgroup is different from the U(1) gauge group of the $\cN=2$ model of section \ref{sec:N=2intoN=8}. Thus, this further truncation provides an alternative embedding of minimal $D=4$ $\cN=2$ gauged supergravity (\ref{minimalN=2}) into $D=4$ $\cN=8$ ISO(7) supergravity (via the $\cN=4$ SO(3)$_\textrm{R}$--invariant subsector of the latter), which is different from the embedding discussed in section \ref{sec:N=2intoN=8}.

\subsection{Embedding into $\cN=8$ supergravity} \label{EqsfromN=8}

The minimal $\cN=2$ and $\cN=3$ supergravities can be finally embedded into the full $\cN=8$ ISO(7) theory by combining their embeddings, given in sections \ref{sec:N=2intoN=8} and \ref{sec:N=3intoN=8} above, into the intermediate $\cN=4$ model, with the embedding \cite{Guarino:2019jef} of the latter into the parent $\cN = 8$ theory. With the metric rescaled as in (\ref{metricrescale}), the two-form potentials in the restricted tensor hierarchy set to zero, and simply keeping an eye on the auxiliary three-form potential that will eventually give rise to the Freund-Rubin term upon uplift to IIA, the problem reduces to keeping track of the gauge fields. The electric vectors (\ref{gaugefields}) of the $\cN=4$ model are embedded into their $\cN=8$ counterparts \cite{Guarino:2015qaa} $\cA^{IJ}$, $\cA^{I}$, $I=1, \ldots, 7$, as \cite{Guarino:2019jef}
\begin{equation} \label{ElectricVectorsinN=8}
\cA^{ij} = \epsilon^{ij}{}_k \, A^{\prime k} \ , \quad
\cA^{ia} = 0 \ , \quad
\cA^{ab} = -\tfrac12  (J_{\hat{\imath} - })^{ab} \, A^{ (\textrm{L}) \hat{\imath}}  \ , \quad
\cA^{i} =  A^{ (\textrm{t}) i} \ ,  \quad
\cA^{a} = 0 \; ,
\end{equation}
and their magnetic duals into the $\cN=8$ magnetic duals as \cite{Guarino:2019jef}
\begin{equation} \label{MagneticVectorsinN=8}
\tilde{\cA}_{ij} = \epsilon_{ij}{}^k \, \tilde{A}^{\prime}_{ k} \ , \quad
\tilde{\cA}_{ia} = 0 \ , \quad
\tilde{\cA}_{ab} = - (J^{\hat{\imath}}_{ - })_{ab} \, \tilde{A}^{ (\textrm{L}) }_{\hat{\imath}}  \ , \quad
\tilde{\cA}_{i} =  \tilde{A}^{ (\textrm{t})}_{ i} \ ,  \quad
\tilde{\cA}_{a} = 0 \; .
\end{equation}
The indices here have been split as $I=(i, a)$, where $i=1,2,3$ and $a=0,1,2,3$ with the index conventions of footnote \ref{fn:IndexConvs}. The antisymmetric, anti-selfdual, quaternionic matrices $(J^{\hat{\imath}}_{ - })_{ab}$ are defined in appendix A of \cite{Guarino:2019jef}. 

The field strengths (\ref{N=4electricFS}), (\ref{N=4magneticFS}) are similarly embedded into the $\cN=8$ ones,  $\cH^{IJ}_\2$, $\cH^{I}_\2$, $\tilde{\cal H}_{\2 IJ}$, $\tilde{\cal H}_{\2 I}$. Bringing (\ref{N=4toSU3invVecsResc1FieldStr}) and (\ref{twoformsembeddingN=2}) to the field strengths version of (\ref{ElectricVectorsinN=8}), (\ref{MagneticVectorsinN=8}), the $\cN=2$ graviphoton is finally embedded into the $\cN=8$ vector field strengths. Note that this embedding involves both the minimal $\cN=2$ graviphoton field strength $F$ and its Hodge dual $*F$. Similarly, bringing (\ref{N=4toN=3VecsFieldStr}), (\ref{twoformsembeddingN=3}) to (\ref{ElectricVectorsinN=8}), (\ref{MagneticVectorsinN=8}) written for the field strengths, the $\cN=3$ graviphoton field strength $F^i$ and its Hodge dual $*F^i$ are embedded into the $\cN=8$ vector field strengths. Equations (\ref{ElectricVectorsinN=8}), (\ref{MagneticVectorsinN=8}) and their analogues for the field strengths  are useful to work out the ten-dimensional uplift of the minimal theories, to which I now turn.

\vspace{12pt}


\section{Minimal $D=4$ truncations of massive IIA supergravity} \label{sec:minIIAtruncs}


Having obtained the consistent embedding of the minimal $\cN=2$ and $\cN=3$ theories into $\cN=8$ ISO(7) supergravity through the intermediate $\cN=4$ SO(3)$_\textrm{R}$--invariant sector, these theories can now be uplifted to  $D=10$. This is done by particularising the general truncation formulae for massive IIA on $S^6$ \cite{Guarino:2015jca,Guarino:2015vca} to the cases at hand through the  $\cN=8$ expressions outlined in section \ref{EqsfromN=8}. The consistency of the truncation from IIA to $D=4$ $\cN=8$, together with the consistency of the $D=4$ subtruncations, guarantee the consistency of the truncation from IIA  to the  minimal $D=4$ $\cN=2$ and $\cN=3$ gauged supergravities. In any case, for further reassurance, the consistency of these minimal truncations is manifestly checked in appendix \ref{app:MinimalintoIIA} at the level of the Bianchi identities and equations of motion of the type IIA supergravity forms. All the type IIA expressions below are written in the Einstein frame conventions of appendix A of \cite{Guarino:2015vca}.

\vspace{6pt}

\subsection{Truncation to minimal $\cN=2$ supergravity} \label{TrunctoN=2}

Particularising the $\cN=8$ consistent truncation formulae of \cite{Guarino:2015jca,Guarino:2015vca} to the relevant $\cN=4$ subsector and on to the minimal $\cN=2$ theory as discussed in sections \ref{sec:N=2intoN=8} and \ref{EqsfromN=8}, a consistent truncation of massive IIA supergravity to $D=4$ $\cN=2$ minimal gauged supergravity is obtained. For this purpose, the SU(3)--invariant truncation formulae of \cite{Varela:2015uca} also come in handy, along with the generic formulae of \cite{Guarino:2015jca,Guarino:2015vca}. This is because the minimal $\cN=2$ theory discussed here is also a subsector of the SU(3)--invariant sector of ISO(7) supergravity. The minimal truncation can be expressed in terms of the same geometric structures on the internal $S^6$ discussed in \cite{Varela:2015uca}. The resulting truncation formulae are formally identical to those found in \cite{Azzurli:2017kxo} for the IIA uplift of a particular $D=4$ black hole. The $D=4$ metric and graviphoton here are generic, though: they are only required to obey the $D=4$ $\cN=2$ field equations (\ref{minimalN=2eoms}). 

\newpage 

Proceeding along these lines, some calculation gives the following formulae for the consistent truncation of massive IIA supergravity to minimal $D=4$ $\cN=2$ gauged supergravity (\ref{minimalN=2}):
{\setlength\arraycolsep{0pt}
\begin{eqnarray} \label{SU3U1lIIASolution}
&& d \hat{s}_{10}^2 =  2^{-\frac{5}{8}} \, 3^{-1} \,  g^{-\frac{1}{12}} \, m^{\frac{1}{12}} \,  \big( 3 + \cos 2\alpha \big)^{1/2} \big( 5 + \cos 2\alpha \big)^{1/8}   \,  d\bar{s}^2_4     + ds^{2}_6  \; , \nonumber  \\[10pt]
&&  \hat H_\3 =    \hat H_\3^0  + \tfrac{1}{2\sqrt{3}} \, g^{-\frac23} \, m^{-\frac13} \, \sin \alpha \, d\alpha \wedge *F \; ,
\nonumber \\[10pt]
&&  \hat F_\4 =   \hat F_\4^0 +  \tfrac{1}{\sqrt{3}} \, g^{\frac23} \, m^{\frac13} \,  \overline{\textrm{vol}}_4   \nonumber \\[4pt]
&& \qquad\qquad\quad  -  \tfrac{1}{4} \, g^{-\frac73} \, m^{\frac13} \,  \sin \alpha \cos \alpha \,  \Big[  \,  \frac{4  \sin \alpha \cos \alpha }{ 3 + \cos 2\alpha } \, \bm{J} + d\alpha \wedge \bm{\hat{\eta}} \,  \Big] \wedge F \nonumber \\[4pt]
&& \qquad\qquad\quad  -  \tfrac{1}{2\sqrt{3}} \, g^{-\frac73} \, m^{\frac13} \,  \sin \alpha \,  \Big[  \,  \frac{2  \sin \alpha  }{ 3 + \cos 2\alpha } \, \bm{J} +  \frac{3  \cos \alpha  }{ 5 + \cos 2\alpha } \, d\alpha \wedge \bm{\hat{\eta}} \,  \Big] \wedge * F  \; .
\nonumber \\[10pt]
&& \hat F_\2 =    \hat F_\2^0  +   g^{-\frac23} \, m^{\frac23} \, \frac{\cos \alpha }{5+ \cos 2\alpha } \, F  - \tfrac{1}{2\sqrt{3}} \, g^{-\frac23} \, m^{\frac23} \,  \cos \alpha  * F     \; .
\end{eqnarray}
}Here, $d\bar{s}^2_4$ and $F$ are the metric and field strength of the $D=4$ $\cN=2$ theory (\ref{minimalN=2}), and $*F$ is the Hodge dual with respect to the former. The $S^6$ angle $\alpha$ ranges as $0 \leq \alpha \leq \pi$, while $\bm{J}$ is the K\"ahler form on the complex projective plane; the one-form $\bm{\hat{\eta}}$ is defined as 
\begin{equation} \label{eq:etashifted}
\bm{\hat{\eta}} \equiv \bm{\eta} +  \tfrac{1}{3}  g  A  \equiv
d\psi +\sigma +  \tfrac{1}{3}  g  A  \; ,
\end{equation}
in terms of the $S^6$ angle $\psi$, with $0 \leq \psi \leq 2\pi$, a one-form $\sigma$ such that $d \sigma = 2 \,  \bm{J}$, and the $D=4$ $\cN=2$  gauge field $A$ affected by the coupling constant $g$. Finally, the metric $ds^{2}_6$ and the forms $\hat F_\4^0$, $\hat H_\3^0$ and $\hat F_\2^0$ correspond to the background values of the $\cN=2$ AdS$_4$ solution of \cite{Guarino:2015jca}, shifted with the $D=4$ gauge potential $A$ through $\bm{\hat{\eta}}$ in (\ref{eq:etashifted}), namely,
{\setlength\arraycolsep{0pt}
\begin{eqnarray} \label{SU3U1lIIASolutionBackground}
&& d s_{6}^2 =  L^2  \big( 3 + \cos 2\alpha \big)^{1/2} \big( 5 + \cos 2\alpha \big)^{1/8}   \Big[ \, \frac32 \,  d\alpha^2 +  \frac{ 6 \sin^2 \alpha}{ 3 + \cos 2\alpha } \, ds^2 ( \mathbb{CP}^2 ) %
+   \frac{ 9 \sin^2 \alpha}{ 5 + \cos 2\alpha } \, \bm{\hat{\eta}}^2  \Big] \; , \nonumber  \\
&& e^{\hat \phi} =  e^{\phi_0} \frac{  \big( 5 + \cos 2\alpha \big)^{3/4} }{ 3 + \cos 2\alpha }  \;  , \nonumber  \\
&&  L^{-2} e^{-\frac12 \phi_0}  \hat H_\3^0 =   24 \sqrt{2} \, \frac{  \sin^3 \alpha }{   \big( 3 + \cos 2\alpha \big)^2   }  \ \bm{J}  \wedge d\alpha  \; ,
\nonumber \\[10pt]
&& L^{-3} e^{\frac14 \phi_0}   \hat F_\4^0 =  12\sqrt{3} \, \frac{  7 +  3\cos 2\alpha }{   \big( 3 + \cos 2\alpha \big)^2   } \, \sin^4 \alpha \ \textrm{vol} (  \mathbb{CP}^2)  \nonumber \\
&& \qquad  \qquad \qquad + 18\sqrt{3}  \, \frac{  (9 +  \cos 2\alpha) \sin^3 \alpha \cos \alpha }{   \big( 3 + \cos 2\alpha \big)  \big( 5 + \cos 2\alpha \big)   }  \ \bm{J}  \wedge d\alpha \wedge \bm{\hat{\eta}}  \; ,
\nonumber \\[10pt]
&&  L^{-1} e^{\frac34 \phi_0}  \hat F_\2^0 =   -4  \sqrt{6} \,  \frac{  \sin^2 \alpha \cos \alpha  }{   \big( 3 + \cos 2\alpha \big)    \big( 5 + \cos 2\alpha \big)   }  \ \bm{J}     -3  \sqrt{6} \,  \frac{ \big( 3 - \cos 2\alpha \big)  }{   \big( 5 + \cos 2\alpha \big)^2   }   \, \sin \alpha \ d\alpha \wedge \bm{\hat{\eta}}   \; , \nonumber \\[10pt]
&& L \, e^{\frac54 \phi_0} \hat F_\0 = 3^{-\frac12} \; .
\end{eqnarray}
 }
 
 \noindent In these expressions, $ds^2 ( \mathbb{CP}^2 )$ and $\textrm{vol} (  \mathbb{CP}^2)$ are the Fubini-Study metric, normalised so that its Ricci tensor equals 6 times the metric, and the corresponding volume form. In this minimal $\cN=2$ truncation, the dilaton $e^{\hat \phi}$ and the Romans mass $\hat F_\0$ take on their exact background values \cite{Guarino:2015jca}. Although the expressions  (\ref{SU3U1lIIASolution}) were given in terms of the $\cN=8$ couplings $g$ and $m$, these should be traded for the classical parameters $L$ and $e^{\frac54 \phi_0}$ that characterise the background geometry, via \cite{Guarino:2015jca} $L^2 \equiv 2^{-\frac{5}{8}} \, 3^{-1} \, g^{-\frac{25}{12}} \,  m^{\frac{1}{12}}$ and $e^{\phi_0} \equiv  2^{\frac{1}{4}} \, g^{\frac{5}{6}} \,  m^{-\frac{5}{6}}$. These parameters are quantised in the full string theory \cite{Varela:2015uca,Guarino:2016ynd}.
 
The general formalism of \cite{Guarino:2015jca,Guarino:2015vca,Varela:2015uca} ensures the consistency of the truncation formulae (\ref{SU3U1lIIASolution})--(\ref{SU3U1lIIASolutionBackground}). I have nevertheless verified, up to an explicit check of the Einstein equation, that consistency does indeed hold: (\ref{SU3U1lIIASolution})--(\ref{SU3U1lIIASolutionBackground}) solve the field equations of massive type IIA supergravity provided the Bianchi identity and the equations of motion (\ref{minimalN=2eoms}) of minimal $D=4$ $\cN=2$ gauged supergravity are imposed. See appendix \ref{app:IIAtoN=2} for the details. 


\subsection{Truncation to minimal $\cN=3$ supergravity} \label{TrunctoN=3}

The uplift of the minimal $\cN=3$ theory to massive type IIA proceeds similarly. Together with the general uplifting formulae of \cite{Guarino:2015jca,Guarino:2015vca}, the SO(4)--invariant uplifting formulae of \cite{DeLuca:2018buk} are useful for this purpose, even though the $\cN=3$ minimal theory is not a subsector of the SO(4)--invariant sector of the full $\cN=8$ supergravity. In any case, the general embedding formulae of \cite{Guarino:2015jca,Guarino:2015vca} provide the contributions of the $\textrm{SO}(3)_\textrm{d}$ Yang-Mills gauged fields on top of the $\cN=3$ background solution as constructed in \cite{DeLuca:2018buk}. I will drop the hat on the $\textrm{SO}(3)_\textrm{L}$ index $\hat{\imath}$ in order to label $\textrm{SO}(3)_\textrm{d}$ triplets as $i=1,2,3$ (see footnote \ref{fn:IndexConvs}), and will follow the conventions of \cite{DeLuca:2018buk} for the quantities that pertain to the background geometry. 

In order to express the result, it is convenient to introduce the $S^6$ angle $\alpha$ (different from the angle $\alpha$ of section \ref{TrunctoN=2}), with range $0 \leq \alpha \leq \frac{\pi}{2}$. It is also useful to introduce constrained coordinates $\tmu^i$, $i=1,2,3$, on an $S^2$, 
\begin{eqnarray} \label{eq:S2}
\delta_{ij} \, \tmu^i \tmu^j = 1 \; , 
\end{eqnarray} 
and right-invariant one-forms $\rho^i$ on an $S^3$, subject the Maurer-Cartan equations
\begin{eqnarray} \label{eq:MC}
d\rho^i = - \tfrac12 \epsilon^i{}_{jk} \,  \rho^j \wedge \rho^k \; .
\end{eqnarray}
In all the expressions below, the right-invariant forms and the covariant derivatives of the $\tmu^i$ appear shifted with the gauge field $A^i$ of $D=4$ $\cN=3$ supergravity as
\begin{eqnarray} \label{eq:rhoshifted}
\hrho^i \equiv \rho^i - g A^i \; , 
\end{eqnarray}
and
\begin{eqnarray} \label{eq:Dmushifted}
\hD \tilde{\mu}^i \equiv  d\tilde{\mu}^i + g\,  \epsilon^i{}_{jk} A^j \tilde{\mu}^k + \epsilon^i{}_{jk} \hat{{\cal A}}^j \tilde{\mu}^k \; , \qquad  \textrm{with} \qquad 
 \hat{{\cal A}}^i =  \frac{\sin^2 \alpha }{3 + \cos 2\alpha} \, \hrho^i \; .
\end{eqnarray}

These definitions allow one to express the massive IIA uplift of minimal $D=4$ $\cN=3$ gauged supergravity (\ref{minimalN=3}). A long calculation yields

\newpage 

{\setlength\arraycolsep{0pt}
\begin{eqnarray} \label{N=3mIIembedding}
&& d \hat{s}_{10}^2 = 2^{-\frac{31}{12}} \cdot 3^{\frac38} \,g^{-\frac{1}{12}} \, m^{\frac{1}{12}} \,   \big( 3 + \cos 2\alpha  \big)^{1/8} \Big( 3 \cos^4 \alpha + 3 \cos^2 \alpha +2 \Big)^{1/4}     \,  d\bar{s}^2_4     + ds^{2}_6  \; , \nonumber  \\[12pt]
&&  \hat H_\3 =    \hat H_\3^0   - 3 \cdot 2^{-\frac53} \,g^{-\frac23} \, m^{-\frac13} \, \frac{\sin^2 \alpha \, \cos^3 \alpha }{3 \cos^4 \alpha +3 \cos^2 \alpha +2  } \hD \tmu_i \wedge   F^i  \nonumber \\[5pt]
 && \qquad + 3 \cdot 2^{-\frac53} \,g^{-\frac23} \, m^{-\frac13} \, \frac{\sin^2 \alpha \, \cos \alpha }{3+ \cos 2\alpha } \, \epsilon_{ijk} \tmu^i \hrho^j \wedge F^k -2^{-\frac53}  \cdot 3^{\frac12}  \,g^{-\frac23} \, m^{-\frac13} \, \sin\alpha \, d\alpha \wedge \tmu_i*F^i  
\nonumber \\[5pt]
 && \qquad   + 2^{-\frac53} \cdot 3^{\frac12} \,g^{-\frac23} \, m^{-\frac13} \,  \cos \alpha \,  \hD \tmu_i \wedge  * F^i  
 + 2^{-\frac53} \cdot 3^{\frac12} \,g^{-\frac23} \, m^{-\frac13} \, \frac{\sin^2 \alpha \, \cos \alpha }{3+ \cos 2\alpha } \, \epsilon_{ijk} \tmu^i \hrho^j \wedge *F^k\; ,
\nonumber \\[12pt]
&&  \hat F_\4 =   \hat F_\4^0 +   2^{-\frac{10}{3}} \cdot 3^{\frac32}   \,  g^{\frac{2}{3}} \, m^{\frac13} \, \overline{\textrm{vol}}_4   +3 \cdot 2^{-\frac{10}{3}} \, g^{-\frac73} \, m^{\frac13} \, \sin \alpha\,  \cos \alpha \, d\alpha \wedge \tmu_i \tmu_j \, \hrho^i \wedge  F^j \nonumber   \\[5pt]
 && \qquad  - 3 \cdot 2^{-\frac{10}{3}} \, g^{-\frac73} \, m^{\frac13} \, \frac{\big( 3 + \cos 2\alpha  \big) \cos^4 \alpha}{ 3 \cos^4 \alpha +3  \cos^2 \alpha +2 } \, \epsilon_{ijk} \hD \tmu^i \wedge \hD \tmu^j \wedge F^k   
\nonumber \\[5pt] && \qquad  
-3 \cdot 2^{-\frac73} \,g^{-\frac73} \, m^{\frac13} \, \frac{\sin^2 \alpha \, \cos^2 \alpha }{3+ \cos 2\alpha } \, \tmu_i \hD \tmu_j \wedge \hrho^i \wedge F^j  
\nonumber \\[5pt] && \qquad  
+3 \cdot 2^{-\frac{10}{3}} \,g^{-\frac73} \, m^{\frac13} \, \frac{\sin^4 \alpha \, \cos^2 \alpha }{ (3+ \cos 2\alpha )^2 } \, \epsilon_{ijk} \hrho^i \wedge \hrho^j \wedge F^k 
\nonumber \\[5pt] && \qquad  
-3 \cdot 2^{-\frac{13}{3}} \,g^{-\frac73} \, m^{\frac13} \, \frac{ 7+ \cos 2\alpha }{ (3+ \cos 2\alpha )^2 } \, \sin^2 \alpha \, \cos^2 \alpha  \,  \epsilon_{ijk} \tmu^i \hrho^j \wedge \hrho^k \wedge \tmu_h F^h 
\nonumber \\[5pt] && \qquad  
+ 2^{-\frac{13}{3}} \cdot 3^{\frac{1}{2}}  \,g^{-\frac73} \, m^{\frac13} \, \frac{ 7 \sin \alpha + 3 \sin 3\alpha  }{ 3+ \cos 2\alpha  } \, \cos \alpha  \, d\alpha \wedge \hrho_i \wedge * F^i
\nonumber \\[5pt]  && \qquad  
 -2^{-\frac{10}{3}} \cdot 3^{\frac{1}{2}}  \,g^{-\frac73} \, m^{\frac13} \, \frac{ 1 + 3 \cos 2\alpha  }{ 3+ \cos 2\alpha  } \, \sin \alpha \, \cos \alpha   \, d\alpha \wedge \tmu_i \tmu_j \, \hrho^i \wedge  * F^j  
\nonumber \\[5pt] && \qquad  + 2^{-\frac{7}{3}} \cdot 3^{\frac{1}{2}}  \,g^{-\frac73} \, m^{\frac13}  \, \sin \alpha \, \cos \alpha   \, d\alpha \wedge \epsilon_{ijk} \tmu^i \hD \tmu^j \wedge *F^k 
\nonumber \\[5pt] && \qquad  -2^{-\frac{10}{3}} \cdot 3^{\frac{1}{2}}  \, g^{-\frac73} \, m^{\frac13} \, \frac{\big( 3 + \cos 2\alpha  \big) \cos^2 \alpha}{ 3 \cos^4 \alpha +3  \cos^2 \alpha +2 } \,  \, \epsilon_{ijk} \hD \tmu^i \wedge \hD \tmu^j \wedge * F^k   
\nonumber \\[5pt] && \qquad  
 -2^{-\frac{7}{3}} \cdot 3^{\frac{1}{2}}  \, g^{-\frac73}  \, m^{\frac13} \, \frac{\sin^2 \alpha \, \cos^2 \alpha }{ 3+ \cos 2\alpha  } \,  \tmu_i \hD \tmu_j \wedge \hrho^i \wedge  * F^j  
\nonumber \\[5pt] && \qquad  
 -2^{-\frac{10}{3}} \cdot 3^{\frac{1}{2}}  \, g^{-\frac73}  \, m^{\frac13} \, \frac{ 3 \cos^4 \alpha +3  \cos^2 \alpha +2 }{ (3+ \cos 2\alpha )^2 } \, \sin^2\alpha \, \epsilon_{ijk} \hrho^i \wedge \hrho^j \wedge * F^k 
\nonumber \\[5pt] && \qquad  
 +2^{-\frac{13}{3}} \cdot 3^{\frac{1}{2}}  \, g^{-\frac73}  \, m^{\frac13} \,   \frac{ 5+ 3\cos 2\alpha }{ (3+ \cos 2\alpha )^2 } \, \sin^2 \alpha \, \cos^2 \alpha  \,  \epsilon_{ijk} \tmu^i \hrho^j \wedge \hrho^k \wedge \tmu_h * F^h
\; ,
\nonumber \\[12pt]
&& \hat F_\2 =    \hat F_\2^0  - 2^{\frac13} \,g^{-\frac23} \, m^{\frac23} \, \frac{\cos \alpha }{3+ \cos 2\alpha } \,  \tmu_i F^i + 2^{-\frac53} \cdot 3^{\frac12}  \,g^{-\frac23} \, m^{\frac23} \cos \alpha \,    \tmu_i *F^i    \; .
\end{eqnarray}
}Here, $F^i$ is the $D=4$ $\cN=3$ field strength (\ref{N=3FielStrengths}) and $*F^i$ its Hodge dual with respect to the four-dimensional metric $d\bar{s}^2_4$. The forms $\hat F_\4^0$, $\hat H_\3^0$ and $\hat F_\2^0$ correspond to the background values of the $\cN=3$ AdS$_4$ solution given in \cite{DeLuca:2018buk}, shifted by the $D=4$ gauge field $A^i$ as in (\ref{eq:rhoshifted}), (\ref{eq:Dmushifted}), namely,
{\setlength{\arraycolsep}{1pt}
\begin{eqnarray} \label{SO4SolN=3}
d s^2_{6} & = & L^2 \, \big( 3 + \cos 2\alpha  \big)^{1/8} \Big( 3 \cos^4 \alpha + 3 \cos^2 \alpha +2 \Big)^{1/4}  \times \nonumber  \\[4pt]
&& \qquad \quad  \Big[ \frac{2 \big( 3 + \cos 2\alpha  \big) \cos^2 \alpha}{ 3 \cos^4 \alpha +3  \cos^2 \alpha +2 } \, \delta_{ij}    \hD \tilde{\mu}^i \hD \tilde{\mu}^j  + 2 \,   d\alpha^2+ \frac{2 \sin^2 \alpha }{3 + \cos 2\alpha} \, \delta_{ij}  \hrho^i \hrho^j  \Big] ,  \nonumber \\[12pt]
\label{DilatonN=3} e^{\hat{\phi}} &=&  e^{\phi_0}  \, \frac{\big( 3 + \cos 2\alpha  \big)^{3/4}}{\big( 3 \cos^4 \alpha +3  \cos^2 \alpha +2 \big)^{1/2}} \; , \nonumber \\[12pt]
L^{-3} e^{\frac{1}{4} \phi_0} \, \hat{F}_\4^0 & = & 
 -\frac{ 4\sqrt{6} \, \big(  2 \cos^4\alpha + 3 \cos^2 \alpha +3 \big)  \sin \alpha  \cos^3 \alpha   }{  \big( 3  + \cos 2\alpha \big) \big( 3 \cos^4\alpha + 3\cos^2 \alpha +2 \big) }  \, d\alpha \wedge \epsilon_{ijk} \, \hD \tmu^i \wedge  \hD \tmu^j \wedge \hrho^k  \nonumber \\[5pt]
&& +\frac{ \sqrt{6} \,  \big( 5  + 3 \cos 2\alpha \big)   \sin^2 \alpha  \cos^2  \alpha   }{ 2 \, \big( 3 \cos^4\alpha + 3\cos^2 \alpha +2 \big) }    \, \hD \tmu_i \wedge  \hD \tmu_j \wedge \hrho^i \wedge \hrho^j   \nonumber \\[5pt]
&& -\frac{ 4 \sqrt{6} \,  \sin^5 \alpha \cos \alpha }{  \big( 3  + \cos 2\alpha \big)^2  }   \, d\alpha \wedge \tmu_i \,  \hD \tmu_j \wedge \hrho^i \wedge \hrho^j    \nonumber \\[5pt]
&&  -\frac{  2 \sqrt{2} \, \big( 5  + 3 \cos 2\alpha \big)  \sin^3 \alpha \cos \alpha}{ \sqrt{3} \, \big( 3  + \cos 2\alpha \big)^3 }   \, d\alpha \wedge  \epsilon_{ijk} \,  \hrho^i \wedge \hrho^j \wedge \hrho^k \, ,  \nonumber \\[12pt] 
L^{-2} e^{-\frac{1}{2} \phi_0} \, \hat{H}_\3^0 & = &  -\frac{ 2\sqrt{3} \, \big( 3 \cos^6\alpha+ 8 \cos^4\alpha + 11 \cos^2 \alpha +2 \big)   }{ \big( 3 \cos^4\alpha + 3\cos^2 \alpha +2 \big)^2  }   \sin \alpha  \cos^2 \alpha \, d\alpha \wedge \epsilon_{ijk} \, \tmu^i  \hD \tmu^j \wedge \hD \tmu^k \nonumber \\[5pt]
&& +\frac{ 8\sqrt{3} \, \big(  \cos^4\alpha + \cos^2 \alpha +2 \big)  \sin \alpha  \cos^2 \alpha   }{  \big( 3  + \cos 2\alpha \big) \big( 3 \cos^4\alpha + 3\cos^2 \alpha +2 \big) }   \, d\alpha \wedge \hD \tmu_i \wedge  \hrho^i  \nonumber \\[5pt]
&& +\frac{ \sqrt{3} \,  \big( 3  + \cos 2\alpha \big)   \sin^2 \alpha  \cos  \alpha   }{ 2 \, \big( 3 \cos^4\alpha + 3\cos^2 \alpha +2 \big) }    \, \epsilon_{ijk} \, \hD \tmu^i \wedge \hrho^j \wedge \hrho^k  \nonumber \\[5pt]
&& -\frac{ 2 \sqrt{3} \,  \sin^5 \alpha }{  \big( 3  + \cos 2\alpha \big)^2  }   \, d\alpha \wedge  \epsilon_{ijk} \, \tmu^i \hrho^j \wedge \hrho^k   \; ,   \nonumber \\[12pt]
L^{-1} e^{\frac{3}{4} \phi_0} \, \hat{F}_\2^0 & = &  \frac{ \sqrt{2} \,  \big( 5  + 3 \cos 2\alpha \big)  \cos^3 \alpha  }{ 4 \,   \big( 3 \cos^4\alpha + 3\cos^2 \alpha +2 \big) }  \, \epsilon_{ijk} \, \tmu^i  \hD \tmu^j \wedge \hD \tmu^k 
 +\frac{ 2 \sqrt{2} \,  \sin^2 \alpha \cos \alpha}{ 3  + \cos 2\alpha  }  \, \hD \tmu_i \wedge \hrho^i \nonumber \\[5pt]  
&& -\frac{ 4 \sqrt{2} \,  \sin^3 \alpha }{ \big( 3  + \cos 2\alpha \big)^2 }  \, d\alpha \wedge \tmu_i \, \hrho^i 
+\frac{ 3  \sin^4 \alpha \cos \alpha }{ \sqrt{2} \,  \big( 3  + \cos 2\alpha \big)^2 }  \, \epsilon_{ijk} \, \tmu^i  \hrho^j \wedge \hrho^k \; , \nonumber \\[12pt]
L \, e^{\frac{5}{4} \phi_0} \, \hat{F}_\0 & = &  \frac{ \sqrt{3}}{2\sqrt{2}} \;.
\end{eqnarray}
Again, the dilaton $e^{\hat \phi}$ and the Romans mass $\hat F_\0$ in (\ref{SO4SolN=3}) correspond to their exact background values \cite{DeLuca:2018buk}. Like in the previous case, the constants $g$ and $m$ that appear in (\ref{N=3mIIembedding}) should be replaced with the constants $L$ and  $e^{\phi_0} $ that characterise the background geometry where, now, $L^2 \equiv 2^{-\frac{31}{12}} \, 3^{\frac{3}{8}} \, g^{-\frac{25}{12}} \,  m^{\frac{1}{12}}$ and $e^{\phi_0} \equiv  2^{-\frac{1}{6}} \, 3^{\frac{1}{4}} \,  g^{\frac{5}{6}} m^{-\frac{5}{6}}$  \cite{DeLuca:2018buk}. 

By the consistency of the general $\cN=8$ truncation of massive IIA on $S^6$ \cite{Guarino:2015jca,Guarino:2015vca}, and the consistency of the further truncation within $D=4$ discussed in sections \ref{sec:N=3intoN=8} and \ref{EqsfromN=8}, the particular subtruncation (\ref{N=3mIIembedding}), (\ref{SO4SolN=3})  must also be consistent. As a further check, I have verified that this is indeed the case at the level of the IIA form field equations: the type IIA configuration (\ref{N=3mIIembedding}), (\ref{SO4SolN=3}) solves the field equations of massive type IIA supergravity when those, (\ref{minimalN=3eoms}), of minimal $D=4$ $\cN=3$ gauged supergravity are imposed. See appendix \ref{app:N=3truncIIA} for the details. Moreover, using the conditions (\ref{N=3toN=2}) under which the minimal $\cN=3$ theory truncates into the $\cN=2$ one, equations (\ref{N=3mIIembedding}), (\ref{SO4SolN=3}) provide a second consistent truncation of massive IIA to minimal $D=4$ $\cN=2$ supergravity (\ref{minimalN=2}). This $\cN=2$ truncation is different from that defined by (\ref{SU3U1lIIASolution}), (\ref{SU3U1lIIASolutionBackground}).

\section{Discussion} \label{sec:discussion}

Consistent truncations have been presented to the pure $\cN=2$ and $\cN=3$ gauged supergravities in $D=4$, both from a larger theory also in $D=4$, and from ten-dimensional massive type IIA supergravity. Similar two-step truncations, from string/M-theory down to a matter-coupled (or even maximal) $D$-dimensional gauged supergravity, followed by a further truncation to a pure gauged supergravity have been constructed in \cite{Gauntlett:2009zw,Cassani:2010uw,Gauntlett:2010vu,Liu:2010sa,Cassani:2011fu,Cassani:2012pj,Donos:2010ax,Malek:2019ucd,Cheung:2019pge,Larios:2019kbw}. The strategy to strip off the matter couplings of a $D$-dimensional supergravity and truncate it to the gravity multiplet consists in fixing the scalars to their vevs at a suitable vacuum, and truncating out the massive gauge fields at that vacuum (possibly along with some massless gauge fields as well). 

In this sense, the truncations constructed in this paper are in the same spirit than the minimal truncation constructed recently in section 3.3 of \cite{Larios:2019kbw}. In that reference, minimal $D=4$ $\cN=2$ gauged supergravity was embedded into $D=4$ $\cN=8$ SO(8)--gauged supergravity \cite{deWit:1982ig} around its $\cN=2$ vacuum \cite{Warner:1983vz}, and then uplifted \cite{deWit:1986iy,Varela:2015ywx} to $D=11$ supergravity \cite{Cremmer:1978km} on the $\cN=2$ solution of \cite{Corrado:2001nv}. Unlike in \cite{Larios:2019kbw}, in the present case some of the massive vector fields to be truncated out gauge dyonically shift symmetries of St\"uckelberg scalars. As shown in section \ref{sec:minISO7truncs}, the procedure to switch those off is not simply to set them to zero. Instead, the vector duality relations (\ref{VectorDualityRelations}) must be employed to write these gauge fields in terms of the surviving R-symmetry gauge fields and their Hodge duals. Essentially the same process was performed in {\it e.g.}~section 6.1 of \cite{Donos:2010ax}, in a symplectic frame where the St\"uckelberg scalars there appeared dualised into tensors. 

These subtruncations of larger theories to pure supergravities do not follow from any obvious symmetry principles. Thus, their consistency must be checked at the level of the field equations. This has been done in section \ref{sec:minISO7truncs} for the truncations presented in this paper. In contrast, the intermediate $\cN=4$ subsector of the $\cN=8$ ISO(7) theory, that was used for convenience, does arise from the latter as a singlet sector truncation. This $\cN=4$ theory corresponds to the sector of the $\cN=8$ supergravity that is invariant under the SU(2) or $\textrm{SO}(3)_\textrm{R}$ in (\ref{embedding1}) or (\ref{embedding2}) \cite{Guarino:2019jef}.

Using the general $\cN=8$ consistent truncation formulae of \cite{Guarino:2015jca,Guarino:2015vca}, the minimal $\cN=2$ and $\cN=3$ theories have been uplifted to ten dimensions. In this way, consistent truncations have been constructed of massive type IIA supergravity on the internal spaces of the $\cN=2$ and $\cN=3$ AdS$_4$ solutions of \cite{Guarino:2015jca} and \cite{Pang:2015vna,DeLuca:2018buk}, to the minimal $D=4$ $\cN=2$ and $\cN=3$ gauged supergravities. The $\cN=2$ truncation formulae of section \ref{TrunctoN=2} are straightforward extensions of the formulae given in \cite{Azzurli:2017kxo} for the IIA embedding of a particular $D=4$ black hole. In section \ref{TrunctoN=2}, the only restriction on the $D=4$ $\cN=2$ fields is that they obey the minimal supergravity field equations (\ref{minimalN=2eoms}). Thus, not only the black hole considered in \cite{Azzurli:2017kxo} uplifts to $D=10$, but in fact any other solution of minimal $\cN=2$ supergravity does through those consistent embedding formulae. For example, the SU(3)--invariant, supersymmetric Reissner-Nordstr\"om black hole with constant scalars discussed in \cite{Guarino:2017eag,Hosseini:2017fjo} can be more economically regarded as a solution of minimal $\cN=2$ gauged supergravity. Also this black hole uplifts to IIA using the same formulae in \cite{Azzurli:2017kxo} and section \ref{TrunctoN=2} here, as any other solution, supersymmetric or otherwise, of minimal $D=4$ $\cN=2$ supergravity does. Explicit calculations in related contexts and in \cite{Azzurli:2017kxo} for their specific black hole, make supersymmetry expected to be preserved in general by the uplifting process. Similar statements apply to the $\cN=3$ uplift. The supersymmetric solutions to minimal $D=4$ $\cN=2$ gauged supergravity have been classified in \cite{Caldarelli:2003pb}.

In addition to the IIA uplift of $\cN=2$ supergravity given in section \ref{TrunctoN=2}, an alternative uplift of this theory can be given. This follows from the results of section \ref{TrunctoN=3} simply by bringing to that section the restrictions (\ref{N=3toN=2}) for the further truncation of the $D=4$ $\cN=3$ theory to $\cN=2$. While various consistent truncations of string or M-theory are known to $D=4$ $\cN=2$ pure gauged supergravity \cite{Pope:1985jg,Gauntlett:2007ma,Larios:2019kbw,Larios:2019lxq}, truncations to the pure $\cN=3$ gauged  theory are less common. One such truncation has been constructed, from $D=11$, in \cite{Pope:1985bu} (see also \cite{Cassani:2011fu}). 

The internal spaces corresponding to the IIA truncations of section \ref{sec:minIIAtruncs} correspond to smooth geometries on topological six-spheres. Some (singular) generalisations for the background geometries can be easily engineered \cite{Guarino:2015jca,DeLuca:2018buk} (see also \cite{Fluder:2015eoa}), that are carried over to the corresponding minimal truncations. For example, the $\cN=2$ truncation of section \ref{TrunctoN=2} is still valid if $\mathbb{CP}^2$ is replaced, along with its related quantities $\bm{J}$ and $\sigma$, with any local K\"ahler-Einstein four-dimensional space of positive curvature. Similarly, the $\cN=3$ truncation of section \ref{TrunctoN=3} still holds if the $S^3$ on which the $\rho^i$ take values is replaced with the lens space $S^3/\mathbb{Z}_p$, with the discrete identification acting on the Hopf fiber. 

In any case, the results of this paper deliver the consistent truncations to minimal gauged supergravities envisaged in general in  \cite{Gauntlett:2007ma,Cassani:2019vcl}, corresponding to the supersymmetric AdS$_4$ solutions of massive type IIA supergravity constructed in \cite{Guarino:2015jca} and \cite{Pang:2015vna,DeLuca:2018buk}.


\section*{Acknowledgements}

Mirjam Cveti\v c, Adolfo Guarino, Gabriel Larios, Praxitelis Ntokos, Tom\'as Ort\'\i n, Chris Pope, Mar\'\i a J.~Rodr\'\i guez and Javier Tarr\'\i o are kindly thanked for discussions. This work is sup\-por\-ted by the NSF grant PHY-1720364 and, partially, by grants SEV-2016-0597, FPA2015-65480-P and PGC2018-095976-B-C21 from MCIU/AEI/FEDER, UE.



\appendix

\addtocontents{toc}{\setcounter{tocdepth}{1}}


\section{Minimal $\cN=2$ and $\cN=3$ gauged supergravities} 


\subsection{Conventions} \label{sec:N=2N=3Conv}

The bosonic sector of minimal $D=4$ $\cN=2$ gauged supergravity \cite{Fradkin:1976xz,Freedman:1976aw} includes the metric\footnote{The metrics $\bar{g}_{\mu\nu}$ of the minimal $\cN=2$ and $\cN=3$ theories are denoted with bars, as they are related to the metric $g_{\mu\nu}$ of the parent $\cN=8$ theory (and the intermediate $\cN=4$ theory of section \ref{sec:N=4subsec}) by the constant rescaling (\ref{metricrescale}). Unlike in \cite{Larios:2019kbw}, bars are omitted for simplicity in the $\cN=2$ graviphoton $F$ and its Hodge dual $*F$ with respect to $\bar{g}_{\mu\nu}$, and similarly for $F^i$ and $*F^i$ in the $\cN=3$ case.} $\bar{g}_{\mu\nu}$, with line element $d\bar{s}_4^2$, and a gauge field $A$ with field strength $F = d A$, subject to the field equations
\begin{equation} \label{minimalN=2eoms}
d F = 0 \; , \qquad 
d * F = 0 \; , \qquad 
\bar{R}_{\mu\nu} = -3 g^2 \bar{g}_{\mu\nu} + \tfrac12 \big( F_{\mu \sigma} F_\nu{}^\sigma -\tfrac14 \bar{g}_{\mu\nu} \, F_{\rho \sigma} F^{\rho \sigma}  \big) \; .
\end{equation}
The latter two derive from the Einstein-Maxwell Lagrangian with a negative cosmological constant $-6 g^2$,
\begin{equation} \label{minimalN=2}
{\cal L}  =  \bar{R} \, \overline{\textrm{vol}}_4 -\tfrac12 \, F \wedge * F + 6 g^2 \, \overline{\textrm{vol}}_4 \; .
\end{equation}
The AdS vacuum is attained for $d\bar{s}^2_4 = g^{-2} \,  ds^2 (\textrm{AdS}_4)$, with $ds^2 (\textrm{AdS}_4)$ the unit-radius anti-de Sitter metric, and $A=0$.

The bosonic field content of minimal $D=4$ $\cN=3$ gauged supergravity \cite{Freedman:1976aw} contains the metric $\bar{g}_{\mu\nu}$, with line element $d\bar{s}_4^2$, and SO(3) gauge fields $A^i$ with field strengths 
\begin{equation} \label{N=3FielStrengths}
F^i = d A^{i } + \tfrac12 \, g \epsilon^i{}_{jk} \, A^{ j } \wedge A^{ k } \; .
\end{equation}
These fields obey the field equations
\begin{equation} \label{minimalN=3eoms}
D F^i = 0 \; , \qquad 
D*F^i = 0 \; , \qquad 
\bar{R}_{\mu\nu} = -3 g^2 \bar{g}_{\mu\nu} + 2 \big( F^i_{\mu \sigma} F_{i \, \nu}{}^\sigma -\tfrac14 \bar{g}_{\mu\nu} \, F^i_{\rho \sigma} F_i^{\rho \sigma}  \big) \; ,
\end{equation}
where the SO(3) indices $i=1,2,3$ are raised and lowered with $\delta_{ij}$, and the covariant derivatives are
\begin{equation} \label{N=3FielStrengthsBianchis}
D F^i \equiv d F^{i } + g \epsilon^i{}_{jk} \, A^{ j } \wedge F^{ k }  \; , \qquad 
D * F^i \equiv d * F^{i } + g \epsilon^i{}_{jk} \, A^{ j } \wedge * F^{ k } \; .
\end{equation}
The equations of motion (the latter two equations in (\ref{minimalN=3eoms})) derive from the Einstein-Yang-Mills Lagrangian with a cosmological constant $ -6 g^2$,
\begin{equation} \label{minimalN=3}
{\cal L}  =  \bar{R} \, \overline{\textrm{vol}}_4 -2 \, F^i \wedge * F_i + 6 g^2 \, \overline{\textrm{vol}}_4 \; .
\end{equation}
The AdS vacuum is attained for $d\bar{s}^2_4 = g^{-2} \,  ds^2 (\textrm{AdS}_4)$, where $ds^2 (\textrm{AdS}_4)$ is again the unit-radius anti-de Sitter metric, and $A^i=0$. The $\cN=3$ theory (\ref{minimalN=3}) can be consistently truncated to the  $\cN=2$ theory (\ref{minimalN=2}) through the identifications
\begin{equation} \label{N=3toN=2}
A^1 = A^2 = 0 \; , \qquad  A \equiv 2 A^3 \; .
\end{equation}

\subsection{SO(7) embedding of the $\cN=2$ and $\cN=3$ gauge groups} \label{sec:GaugeGroupEmbedding}

It is interesting to determine the generators of SO(7), and ultimately of $\textrm{SL}(8) \subset \textrm{E}_{7(7)}$, corresponding to the U(1) and $\textrm{SO}(3)_\textrm{d}$ subgroups that are gauged in the $\cN=2$ and $\cN=3$ minimal models. This U(1) is the commutant of the SU(3) of (\ref{embedding1}) inside SO(7). The group $\textrm{SO}(3)_\textrm{d}$ is the one that appears in (\ref{embedding2}), and its generators have already been given in (A.5) of \cite{Guarino:2019jef}. In any case, the relevant generators can be found by bringing (\ref{ElectricVectorsinN=8}), to the general expression for the ISO(7) gauge fields coupled to $\textrm{SL}(8)$ generators \cite{Guarino:2015qaa}. If $t_A{}^B$, $A=1, \ldots, 8$, are the generators of the $\textrm{SL}(8)$ subalgebra of $\textrm{E}_{7(7)}$, then SO(7) is generated by $T_{IJ} \, \equiv 2  \, t_{[I}{}^K \delta_{J]K} $, with $A = (I, 8)$, $I= 1, \ldots, 7$. The U(1) and $\textrm{SO}(3)_\textrm{d}$ R-symmetry groups of the minimal $\cN=2$ and $\cN=3$ supergravities turn out to be generated by the generators $T$ and $T_k$, $k=1,2,3$, of SO(7) given by:
\begin{equation}
\textrm{U}(1) \, : \; T \equiv \epsilon_3{}^{ij} \, T_{ij} + (J_{\hat{3}-})^{ab} \, T_{ab} \; , \qquad \quad 
\textrm{SO}(3)_\textrm{d} \, : \; T_k \equiv \epsilon_k{}^{ij} \, T_{ij} -\tfrac12 \, \delta^{\hat{\imath}}_k \,   (J_{\hat{\imath} - })^{ab} \, T_{ab} \; ,
\end{equation}
with the index $I=(i, a)$ split as below (\ref{MagneticVectorsinN=8}) (see also footnone \ref{fn:IndexConvs}), and the antisymmetric, anti-selfdual matrices $(J_{\hat{\imath} - })^{ab}$ as defined in appendix A of \cite{Guarino:2019jef}. In particular this U(1) is different from the U(1) Cartan subgroup of $\textrm{SO}(3)_\textrm{d}$, since $T \neq T_3$.

\section{Consistency proof for the IIA truncations} \label{app:MinimalintoIIA}

The truncations of massive type IIA down to minimal $\cN=2$ and $\cN=3$ gauged supergravities given in the main text are consistent by construction. First of all, the truncation within $D=4$ from $\cN=8$ to $\cN=2$ and $\cN=3$ was checked to be consistent at the level of the $D=4$ bosonic equations of motion, including Einstein, in section \ref{sec:minISO7truncs}. Then, the minimal subsectors were uplifted using the general consistent truncation formulae of \cite{Guarino:2015jca,Guarino:2015vca,Varela:2015uca} in section \ref{sec:minIIAtruncs}. Nevertheless, it does not hurt to check explicitly the consistency of the $D=10$ to $D=4$ truncations of section \ref{sec:minIIAtruncs}. Here I present a lengthy, but worthwhile, explicit consistency proof at the level of the bosonic field equations of the type IIA supergravity forms. Short of explicitly checking the Einstein equation, these results confirm the consistency of the minimal $D=4$ truncations of type IIA presented in the main text.

\subsection{Truncation to minimal $\cN=2$ supergravity} \label{app:IIAtoN=2}

The massive type IIA configuration (\ref{SU3U1lIIASolution}), (\ref{SU3U1lIIASolutionBackground}) has the local form
{\setlength\arraycolsep{0pt}
\begin{eqnarray} \label{SU3IIAconfig}
&& d\hat{s}_{10}^2 = e^{2X(\alpha)} d\bar{s}^2_4 + e^{2A(\alpha)} d\alpha^2 + e^{2B(\alpha)} ds^2 (\mathbb{CP}_2) + e^{2C(\alpha)} \bm{\hat{\eta}}^2 \; , 
 \qquad \hat \phi = \phi(\alpha) \; ,  \nonumber \\[8pt]
&& \hat F_\4 = \mu_0 \overline{\textrm{vol}}_4 + A_4(\alpha) \,  \textrm{vol} (\mathbb{CP}_2) +B_4(\alpha) \, \bm{J} \wedge d\alpha \wedge  \bm{\hat{\eta}}
\nonumber \\
&& \qquad \; + \big( P (\alpha) \, \bm{J} + Q(\alpha) \,  d\alpha \wedge  \bm{\hat{\eta}}  \big) \wedge F + \big( S (\alpha) \, \bm{J} + T(\alpha) \,  d\alpha \wedge  \bm{\hat{\eta}}  \big) \wedge *F   \; ,   \nonumber \\[8pt]
&& \hat H_\3 = B_3(\alpha) \, \bm{J} \wedge d\alpha + N_3(\alpha) *F \wedge d\alpha  \; ,  \nonumber \\[8pt]
&& \hat F_\2 = A_2(\alpha)  \, \bm{J }  +B_2(\alpha) \, d\alpha \wedge  \bm{\hat{\eta}}  + M_2 (\alpha) \, F + N_2 (\alpha) \, * F  \; ,
\end{eqnarray}
}with $\bm{\hat{\eta}} = \bm{\eta} +\tfrac13 g A = d\psi + \sigma + \tfrac13 g A$. In these expressions, $d\bar{s}_4^2$ and $F= dA$ are the fields of minimal $\cN=2$ gauged supergravity (\ref{minimalN=2}), $\mu_0$ is a constant, $X(\alpha)$, etc., are functions of the angle $\alpha$ that can be read off from (\ref{SU3U1lIIASolution}), (\ref{SU3U1lIIASolutionBackground}), and $\bm{J} = \frac12 d\sigma$ is the K\"ahler form on $\mathbb{CP}_2$. This configuration, with the explicit functions $X(\alpha)$, etc. that can be read off from  (\ref{SU3U1lIIASolution}), (\ref{SU3U1lIIASolutionBackground}) of the main text, was obtained in section \ref{TrunctoN=2} by consistent uplift using the formulae of \cite{Guarino:2015jca,Guarino:2015vca,Varela:2015uca}. Here, I will explicitly verify at the level of the field equations of the ten-dimensional form fields that these expressions do indeed define a consistent truncation of massive IIA supergravity down to minimal $\cN=2$ gauged supergravity in four dimensions. 

Imposing the Bianchi identities of massive type IIA, (A.4) of \cite{Guarino:2015vca} in our conventions, on the configuration (\ref{SU3IIAconfig}), and using the Bianchi identity, $d F = 0 $, and Maxwell equation, $d*F=0$, of the $D=4$ graviphoton, the relations (B.4) of \cite{Varela:2015uca} (with $C_4 = D_4 = C_3 = D_3 = E_3 = F_3 = C_2 = D_2 = 0$ there) are recovered, together with 
{\setlength\arraycolsep{0pt}
\begin{eqnarray} \label{SU3IIABianchis}
&& P^\prime - g B_4 - M_2 B_3 -2 Q  =0 \;, \nonumber \\
&& S^\prime - A_2N_3 -B_3 N_2 -2 T=0 \;, \nonumber \\
&& gQ-N_2 N_3  = 0 \; ,  \nonumber \\
&& g T + M_2 N_3  = 0 \; , \nonumber \\
&& M_2^\prime - gB_2 = 0  \; , \nonumber \\
&& N_2^\prime - m N_3 = 0  \; .
\end{eqnarray}
}Here and subsequently, a prime denotes derivative with respect to $\alpha$, and the explicit $\alpha$ dependence is suppressed from the functions. These and the following relations are obtained by assuming that $F$, $*F$, $F \wedge F$ and $F \wedge *F$ are non-vanishing and independent.  

Next, I turn to check the equations of motion of the type IIA form fields, collected in (A.5) of \cite{Guarino:2015vca}. The $\hat F_\4$ equation of motion gives (B.5) of \cite{Varela:2015uca} along with
{\setlength\arraycolsep{0pt}
\begin{eqnarray} \label{SU3IIAF4eom}
&&  \big( e^{ \frac12 \phi -A +4B -C }  Q \big)^\prime - 4  \, e^{ \frac12 \phi + A +C }  P + 2 B_3 S +N_3 A_4  =0 \;, \nonumber \\
&&  \big( e^{ \frac12 \phi -A +4B -C }  T \big)^\prime - 4  \, e^{ \frac12 \phi + A +C }  S  - 2 P B_3 - g  \mu_0 e^{ \frac12 \phi -4X +A +4B +C }   =0 \;, \nonumber \\
&&  N_3 = g \,  e^{ \frac12 \phi +A +C }    \; ;
\end{eqnarray}
}the $\hat H_\3$ equation of motion gives (B.6) of \cite{Varela:2015uca} and
{\setlength\arraycolsep{0pt}
\begin{eqnarray} \label{SU3IIAH3eom}
&&  \big( e^{ - \phi -A+4B+C }  N_3 \big)^\prime -  e^{ \frac12 \phi - A +4B-C }  B_2 T -  2 e^{ \frac12 \phi + A +C }  S A_2 + 2 P B_4 +A_4 Q  \nonumber \\
&& \qquad \qquad  - m  \, e^{ \frac32 \phi +A +4B+C } N_2  - \mu_0 \, e^{ \frac12 \phi -4X +A +4B+C } M_2   =0 \;, \nonumber \\[10pt]
&&    e^{ \frac12 \phi - A +4B-C }  B_2  Q +  2 e^{ \frac12 \phi + A +C }  P A_2 + 2 S B_4 +A_4 T  \nonumber \\
&& \qquad \qquad  + m  \, e^{ \frac32 \phi +A +4B+C } M_2  - \mu_0 \, e^{ \frac12 \phi -4X +A +4B+C } N_2   =0 \;, \nonumber \\[10pt]
&&  g\, e^{ - \phi -A+4B+C }  N_3  -  e^{ \frac12 \phi - A +4B-C }  \big( M_2 T + N_2 Q \big) +P^2 - S^2    =0 \;, \nonumber \\[10pt]
&&   e^{ \frac12 \phi + A +C } \big(  M_2 S +   N_2 P \big)  -PQ+ST = 0    \;, \nonumber \\[10pt]
&&   e^{ \frac12 \phi - A +4B -C }  \big( M_2 Q -   N_2 T \big)  +2PS  = 0    \;, \nonumber \\[10pt]
&&   e^{ \frac12 \phi + A +C }  \big( M_2 P -  N_2 S \big) +PT+SQ  = 0    \; ;
\end{eqnarray}
}the $\hat F_\2$ equation of motion gives (B.7) of \cite{Varela:2015uca} together with
{\setlength\arraycolsep{0pt}
\begin{eqnarray} \label{SU3IIAF2eom}
&& g \, e^{ \frac32 \phi + A +C }  M_2  +  e^{ \frac12 \phi -A -C }  N_3 T = 0  \; ,   \nonumber \\[10pt]
&&  g \, e^{ \frac32 \phi + A +C }  N_2  -  e^{ \frac12 \phi -A -C }  N_3 Q = 0   \; ;
\end{eqnarray}
}and the dilaton equation of motion gives (B.8) of \cite{Varela:2015uca} along with
{\setlength\arraycolsep{0pt}
\begin{eqnarray} \label{SU3IIADilaton}
&&  \tfrac32 \,  e^{ \frac32 \phi + A +4B +C } \big( M_2^2 - N_2^2 \big)    +  e^{ -\phi - A +4B +C } N_3^2 +  e^{ \frac12 \phi + A +C } \big( P^2 - S^2  \big)  \nonumber \\[5pt]
&&  \qquad \qquad  + \tfrac12  e^{ \frac12 \phi - A +4B-C } \big( Q^2 - T^2  \big)   = 0 \; ,
 \nonumber \\[10pt]
&& \tfrac32 \,  e^{ \frac32 \phi + A +4B +C } M_2 N_2   +  e^{ \frac12 \phi + A +C } P S + \tfrac12  e^{ \frac12 \phi - A +4B -C } QT   = 0 \; . 
\end{eqnarray}
}

\vspace{-10pt}

Equations (\ref{SU3IIABianchis})--(\ref{SU3IIADilaton}) can be shown to be identically satisfied with the functions that can be read off from (\ref{N=3mIIembedding}), (\ref{SO4SolN=3}) of the main text. This shows that the massive IIA field equations are satisfied on the field equations of $D=4$ $\cN=2$ minimal gauged supergravity (\ref{minimalN=2}). Up to a check of the $D=10$ Einstein equation, the truncation is thus consistent.

\subsection{Truncation to minimal $\cN=3$ supergravity} \label{app:N=3truncIIA}

The massive type IIA configuration (\ref{N=3mIIembedding}), (\ref{SO4SolN=3}) is of the (also massive) local form 
{\setlength\arraycolsep{0pt}
\begin{eqnarray} \label{SO4IIAconfig}
&& d\hat{s}_{10}^2 = e^{2X(\alpha)} d\bar{s}^2_4   + e^{2B(\alpha)} \, \delta_{ij}    \hD \tilde{\mu}^i \hD \tilde{\mu}^j  + e^{2A(\alpha)} d\alpha^2  + \tfrac14 \, e^{2C(\alpha)} \, \delta_{ij} \hrho^i \hrho^j  \; , 
 \qquad \hat \phi = \phi(\alpha) \; ,  \nonumber \\[10pt]
&& \hat F_\4 = \mu_0 \overline{\textrm{vol}}_4 
+C_1(\alpha)  \, d\alpha \wedge \epsilon_{ijk} \, \hD \tmu^i \wedge  \hD \tmu^j \wedge \hrho^k 
+C_2(\alpha)    \, \hD \tmu_i \wedge  \hD \tmu_j \wedge \hrho^i \wedge \hrho^j   \nonumber \\[5pt]
&&\qquad \; +C_3(\alpha)  \, d\alpha \wedge \tmu_i \,  \hD \tmu_j \wedge \hrho^i \wedge \hrho^j   
+C_4(\alpha)    \, d\alpha \wedge  \epsilon_{ijk} \,  \hrho^i \wedge \hrho^j \wedge \hrho^k \,   \nonumber \\[5pt]
&& \qquad \; + M_4 ( \alpha ) \, d\alpha \wedge \hrho_i \wedge  F^i
+ N_4 ( \alpha ) \, d\alpha \wedge \tmu_i \tmu_j  \, \hrho^i \wedge  F^j  
+ P_4 ( \alpha ) \, d\alpha \wedge \epsilon_{ijk} \tmu^i \hD \tmu^j \wedge F^k \nonumber   \\[5pt]
&& \qquad  \; + Q_4 ( \alpha ) \, \epsilon_{ijk} \hD \tmu^i \wedge \hD \tmu^j \wedge F^k   
+ R_4 ( \alpha )  \, \tmu_i \hD \tmu_j \wedge \hrho^i \wedge F^j  \nonumber \\[5pt]
 && \qquad  
+ T_4 ( \alpha )   \, \epsilon_{ijk} \hrho^i \wedge \hrho^j \wedge F^k  
+ U_4 ( \alpha )     \,  \epsilon_{ijk} \tmu^i \hrho^j \wedge \hrho^k \wedge \tmu_h F^h \nonumber  \\[5pt]
&& \qquad \; + \overline{M}_4 ( \alpha ) \, d\alpha \wedge \hrho_i \wedge  * F^i
+ \overline{N}_4 ( \alpha ) \, d\alpha \wedge \tmu_i \tmu_j  \, \hrho^i \wedge  * F^j  
+ \overline{P}_4 ( \alpha ) \, d\alpha \wedge \epsilon_{ijk} \tmu^i \hD \tmu^j \wedge * F^k \nonumber   \\[5pt]
&& \qquad  \; + \overline{Q}_4 ( \alpha ) \, \epsilon_{ijk} \hD \tmu^i \wedge \hD \tmu^j \wedge * F^k   
+ \overline{R}_4 ( \alpha )  \, \tmu_i \hD \tmu_j \wedge \hrho^i \wedge * F^j  \nonumber \\[5pt]
 && \qquad  
+ \overline{T}_4 ( \alpha )   \, \epsilon_{ijk} \hrho^i \wedge \hrho^j \wedge * F^k  
+ \overline{U}_4 ( \alpha )     \,  \epsilon_{ijk} \tmu^i \hrho^j \wedge \hrho^k \wedge \tmu_h *  F^h \; , \nonumber  \\[10pt]
&& \hat H_\3 =  B_1(\alpha)   \, d\alpha \wedge \epsilon_{ijk} \, \tmu^i  \hD \tmu^j \wedge \hD \tmu^k 
 +B_2(\alpha)  \, d\alpha \wedge \hD \tmu_i \wedge  \hrho^i  +B_3(\alpha)    \, \epsilon_{ijk} \, \hD \tmu^i \wedge \hrho^j \wedge \hrho^k  \nonumber \\[5pt]
&& \qquad \;  + B_4 (\alpha)   \, d\alpha \wedge  \epsilon_{ijk} \, \tmu^i \hrho^j \wedge \hrho^k \nonumber \\[5pt]
&& \qquad \;  
+  M_3 (\alpha) \, d\alpha \wedge \tmu_i F^i  
+  P_3 (\alpha) \, \hD \tmu_i \wedge F^i  
+  R_3 (\alpha) \,  \epsilon_{ijk} \, \tmu^i \hrho^j \wedge F^k
\nonumber \\[5pt]
&& \qquad \;  
+  \overline{M}_3 (\alpha) \, d\alpha \wedge \tmu_i *F^i  
+  \overline{P}_3 (\alpha) \, \hD \tmu_i \wedge *F^i  
+  \overline{R}_3 (\alpha) \,  \epsilon_{ijk} \, \tmu^i \hrho^j \wedge *F^k
  \; ,   \nonumber \\[10pt]
&& \hat F_\2 =  A_1 (\alpha)  \, \epsilon_{ijk} \, \tmu^i  \hD \tmu^j \wedge \hD \tmu^k 
+A_2 (\alpha)   \, \hD \tmu_i \wedge \hrho^i 
+A_3 (\alpha)    \, d\alpha \wedge \tmu_i \, \hrho^i 
+A_4 (\alpha)   \, \epsilon_{ijk} \, \tmu^i  \hrho^j \wedge \hrho^k  \nonumber \\[5pt]
&& \qquad + M_2 (\alpha) \, \tmu_i F^i + \overline{M}_2 (\alpha) \, \tmu_i * F^i \; , 
\end{eqnarray}
}where $d\bar{s}^2_4$ and $F^i$ are the fields of $D=4$ $\cN=3$ gauged supergravity (\ref{minimalN=3}), $\mu_0$ is a constant, $X(\alpha)$, etc., are all real functions of the angle $\alpha$, the $\tmu^i$, $i=1,2,3$, are constrained coordinates that define a unit radius $S^2$ through (\ref{eq:S2}), and $\hrho^i$ correspond to the right-invariant Maurer-Cartan forms on $S^3$, shifted by the $D=4$ $\cN=3$ SO(3) gauge field $A^i$, as in (\ref{eq:rhoshifted}). In this appendix, I will denote the covariant derivative of $\tmu^i$ for convenience as
\begin{eqnarray} \label{covderapp}
\hD \tilde{\mu}^i =  d\tilde{\mu}^i + g \epsilon^i{}_{jk} A^j \tilde{\mu}^k + \epsilon^i{}_{jk} \hat{{\cal A}}^j \tilde{\mu}^k  \; , \qquad  \textrm{with} \qquad 
\hat{{\cal A}}^i =  A_0 (\alpha)  \, \hrho^i \; .
\end{eqnarray}
The function $A_0 (\alpha)$, as well as all other functions of $\alpha$ in (\ref{SO4IIAconfig}), can be read off from the concrete expressions given section \ref{TrunctoN=3} of the main text. It is worth emphasising that all of the above functions are real. Bars have been used as a mere notational device, in order to neutralise the real threat of running out of symbols. For example, $M_2 (\alpha)$ and $\overline{M}_2 (\alpha)$, etc., are real and unrelated. 

Let us now find the equations that these functions must obey for the configuration (\ref{SO4IIAconfig}), (\ref{covderapp}) to solve the massive type IIA field equations, assuming that the $D=4$ Yang-Mills field strength $F^i$ is covariantly closed and co-closed as dictated by its Bianchi identity and equation of motion, (\ref{minimalN=3eoms}). For that purpose, one needs to use the identities recorded in equation (A.10) of \cite{DeLuca:2018buk} as well as the following ones involving the $D=4$ gauge field strength:
\begin{eqnarray} \label{UsefulIds}
&& \tmu_h \hrho^h \wedge \epsilon_{ijk} \tmu^i \hrho^j \wedge F^k = \tfrac12 \epsilon_{ijk} \hrho^i \wedge \hrho^j \wedge F^k -\tfrac12 \epsilon_{ijk} \tmu^i \hrho^j \wedge \hrho^k \wedge \tmu_h F^h \; , \nonumber  \\[5pt]
&& \epsilon_{ijk} \tmu^i \hD \tmu^j \wedge \hD \tmu^k \wedge \tmu_h F^h =  \epsilon_{ijk} \hD \tmu^i \wedge \hD \tmu^j \wedge F^k \; , \nonumber  \\[5pt]
&& \hD \tmu_h \wedge \hrho^h \wedge \epsilon_{ijk} \tmu^i \hrho^j \wedge F^k = -\tfrac12 \epsilon_{ijk} \tmu^i \hrho^j \wedge \hrho^k \wedge \hD \tmu_h \wedge  F^h \; , \nonumber  \\[5pt]
&&  \tmu_h  \hrho^h \wedge \epsilon_{ijk} \hD \tmu^i \wedge \hrho^j \wedge F^k = \tfrac12 \epsilon_{ijk} \hD \tmu^i \wedge \hrho^j \wedge \hrho^k \wedge \tmu_h   F^h \; , \nonumber  \\[5pt]
&& \epsilon_{ijk} \tmu^i \hD \tmu^j \wedge \hD \tmu^k \wedge \epsilon_{hlm} \tmu^h \hrho^l \wedge F^m  = 2 \hD \tmu_i \wedge \hD \tmu_j \wedge \hrho^i \wedge F^j \; , \nonumber  \\[5pt]
&& \hrho_h \wedge \epsilon_{ijk} \tmu^i \hrho^j \wedge F^k \wedge F^h  = \tfrac12 \epsilon_{ijk} \hrho^i \wedge \hrho^j \wedge F^k \wedge F^h \tmu_h -\tfrac12 \epsilon_{ijk} \tmu^i  \hrho^j \wedge \hrho^k \wedge F^h \wedge F_h  \; , \nonumber  \\[5pt]
&& \tmu_m \hrho^m \wedge \epsilon_{ijk} \tmu^i \hrho^j \wedge F^k \wedge F^h \tmu_h = \tfrac12 \epsilon_{ijk} \hrho^i \wedge \hrho^j \wedge F^k \wedge F^h \tmu_h \nonumber \\
&& \qquad\qquad\qquad\qquad\qquad\qquad\qquad -\tfrac12 \epsilon_{ijk} \tmu^i  \hrho^j \wedge \hrho^k \wedge F^m \wedge F^n \tmu_m \tmu_n   \; , \nonumber  \\[5pt]
&& \hD \tmu^h \wedge \epsilon_{ijk} \tmu^i \hD \tmu^j \wedge F^k \wedge F_h = \tfrac12 \epsilon_{ijk} \tmu^i \hD \tmu^j \wedge \hD \tmu^k \wedge  \Big( F^h \wedge F^l \tmu_h \tmu_l - F^h \wedge F_h \Big)  \; , \nonumber  \\[5pt]
&& \epsilon_{ijk} \tmu^i \hD \tmu^j \wedge F^k \wedge \epsilon_{mnp} \hD\tmu^m \wedge \hrho^n \wedge \hrho^p  \nonumber \\
&& \qquad\qquad\qquad\qquad\qquad\qquad\qquad = \epsilon_{ijk} \hD\tmu^i \wedge \hD \tmu^j \wedge \hrho^k \wedge \epsilon_{mnp} \tmu^m \hrho^n \wedge F^p
\nonumber \\
&& \qquad\qquad\qquad\qquad\qquad\qquad\qquad = 2 \tmu_i \hD\tmu_j \wedge \hD \tmu_h \wedge \hrho^h \wedge \hrho^i \wedge F^j \nonumber \\
&& = \tfrac12 \epsilon_{ijk} \tmu^i \hD \tmu^j \wedge \hD \tmu^k \wedge \epsilon_{mnp} \hrho^m \wedge \hrho^n \wedge F^p - \hD \tmu_i \wedge \hD \tmu_j \wedge \hrho^i \wedge \hrho^j \wedge F^k \tmu_k  \; , \nonumber  \\[5pt]
&& \epsilon_{ijk} \tilde{\mu}^i \hat{\rho}^j \wedge \hat{\rho}^k \wedge \epsilon_{mnp} \,  \tilde{\mu}^m \hat{\rho}^n \wedge F^p = 0 \; .
\end{eqnarray}

Equipped with these identities, and requiring that $F^i$, $*F^i$, $F^i \wedge F_i$, $\tmu_i \tmu_j F^i \wedge F^j$, $F^i \wedge * F_i$ and $\tmu_i \tmu_j F^i \wedge * F^j$ be independent quantities, a very lengthy calculation allows one to obtain the set of algebraic and differential relations that the functions of $\alpha$ in (\ref{SO4IIAconfig}) must obey for that configuration to solve the type IIA field equations. The $\hat F_\4$ Bianchi identity gives the first equation in (C.7) of \cite{DeLuca:2018buk} and

\newpage

{\setlength\arraycolsep{0pt}
\begin{eqnarray} \label{SO4IIAF4Bianchi}
&& Q_4^\prime -P_4 +g C_1 -A_1 M_3 - B_1 M_2 =0 \; , \nonumber  \\[4pt]
&& R_4^\prime -N_4 +A_0 P_4 +g C_3 -2g (1-A_0) C_1 + A_3 P_3 = 0 \; , \nonumber  \\[4pt]
&& N_4 +A_0 P_4 +g C_3 +2 Q_4 A_0^\prime + A_2 M_3  + B_2 M_2 = 0 \; , \nonumber  \\[4pt]
&& U_4^\prime +\tfrac12 N_4 -\tfrac32 N_4 A_0 -\tfrac12 A_0 (1- A_0) P_4 + \tfrac12 g C_3 (1-A_0) -\tfrac12 R_4 A_0^\prime + \tfrac12 A_3 R_3 - A_4 M_3 - B_4 M_2  = 0 \; , \nonumber  \\[4pt]
&& T_4^\prime +\tfrac12 M_4 +\tfrac12 N_4 A_0 +\tfrac12 A_0 (1- A_0) P_4 - \tfrac12 g C_3 (1-A_0) +3 g C_4+\tfrac12 R_4 A_0^\prime - \tfrac12 A_3 R_3   = 0 \; , \nonumber  \\[4pt]
&& R_4 + 2 A_0 Q_4 + 2g C_2 -2A_1 R_3 + A_2 P_3 = 0  \; , \nonumber  \\[4pt]
&& (1-2A_0) R_4 + 2g (1-A_0) C_2 +2U_4 +A_2 R_3 -2 A_4 P_3 = 0  \; , \nonumber  \\[4pt]
&& A_0 R_4 + 2 A_0 (1-A_0) Q_4 -2U_4 +2B_3 M_2 = 0  \; , \nonumber  \\[4pt]
&& M_4 - P_4 (1-A_0) = 0 \; , \nonumber  \\[4pt]
&& gN_4 +g  P_4 (1-A_0) -M_2 M_3 + \overline{M}_2 \overline{M}_3 = 0 \; , \nonumber  \\[4pt]
&& 2 g (1-A_0) Q_4 +g  R_4  -M_2 P_3 + \overline{M}_2 \overline{P}_3 = 0 \; , \nonumber  \\[4pt]
&& 2 g U_4 -M_2 R_3 + \overline{M}_2 \overline{R}_3 = 0 \; , \nonumber  \\[4pt]
&& \overline{Q}_4^\prime -\overline{P}_4  -A_1 \overline{M}_3 - B_1 \overline{M}_2 =0 \; , \nonumber  \\[4pt]
&& \overline{R}_4^\prime -\overline{N}_4 +A_0 \overline{P}_4  + A_3 \overline{P}_3 = 0 \; , \nonumber  \\[4pt]
&& \overline{N}_4 +A_0 \overline{P}_4  +2 \overline{Q}_4 A_0^\prime + A_2 \overline{M}_3  + B_2 \overline{M}_2 = 0 \; , \nonumber  \\[4pt]
&& \overline{U}_4^\prime +\tfrac12 \overline{N}_4 -\tfrac32 \overline{N}_4 A_0 -\tfrac12 A_0 (1- A_0) \overline{P}_4  -\tfrac12 \overline{R}_4 A_0^\prime + \tfrac12 A_3 \overline{R}_3 - A_4 \overline{M}_3 - B_4 \overline{M}_2  = 0 \; , \nonumber  \\[4pt]
&& \overline{T}_4^\prime +\tfrac12 \overline{M}_4 +\tfrac12 \overline{N}_4 A_0 +\tfrac12 A_0 (1- A_0) \overline{P}_4  +\tfrac12 \overline{R}_4 A_0^\prime - \tfrac12 A_3 \overline{R}_3   = 0 \; , \nonumber  \\[4pt]
&& \overline{R}_4 + 2 A_0 \overline{Q}_4  -2A_1 \overline{R}_3 + A_2 \overline{P}_3 = 0  \; , \nonumber  \\[4pt]
&& (1-2A_0) \overline{R}_4  +2\overline{U}_4 +A_2 \overline{R}_3 -2 A_4 \overline{P}_3 = 0  \; , \nonumber  \\[4pt]
&& A_0 \overline{R}_4 + 2 A_0 (1-A_0) \overline{Q}_4 -2\overline{U}_4 +2B_3 \overline{M}_2 = 0  \; , \nonumber  \\[4pt]
&& \overline{M}_4 - \overline{P}_4 (1-A_0) = 0 \; , \nonumber  \\[4pt]
&& g\overline{N}_4 +g  \overline{P}_4 (1-A_0) -M_2 \overline{M}_3 - \overline{M}_2 M_3 = 0 \; , \nonumber  \\[4pt]
&& 2 g (1-A_0) \overline{Q}_4 +g  \overline{R}_4  -M_2 \overline{P}_3 - \overline{M}_2 P_3 = 0 \; , \nonumber  \\[4pt]
&& 2 g \overline{U}_4 -M_2 \overline{R}_3 - \overline{M}_2 R_3 = 0 \; ;
\end{eqnarray}
}the $\hat H_\3$ Bianchi identity gives the second equation in (C.7) of \cite{DeLuca:2018buk}, together with
{\setlength\arraycolsep{0pt}
\begin{eqnarray} \label{SO4IIAH3Bianchi}
&& P_3^\prime -M_3 -2g B_1 (1-A_0) -g B_2  =0 \; , \nonumber  \\[4pt]
&& R_3^\prime - A_0^\prime P_3 -A_0 M_3 -g B_2 (1-A_0) -2 g B_4 =0  \; , \nonumber  \\[4pt]
&& R_3 - A_0 P_3 -2 g B_3  =0 \; , \nonumber  \\[4pt]
&& \overline{P}_3^\prime - \overline{M}_3   =0 \; , \nonumber  \\[4pt]
&& \overline{R}_3^\prime - A_0^\prime \overline{P}_3 -A_0 \overline{M}_3 =0 \; , \nonumber  \\[4pt]
&& \overline{R}_3 - A_0 \overline{P}_3 =0 \; ;
\end{eqnarray}
}and the $\hat F_\2$ Bianchi identity gives the third through sixth equations in (C.7) of \cite{DeLuca:2018buk} and
{\setlength\arraycolsep{0pt}
\begin{eqnarray} \label{SO4IIAH3Bianchi}
&& M_2^\prime -m M_3 +g A_3  =0 \; , \nonumber  \\[4pt]
&& M_2 - mP_3 +gA_2 + 2g (1-A_0) A_1 =0  \; , \nonumber  \\[4pt]
&& M_2 A_0 - m R_3 +2g A_4 + g (1-A_0) A_2=0 \; , \nonumber  \\[4pt]
&& \overline{M}_2^\prime -m \overline{M}_3  =0 \; , \nonumber  \\[4pt]
&& \overline{M}_2 - m \overline{P}_3 =0  \; , \nonumber  \\[4pt]
&& \overline{M}_2 A_0 - m \overline{R}_3  =0 \; .
\end{eqnarray}
}In these expressions, I have dropped the explicit $\alpha$ dependence and have denoted with a prime the derivative with respect to it.

Turning now to the equations of motion, the $\hat F_\4$ equation of motion gives (C.8) of  \cite{DeLuca:2018buk} together with 
{\setlength\arraycolsep{0pt}
\begin{eqnarray} \label{SO4IIAF4eom}
&&  \big( e^{ \frac12 \phi -A+3C }  P_4 \big)^\prime  -4  \, e^{ \frac12 \phi - A +2B +C }  M_4 A_0^\prime -2 \, e^{ \frac12 \phi + A -2B +3 C }  Q_4 +4 \, e^{ \frac12 \phi + A  +C }  R_4 A_0  \nonumber \\
&& \qquad  +32 \, e^{ \frac12 \phi + A +2B -C }  T_4 A_0   (1-A_0)  +16 B_2 \overline{T}_4  -16 B_3 \overline{M}_4 +16 B_4 \overline{R}_4 \nonumber \\
&& \qquad -48 C_4 \overline{P}_3  -8 C_3 \overline{R}_3   =0 \;, \nonumber \\[12pt]
&&  \big( e^{ \frac12 \phi -A +2B +C }  M_4 \big)^\prime  +  \, e^{ \frac12 \phi +A +C }  R_4  +8 \, e^{ \frac12 \phi + A +2B -C }  \big( T_4 + U_4 A_0 \big)    \nonumber \\
&& \qquad  +8 B_1 \overline{T}_4  +2 B_2 \overline{R}_4 +4 B_3 \overline{P}_4 +2 C_3 \overline{P}_3  +4 C_1 \overline{R}_3   =0 \;, \nonumber \\[12pt]
&&  \big( e^{ \frac12 \phi -A +2B +C }  N_4 \big)^\prime  -3  \, e^{ \frac12 \phi +A +C }  R_4  +8 \, e^{ \frac12 \phi + A +2B -C }  \big( 1 -3 A_0 \big)  U_4  \nonumber \\
&& \qquad  +8 B_1 \overline{U}_4  -2 B_2 \overline{R}_4 -4 B_3 \overline{P}_4 +8 B_4 \overline{Q}_4  +4 C_2 \overline{M}_3 -2 C_3 \overline{P}_3  -4 C_1 \overline{R}_3  =0 \;, \nonumber \\[12pt]
&&  4g \, e^{ \frac12 \phi + A +2B -C }  T_4  +P_3 \overline{P}_4 + \overline{P}_3 P_4 = 0   \;, \nonumber \\[12pt]
&&  4g \, e^{ \frac12 \phi + A +2B -C }  U_4 + 2\big( M_3 \overline{Q}_4 +\overline{M}_3 Q_4 \big) -  \big( P_3 \overline{P}_4 +  \overline{P}_3 P_4 \big) =0 \;, \nonumber \\[12pt]
&&   g \, e^{ \frac12 \phi + A +C }  R_4 + 2 \big( R_3 \overline{P}_4 +\overline{R}_3 P_4 \big)  =0 \;, \nonumber \\[12pt]
&&  4g \, e^{ \frac12 \phi + A +2B -C }  T_4 (1-A_0)   + P_3 \overline{M}_4 + \overline{P}_3 M_4 = 0   \;, \nonumber \\[12pt]
&&  4g \, e^{ \frac12 \phi + A +2B -C }  U_4 (1-A_0)    - \big( M_3 \overline{R}_4 + \overline{M}_3 R_4 \big)  + P_3 \overline{N}_4 + \overline{P}_3 N_4 = 0   \;, \nonumber \\[12pt]
&&  g \, e^{ \frac12 \phi + A -2B +3C }  Q_4 +  8 \big( M_3 \overline{T}_4  +\overline{M}_3 T_4 \big) +4\big(   R_3  \overline{M}_4  + \overline{R}_3 M_4 \big)   +4 \big(   R_3  \overline{N}_4   +  \overline{R}_3 N_4 \big)   = 0   \;, \nonumber \\[12pt]
&&   g \, e^{ \frac12 \phi + A +C }  R_4 (1-A_0)   + 2 \big( R_3 \overline{M}_4 + \overline{R}_3 M_4 \big) = 0   \;, \nonumber \\[12pt]
&&   g \, e^{ \frac12 \phi + A +C }  R_4 (1-A_0)   +4 \big(  M_3 \overline{U}_4 +  \overline{M}_3 U_4 \big) - 2\big(  R_3 \overline{N}_4 + \overline{R}_3 N_4 \big) = 0   \;, \nonumber \\[12pt]
&&   g \, e^{ \frac12 \phi - A +2B +C }  N_4  + 4\big(  R_3 \overline{Q}_4 + \overline{R}_3 Q_4 \big)  = 0   \;, \nonumber \\[12pt]
&&  4g \, e^{ \frac12 \phi - A +2B +C }  M_4 (1-A_0)  + g \, e^{ \frac12 \phi - A +3C }  P_4   + 16\big( P_3 \overline{T}_4 + \overline{P}_3 T_4 \big) + 8\big(  R_3 \overline{R}_4 +  \overline{R}_3 R_4 \big) = 0   \;, \nonumber \\[12pt]
&&   g \, e^{ \frac12 \phi - A +2B +C }  N_4 (1-A_0)   + 4\big( P_3 \overline{U}_4 + \overline{P}_3 U_4 \big) - 2\big(  R_3 \overline{R}_4 + \overline{R}_3 R_4 \big) = 0   \;, \nonumber \\[12pt]
&&  \big( e^{ \frac12 \phi -A+3C }  \overline{P}_4 \big)^\prime  -4  \, e^{ \frac12 \phi - A +2B +C }  \overline{M}_4 A_0^\prime -2 \, e^{ \frac12 \phi + A -2B +3 C }  \overline{Q}_4 +4 \, e^{ \frac12 \phi + A  +C }  \overline{R}_4 A_0  \nonumber \\
&& \qquad  +32 \, e^{ \frac12 \phi + A +2B -C }  \overline{T}_4 A_0   (1-A_0)  -16 B_2 T_4  + 16  B_3 M_4 -16 B_4 R_4 +48 C_4 P_3  +8 C_3 R_3  \nonumber \\
&& \qquad -g \mu_0 \, e^{ \frac12 \phi - 4X +A +2B+3C } (1-A_0) =0    \;, \nonumber \\[12pt]
&&  \big( e^{ \frac12 \phi -A +2B +C }  \overline{M}_4 \big)^\prime  +  \, e^{ \frac12 \phi +A +C }  \overline{R}_4  +8 \, e^{ \frac12 \phi + A +2B -C }  \big(\overline{T}_4 + \overline{U}_4 A_0 \big)  +\tfrac 14 g \mu_0 \, e^{ \frac12 \phi - 4X +A +2B+3C }    \nonumber \\
&& \qquad  -8 B_1 T_4  -2 B_2 R_4 -4 B_3 P_4 -2 C_3 P_3  -4 C_1 R_3   =0 \;, \nonumber \\[12pt]
&&  \big( e^{ \frac12 \phi -A +2B +C }  \overline{N}_4 \big)^\prime  -3  \, e^{ \frac12 \phi +A +C }  \overline{R}_4  +8 \, e^{ \frac12 \phi + A +2B -C }  \big( 1 -3 A_0 \big)  \overline{U}_4  \nonumber \\
&& \qquad  -8 B_1 U_4  +2 B_2 R_4 +4 B_3 P_4 -8 B_4 Q_4  -4 C_2 M_3 +2 C_3 P_3  +4 C_1 R_3  =0 \;, \nonumber \\[12pt]
&&  4g \, e^{ \frac12 \phi + A +2B -C }  \overline{T}_4  -\big(  P_3 P_4 -  \overline{P}_3 \overline{P}_4 \big)  = 0   \;, \nonumber \\[12pt]
&&  4g \, e^{ \frac12 \phi + A +2B -C }  \overline{U}_4 - 2 \big( M_3 Q_4 -\overline{M}_3 \overline{Q}_4 \big)  + \big(  P_3 P_4 -  \overline{P}_3 \overline{P}_4 \big)  =0 \;, \nonumber \\[12pt]
&&   g \, e^{ \frac12 \phi + A +C }  \overline{R}_4 - 2\big( R_3 P_4 -\overline{R}_3 \overline{P}_4  \big) =0 \;, \nonumber \\[12pt]
&&  4g \, e^{ \frac12 \phi + A +2B -C }  \overline{T}_4 (1-A_0)   - P_3 M_4 + \overline{P}_3 \overline{M}_4 = 0   \;, \nonumber \\[12pt]
&&  4g \, e^{ \frac12 \phi + A +2B -C }  \overline{U}_4 (1-A_0)   +M_3 R_4 - \overline{M}_3 \overline{R}_4 - P_3 N_4 + \overline{P}_3 \overline{N}_4 = 0   \;, \nonumber \\[12pt]
&&   g \, e^{ \frac12 \phi + A -2B +3C }  \overline{Q}_4 - 8\big(  M_3 T_4  - \overline{M}_3 \overline{T}_4 \big) -4\big(   R_3  M_4  - \overline{R}_3 \overline{M}_4 \big)   -4\big(   R_3 N_4   -  \overline{R}_3 \overline{N}_4 \big)   = 0   \;, \nonumber \\[12pt]
&&   g \, e^{ \frac12 \phi + A +C }  \overline{R}_4 (1-A_0)   -2\big(  R_3 M_4 - \overline{R}_3 \overline{M}_4 \big)  = 0   \;, \nonumber \\[12pt]
&&   g \, e^{ \frac12 \phi + A +C }  \overline{R}_4 (1-A_0)   -4 \big(  M_3 U_4 - \overline{M}_3 \overline{U}_4 \big)  + 2\big( R_3 N_4 - \overline{R}_3 \overline{N}_4 \big)  = 0   \;, \nonumber \\[12pt]
&&   g \, e^{ \frac12 \phi - A +2B +C }  \overline{N}_4  - 4\big( R_3 Q_4 - \overline{R}_3 \overline{Q}_4 \big) = 0   \;, \nonumber \\[12pt]
&&  4 g \, e^{ \frac12 \phi - A +2B +C }  \overline{M}_4 (1-A_0)  + g \, e^{ \frac12 \phi - A +3C }  \overline{P}_4   - 16 \big( P_3 T_4 - \overline{P}_3 \overline{T}_4 \big)  - 8\big( R_3 R_4 -  \overline{R}_3 \overline{R}_4 \big) = 0   \;, \nonumber \\[12pt]
&&   g \, e^{ \frac12 \phi - A +2B +C }  \overline{N}_4 (1-A_0)   - 4\big( P_3 U_4 - \overline{P}_3 \overline{U}_4 \big) + 2\big(  R_3 R_4 -  \overline{R}_3 \overline{R}_4 \big) = 0   \; ;
\end{eqnarray}
}the $\hat H_\3$ equation of motion gives (C.9) of \cite{DeLuca:2018buk}, along with
{\setlength\arraycolsep{0pt}
\begin{eqnarray} \label{SO4IIAH3eom}
&&  \big( e^{ - \phi -A+2B+3C }  M_3 \big)^\prime -2 e^{ - \phi +A+ 3C }  P_3 -8  e^{ - \phi +A +2B+ C }  A_0  R_3  \nonumber \\
&& \qquad  -32  C_1 \overline{T}_4 -32  C_1 \overline{U}_4 -16 C_2 \overline{M}_4-16 C_2 \overline{N}_4 - 96 C_4 \overline{Q}_4  - m  \, e^{ \frac32 \phi + A +2B +3C } M_2  \nonumber \\
&&  \qquad - 4 e^{ \frac12 \phi +A-2B+3C }  A_1 Q_4 - 4 e^{ \frac12 \phi -A+2B+C }  A_3 M_4 - 4 e^{ \frac12 \phi -A+2B+C }  A_3 N_4   \nonumber \\
&&  \qquad -64  e^{ \frac12 \phi +A+2B-C } A_4 T_4  -64  e^{ \frac12 \phi +A+2B-C } A_4 U_4 + \mu_0 \overline{M}_2 \,  e^{ \frac12 \phi -4X +A+2B +3C } = 0   \;, \nonumber \\[12pt]
&&  g \, e^{ -\phi + A +2B +C }  R_3  + 2 \big( P_4 \overline{R}_4 +  \overline{P}_4 R_4 \big) = 0   \;, \nonumber \\[12pt]
&& g \, e^{ -\phi + A +2B +C }  R_3  -4\big( M_4 \overline{Q}_4 + \overline{M}_4 Q_4 \big) - 8 e^{ \frac12 \phi + A +2B -C } \big( M_2 T_4 - \overline{M}_2 \overline{T}_4 )   = 0   \;, \nonumber \\[12pt]
&& N_4 \overline{Q}_4 + \overline{N}_4 Q_4 -  \tfrac12 \big( P_4 \overline{R}_4 +\overline{P}_4 R_4 \big) +2 e^{ \frac12 \phi + A +2B -C } \big( M_2 U_4 - \overline{M}_2 \overline{U}_4 )   = 0   \;, \nonumber \\[12pt]
&&  g \, e^{ -\phi + A +3C }  P_3  + 16 \big( P_4 \overline{T}_4 +  \overline{P}_4 T_4 \big) = 0   \;, \nonumber \\[12pt]
&& g \, e^{ -\phi + A +2B +C }  R_3 (1-A_0)   + 2\big( M_4 \overline{R}_4 + \overline{M}_4 R_4  \big) = 0   \;, \nonumber \\[12pt]
&& P_4 \overline{U}_4 + \overline{P}_4 U_4  + \tfrac14  e^{ \frac12 \phi + A +C } \big( M_2 R_4 - \overline{M}_2 \overline{R}_4 ) = 0   \;, \nonumber \\[12pt]
&&  g \, e^{ -\phi + A +3C }  P_3  (1-A_0) + 16 \big( M_4 \overline{T}_4 + \overline{M}_4 T_4 \big) = 0   \;, \nonumber \\[12pt]
&&  g \, e^{ -\phi + A +3C }  P_3  (1-A_0) - 16 \big( M_4 \overline{U}_4 +  \overline{M}_4 U_4 \big)  - 16 \big(  N_4 \overline{U}_4 +  \overline{N}_4 U_4 \big)  - 16 \big( N_4 \overline{T}_4 +  \overline{N}_4 T_4 \big)   \nonumber \\
&& \qquad  -2 e^{ \frac12 \phi + A -2B +3C } \big( M_2 Q_4 - \overline{M}_2 \overline{Q}_4 ) = 0     \;, \nonumber \\[12pt]
&&  g \, e^{ -\phi - A +2B +3C }  M_3  + 32 \big( Q_4 \overline{T}_4 +   \overline{Q}_4 T_4 \big)  + 4  e^{ \frac12 \phi - A +2B +C } \big( M_2 M_4 - \overline{M}_2 \overline{M}_4 ) = 0     \;, \nonumber \\[12pt]
&& Q_4 \overline{U}_4 +  \overline{Q}_4 U_4 +\tfrac18  e^{ \frac12 \phi - A +2B +C } \big( M_2 N_4 - \overline{M}_2 \overline{N}_4 ) = 0     \;, \nonumber \\[12pt]
&&  g \, e^{ -\phi - A +2B +3C }  M_3  (1-A_0) -16 \big( R_4 \overline{U}_4 + \overline{R}_4 U_4 \big)  -16 \big(  R_4 \overline{T}_4 + \overline{R}_4 T_4 \big)  \nonumber \\
&& \qquad -  e^{ \frac12 \phi - A +3C } \big( M_2 P_4 - \overline{M}_2 \overline{P}_4 ) = 0     \;, \nonumber \\[12pt]
&&  \big( e^{ - \phi -A+2B+3C }  \overline{M}_3 \big)^\prime -2 e^{ - \phi +A+ 3C }  \overline{P}_3 -8  e^{ - \phi  +A +2B+ C }  A_0  \overline{R}_3  \nonumber \\
&& \qquad  +32  C_1 T_4 +32  C_1 U_4 +16 C_2 M_4 +16 C_2 N_4 + 96 C_4 Q_4  - m  \, e^{ \frac32 \phi + A +2B +3C } \overline{M}_2  \nonumber \\
&&  \qquad - 4 e^{ \frac12 \phi +A-2B+3C }  A_1 \overline{Q}_4 - 4 e^{ \frac12 \phi -A+2B+C }  A_3 \overline{M}_4 - 4 e^{ \frac12 \phi -A+2B+C }  A_3 \overline{N}_4   \nonumber \\
&&  \qquad -64  e^{ \frac12 \phi +A+2B-C } A_4 \overline{T}_4  -64  e^{ \frac12 \phi +A+2B-C } A_4 \overline{U}_4 - \mu_0 M_2 \,  e^{ \frac12 \phi -4X +A+2B +3C } = 0   \;, \nonumber \\[12pt]
&&  g \, e^{ -\phi + A +2B +C }  \overline{R}_3  - 2 \big( P_4 R_4 - \overline{P}_4 \overline{R}_4 \big) = 0   \;, \nonumber \\[12pt]
&& g \, e^{ -\phi + A +2B +C }  \overline{R}_3  + 4\big( M_4 Q_4 - \overline{M}_4 \overline{Q}_4 \big)  - 8 e^{ \frac12 \phi + A +2B -C } \big( M_2 \overline{T}_4 - \overline{M}_2 T_4 )   = 0   \;, \nonumber \\[12pt]
&& N_4 Q_4 - \overline{N}_4 \overline{Q}_4 -  \tfrac12 \big( P_4 R_4 - \overline{P}_4 \overline{R}_4 \big)  -2 e^{ \frac12 \phi + A +2B -C } \big( M_2 \overline{U}_4 + \overline{M}_2 U_4 )   = 0   \;, \nonumber \\[12pt]
&&  g \, e^{ -\phi + A +3C }  \overline{P}_3  - 16 \big( P_4 T_4 - \overline{P}_4 \overline{T}_4 \big)  = 0   \;, \nonumber \\[12pt]
&& g \, e^{ -\phi + A +2B +C }  \overline{R}_3 (1-A_0)   - 2\big( M_4 R_4 - \overline{M}_4 \overline{R}_4  \big) = 0   \;, \nonumber \\[12pt]
&& P_4 U_4 - \overline{P}_4 \overline{U}_4  - \tfrac14  e^{ \frac12 \phi + A +C } \big( M_2 \overline{R}_4 + \overline{M}_2 R_4 ) = 0   \;, \nonumber \\[12pt]
&&  g \, e^{ -\phi + A +3C }  \overline{P}_3  (1-A_0) - 16 \big( M_4 T_4 -  \overline{M}_4 \overline{T}_4 \big)  = 0   \;, \nonumber \\[12pt]
&&  g \, e^{ -\phi + A +3C }  \overline{P}_3  (1-A_0) + 16 \big( M_4 U_4 - \overline{M}_4 \overline{U}_4 \big)  + 16 \big( N_4 U_4 -  \overline{N}_4 \overline{U}_4 \big)  + 16 \big( N_4 T_4 -  \overline{N}_4 \overline{T}_4 \big)   \nonumber \\
&& \qquad  -2 e^{ \frac12 \phi + A -2B +3C } \big( M_2 \overline{Q}_4 + \overline{M}_2 Q_4 ) = 0     \;, \nonumber \\[12pt]
&&  g \, e^{ -\phi - A +2B +3C }  \overline{M}_3  - 32 \big(  Q_4 T_4 - \overline{Q}_4 \overline{T}_4 \big)  + 4  e^{ \frac12 \phi - A +2B +C } \big( M_2 \overline{M}_4 + \overline{M}_2 M_4 ) = 0     \;, \nonumber \\[12pt]
&& Q_4 U_4 +  \overline{Q}_4 \overline{U}_4 -\tfrac18  e^{ \frac12 \phi - A +2B +C } \big( M_2 \overline{N}_4 + \overline{M}_2 N_4 ) = 0     \;, \nonumber \\[12pt]
&&  g \, e^{ -\phi - A +2B +3C }  \overline{M}_3  (1-A_0) +16 \big( R_4 U_4  - \overline{R}_4 \overline{U}_4 \big) +  16 \big( R_4 T_4  -  \overline{R}_4 \overline{T}_4 \big)  \nonumber \\
&& \qquad -  e^{ \frac12 \phi - A +3C } \big( M_2 \overline{P}_4 + \overline{M}_2 P_4 ) = 0     \; ; 
\end{eqnarray}
}

\vspace{-8pt}
\noindent the $\hat F_\2$ equation of motion gives (C.10) of \cite{DeLuca:2018buk} as well as
{\setlength\arraycolsep{0pt}
\begin{eqnarray} \label{SU3IIAF2eom}
&& g\, e^{ \frac32 \phi +A+2B + 3C }  M_2  (1-A_0)  -   e^{ \frac12 \phi  -A + 3C } \big( M_3 P_4 - \overline{M}_3 \overline{P}_4  \big)  \nonumber \\
&& \qquad +2 \,  e^{ \frac12 \phi + A -2B +3C } \big( P_3 Q_4 -  \overline{P}_3 \overline{Q}_4  \big)  =0  \; ,  \nonumber \\[12pt]
&& g\, e^{ \frac32 \phi +A+2B + 3C }  M_2  +4   e^{ \frac12 \phi  -A +2B + C } \big( M_3 M_4 - \overline{M}_3 \overline{M}_4  \big)  +32   \, e^{ \frac12 \phi  +A +2B - C } \big( R_3 T_4 - \overline{R}_3 \overline{T}_4  \big)    \nonumber \\
&& \qquad +32   \, e^{ \frac12 \phi  +A +2B - C } \big( R_3 U_4 - \overline{R}_3 \overline{U}_4  \big)    =0  \; ,  \nonumber \\[12pt]
&&  e^{ \frac12 \phi  -A +2B + C } \big( M_3 N_4 - \overline{M}_3 \overline{N}_4  \big)  - e^{ \frac12 \phi  +A +C } \big( P_3 R_4 - \overline{P}_3 \overline{R}_4  \big)  \nonumber \\
&& \qquad   - 8e^{ \frac12 \phi  +A +2B - C } \big( R_3 U_4 - \overline{R}_3 \overline{U}_4  \big)    =0  \; ,  \nonumber \\[12pt]
&&  e^{ \frac12 \phi  +A + C } \big( P_3 R_4 - \overline{P}_3 \overline{R}_4  \big)  - 8\,  e^{ \frac12 \phi  +A +2B - C } \big( R_3 T_4 - \overline{R}_3 \overline{T}_4  \big)  =0  \; ,  \nonumber \\[12pt]
&& g\, e^{ \frac32 \phi +A+2B + 3C }  \overline{M}_2  (1-A_0)  -   e^{ \frac12 \phi  -A + 3C } \big( M_3 \overline{P}_4 + \overline{M}_3 P_4  \big)  \nonumber \\
&& \qquad +2 \,  e^{ \frac12 \phi + A -2B +3C } \big( P_3 \overline{Q}_4 +  \overline{P}_3 Q_4  \big)  =0  \; ,  \nonumber \\[12pt]
&& g\, e^{ \frac32 \phi +A+2B + 3C }  \overline{M}_2  +4   e^{ \frac12 \phi  -A +2B + C } \big( M_3 \overline{M}_4 + \overline{M}_3 M_4  \big)  +32   \, e^{ \frac12 \phi  +A +2B - C } \big( R_3 \overline{T}_4 + \overline{R}_3 T_4  \big)    \nonumber \\
&& \qquad +32   \, e^{ \frac12 \phi  +A +2B - C } \big( R_3 \overline{U}_4 + \overline{R}_3 U_4  \big)    =0  \; ,  \nonumber \\[12pt]
&&  e^{ \frac12 \phi  -A +2B + C } \big( M_3 \overline{N}_4 + \overline{M}_3 N_4  \big)  - e^{ \frac12 \phi  +A +C } \big( P_3 \overline{R}_4 + \overline{P}_3 R_4  \big)  \nonumber \\
&& \qquad   - 8e^{ \frac12 \phi  +A +2B - C } \big( R_3 \overline{U}_4 + \overline{R}_3 U_4  \big)    =0  \; ,  \nonumber \\[12pt]
&&  e^{ \frac12 \phi  +A + C } \big( P_3 \overline{R}_4 + \overline{P}_3 R_4  \big)  - 8\,  e^{ \frac12 \phi  +A +2B - C } \big( R_3 \overline{T}_4 + \overline{R}_3 T_4  \big)  =0  \; ;
\end{eqnarray}
}

\noindent and, finally, the dilaton equation of motion gives (C.11) of \cite{DeLuca:2018buk} together with
{\setlength\arraycolsep{0pt}
\begin{eqnarray} \label{SO4IIADilaton}
&& 3  \,  e^{ \frac32 \phi + A +2 B +3C } M_2 \overline{M}_2 - 2  \,  e^{ -\phi - A +2 B +3C } M_3 \overline{M}_3 + 2  \,  e^{ -\phi + A  +3C } P_3 \overline{P}_3  + 8 \, e^{ -\phi + A +2 B +C } R_3 \overline{R}_3 \nonumber \\ 
&& \qquad +4 \, e^{ \frac12 \phi - A +2 B +C } \big( N_4 \overline{N}_4 + M_4 \overline{N}_4 + N_4 \overline{M}_4  \big) -  e^{ \frac12 \phi - A +3C } P_4 \overline{P}_4  +4 \, e^{ \frac12 \phi + A -2 B +3C } Q_4 \overline{Q}_4   \nonumber \\ 
&& \qquad   -4 \, e^{ \frac12 \phi + A +C } R_4 \overline{R}_4  +64 \, e^{ \frac12 \phi + A +2 B -C } \big( U_4 \overline{U}_4 + T_4 \overline{U}_4 + U_4 \overline{T}_4  \big)  = 0 \; ,  \nonumber \\[12pt]
&&  2  \,  e^{ -\phi + A  +3C } P_3 \overline{P}_3  + 8 \, e^{ -\phi + A +2 B +C } R_3 \overline{R}_3 -4 \, e^{ \frac12 \phi - A +2 B +C } M_4 \overline{M}_4   -  e^{ \frac12 \phi - A +3C } P_4 \overline{P}_4 \nonumber \\ 
&& \qquad    -4 \, e^{ \frac12 \phi + A +C } R_4 \overline{R}_4  - 64 \, e^{ \frac12 \phi + A +2 B -C }  T_4 \overline{T}_4 = 0 \; ,  \nonumber \\[12pt]
&& 3  \,  e^{ \frac32 \phi + A +2 B +3C } \big( M_2^2  -  \overline{M}_2^2  \big) - 2  \,  e^{ -\phi - A +2 B +3C }  \big( M_3^2 - \overline{M}_3^2  \big) + 2  \,  e^{ -\phi + A  +3C } \big( P_3^2- \overline{P}_3^2 \big)  \nonumber \\ 
&& \qquad + 8 \, e^{ -\phi + A +2 B +C } \big( R_3^2 -  \overline{R}_3^2  \big) +4 \, e^{ \frac12 \phi - A +2 B +C } \big( N_4^2 - \overline{N}_4^2 + 2 M_4 N_4 -2 \overline{M}_4 \overline{N}_4  \big)   \nonumber \\ 
&& \qquad  -  e^{ \frac12 \phi - A +3C } \big(  P_4^2 -  \overline{P}_4^2 \big)   +4 \, e^{ \frac12 \phi + A -2 B +3C } \big( Q_4^2 -  \overline{Q}_4^2   \big)   -4 \, e^{ \frac12 \phi + A +C } \big( R_4^2 -  \overline{R}_4^2  \big)  \nonumber \\ 
&& \qquad    +64 \, e^{ \frac12 \phi + A +2 B -C } \big( U_4^2 -  \overline{U}_4^2  + 2 T_4 U_4 - 2 \overline{T}_4 \overline{U}_4  \big)  = 0 \; ,   \\[12pt]
&&  2  \,  e^{ -\phi + A  +3C } \big( P_3^2 -  \overline{P}_3^2 \big) + 8 \, e^{ -\phi + A +2 B +C } \big( R_3^2 -\overline{R}_3^2 \big) -4 \, e^{ \frac12 \phi - A +2 B +C } \big( M_4^2 -  \overline{M}_4^2 \big) \nonumber \\ 
&& \qquad   -  e^{ \frac12 \phi - A +3C } \big( P_4^2 -  \overline{P}_4^2 \big)  -4 \, e^{ \frac12 \phi + A +C } \big( R_4^2 -  \overline{R}_4^2 \big)  - 64 \, e^{ \frac12 \phi + A +2 B -C } \big( T_4^2 -  \overline{T}_4^2 \big) = 0 \; .  \nonumber 
\end{eqnarray}

I have explicitly checked that equations (\ref{SO4IIAF4Bianchi})--(\ref{SO4IIADilaton}) are identically  satisfied for the functions that can be read off from (\ref{eq:Dmushifted})--(\ref{SO4SolN=3}). In other words, the bosonic field equations of massive IIA supergravity are fulfilled on the field equations of $D=4$ $\cN=3$ minimal gauged supergravity (\ref{minimalN=3}). This shows the consistency of the truncation, up to a check of the $D=10$ Einstein equation.

\bibliography{references}

\providecommand{\href}[2]{#2}\begingroup\raggedright\begin{thebibliography}{10}

\bibitem{Liu:2009dm}
H.~Liu, J.~McGreevy, and D.~Vegh, {\it {Non-Fermi liquids from holography}},
  {\em Phys. Rev.} {\bf D83} (2011) 065029,
  [\href{http://arxiv.org/abs/0903.2477}{{\tt arXiv:0903.2477}}].

\bibitem{DHoker:2009mmn}
E.~D'Hoker and P.~Kraus, {\it {Magnetic Brane Solutions in AdS}},  {\em JHEP}
  {\bf 10} (2009) 088, [\href{http://arxiv.org/abs/0908.3875}{{\tt
  arXiv:0908.3875}}].

\bibitem{Gauntlett:2011mf}
J.~P. Gauntlett, J.~Sonner, and D.~Waldram, {\it {Universal fermionic spectral
  functions from string theory}},  {\em Phys. Rev. Lett.} {\bf 107} (2011)
  241601, [\href{http://arxiv.org/abs/1106.4694}{{\tt arXiv:1106.4694}}].

\bibitem{Martelli:2012sz}
D.~Martelli, A.~Passias, and J.~Sparks, {\it {The supersymmetric NUTs and bolts
  of holography}},  {\em Nucl. Phys.} {\bf B876} (2013) 810--870,
  [\href{http://arxiv.org/abs/1212.4618}{{\tt arXiv:1212.4618}}].

\bibitem{Davison:2013bxa}
R.~A. Davison and A.~Parnachev, {\it {Hydrodynamics of cold holographic
  matter}},  {\em JHEP} {\bf 06} (2013) 100,
  [\href{http://arxiv.org/abs/1303.6334}{{\tt arXiv:1303.6334}}].

\bibitem{Martelli:2013aqa}
D.~Martelli and A.~Passias, {\it {The gravity dual of supersymmetric gauge
  theories on a two-parameter deformed three-sphere}},  {\em Nucl. Phys.} {\bf
  B877} (2013) 51--72, [\href{http://arxiv.org/abs/1306.3893}{{\tt
  arXiv:1306.3893}}].

\bibitem{Cassani:2014zwa}
D.~Cassani and D.~Martelli, {\it {The gravity dual of supersymmetric gauge
  theories on a squashed S$^{1}$ x S$^{3}$}},  {\em JHEP} {\bf 08} (2014) 044,
  [\href{http://arxiv.org/abs/1402.2278}{{\tt arXiv:1402.2278}}].

\bibitem{Benini:2015bwz}
F.~Benini, N.~Bobev, and P.~M. Crichigno, {\it {Two-dimensional SCFTs from
  D3-branes}},  {\em JHEP} {\bf 07} (2016) 020,
  [\href{http://arxiv.org/abs/1511.09462}{{\tt arXiv:1511.09462}}].

\bibitem{Genolini:2016ecx}
P.~Benetti~Genolini, D.~Cassani, D.~Martelli, and J.~Sparks, {\it {Holographic
  renormalization and supersymmetry}},  {\em JHEP} {\bf 02} (2017) 132,
  [\href{http://arxiv.org/abs/1612.06761}{{\tt arXiv:1612.06761}}].

\bibitem{Ammon:2017ded}
M.~Ammon, M.~Kaminski, R.~Koirala, J.~Leiber, and J.~Wu, {\it {Quasinormal
  modes of charged magnetic black branes \& chiral magnetic transport}},  {\em
  JHEP} {\bf 04} (2017) 067, [\href{http://arxiv.org/abs/1701.05565}{{\tt
  arXiv:1701.05565}}].

\bibitem{Blazquez-Salcedo:2017cqm}
J.~L. Blázquez-Salcedo, J.~Kunz, F.~Navarro-Lérida, and E.~Radu, {\it {AdS$_5$
  magnetized solutions in minimal gauged supergravity}},  {\em Phys. Lett.}
  {\bf B771} (2017) 52--58, [\href{http://arxiv.org/abs/1703.04163}{{\tt
  arXiv:1703.04163}}].

\bibitem{Azzurli:2017kxo}
F.~Azzurli, N.~Bobev, P.~M. Crichigno, V.~S. Min, and A.~Zaffaroni, {\it {A
  universal counting of black hole microstates in AdS$_{4}$}},  {\em JHEP} {\bf
  02} (2018) 054, [\href{http://arxiv.org/abs/1707.04257}{{\tt
  arXiv:1707.04257}}].

\bibitem{Bobev:2017uzs}
N.~Bobev and P.~M. Crichigno, {\it {Universal RG Flows Across Dimensions and
  Holography}},  {\em JHEP} {\bf 12} (2017) 065,
  [\href{http://arxiv.org/abs/1708.05052}{{\tt arXiv:1708.05052}}].

\bibitem{Blazquez-Salcedo:2017kig}
J.~L. Blázquez-Salcedo, J.~Kunz, F.~Navarro-Lérida, and E.~Radu, {\it {New
  black holes in $D=5$ minimal gauged supergravity: Deformed boundaries and
  frozen horizons}},  {\em Phys. Rev.} {\bf D97} (2018), no.~8 081502,
  [\href{http://arxiv.org/abs/1711.08292}{{\tt arXiv:1711.08292}}].

\bibitem{Cabo-Bizet:2018ehj}
A.~Cabo-Bizet, D.~Cassani, D.~Martelli, and S.~Murthy, {\it {Microscopic origin
  of the Bekenstein-Hawking entropy of supersymmetric AdS$_{5}$ black holes}},
  {\em JHEP} {\bf 10} (2019) 062, [\href{http://arxiv.org/abs/1810.11442}{{\tt
  arXiv:1810.11442}}].

\bibitem{BenettiGenolini:2019jdz}
P.~Benetti~Genolini, J.~M. Pérez~Ipiña, and J.~Sparks, {\it {Localization of
  the action in AdS/CFT}},  \href{http://arxiv.org/abs/1906.11249}{{\tt
  arXiv:1906.11249}}.

\bibitem{Gang:2019uay}
D.~Gang, N.~Kim, and L.~A. Pando~Zayas, {\it {Precision Microstate Counting for
  the Entropy of Wrapped M5-branes}},
  \href{http://arxiv.org/abs/1905.01559}{{\tt arXiv:1905.01559}}.

\bibitem{Romans:1985tz}
L.~Romans, {\it {Massive N=2a Supergravity in Ten-Dimensions}},  {\em
  Phys.Lett.} {\bf B169} (1986) 374.

\bibitem{Fradkin:1976xz}
E.~S. Fradkin and M.~A. Vasiliev, {\it {Model of Supergravity with Minimal
  Electromagnetic Interaction}},  {\em LEBEDEV-76-197} (1976).

\bibitem{Freedman:1976aw}
D.~Z. Freedman and A.~K. Das, {\it {Gauge Internal Symmetry in Extended
  Supergravity}},  {\em Nucl. Phys.} {\bf B120} (1977) 221--230.

\bibitem{Guarino:2015jca}
A.~Guarino, D.~L. Jafferis, and O.~Varela, {\it {The string origin of dyonic
  N=8 supergravity and its simple Chern-Simons duals}},  {\em Phys. Rev. Lett.}
  {\bf 115} (2015), no.~9 091601, [\href{http://arxiv.org/abs/1504.08009}{{\tt
  arXiv:1504.08009}}].

\bibitem{Pang:2015vna}
Y.~Pang and J.~Rong, {\it {N=3 solution in dyonic ISO(7) gauged maximal
  supergravity and its uplift to massive type IIA supergravity}},  {\em Phys.
  Rev.} {\bf D92} (2015), no.~8 085037,
  [\href{http://arxiv.org/abs/1508.05376}{{\tt arXiv:1508.05376}}].

\bibitem{DeLuca:2018buk}
G.~B. De~Luca, G.~L. Monaco, N.~T. Macpherson, A.~Tomasiello, and O.~Varela,
  {\it {The geometry of $ \mathcal{N}=3 $ AdS$_{4}$ in massive IIA}},  {\em
  JHEP} {\bf 08} (2018) 133, [\href{http://arxiv.org/abs/1805.04823}{{\tt
  arXiv:1805.04823}}].

\bibitem{Gauntlett:2007ma}
J.~P. Gauntlett and O.~Varela, {\it {Consistent Kaluza-Klein reductions for
  general supersymmetric AdS solutions}},  {\em Phys.Rev.} {\bf D76} (2007)
  126007, [\href{http://arxiv.org/abs/0707.2315}{{\tt arXiv:0707.2315}}].

\bibitem{Cassani:2019vcl}
D.~Cassani, G.~Josse, M.~Petrini, and D.~Waldram, {\it {Systematics of
  consistent truncations from generalised geometry}},
  \href{http://arxiv.org/abs/1907.06730}{{\tt arXiv:1907.06730}}.

\bibitem{Duff:1985jd}
M.~J. Duff and C.~N. Pope, {\it {Consistent truncations in Kaluza-Klein
  theories}},  {\em Nucl. Phys.} {\bf B255} (1985) 355--364.

\bibitem{Pope:1987ad}
C.~N. Pope and K.~S. Stelle, {\it {Zilch Currents, Supersymmetry and
  {Kaluza-Klein} Consistency}},  {\em Phys. Lett.} {\bf B198} (1987) 151.

\bibitem{Gauntlett:2002sc}
J.~P. Gauntlett, D.~Martelli, S.~Pakis, and D.~Waldram, {\it {G structures and
  wrapped NS5-branes}},  {\em Commun. Math. Phys.} {\bf 247} (2004) 421--445,
  [\href{http://arxiv.org/abs/hep-th/0205050}{{\tt hep-th/0205050}}].

\bibitem{Pope:1985jg}
C.~N. Pope, {\it {Consistency of truncations in Kaluza-Klein}},  {\em Conf.
  Proc.} {\bf C841031} (1984) 429--431.

\bibitem{Pope:1985bu}
C.~N. Pope, {\it {The Embedding of the Einstein {Yang-Mills} Equations in
  $d=11$ Supergravity}},  {\em Class. Quant. Grav.} {\bf 2} (1985) L77.

\bibitem{Tsikas:1986rx}
T.~T. Tsikas, {\it {Consistent Truncations of Chiral $N=2 D=10$ Supergravity on
  the Round Five Sphere}},  {\em Class. Quant. Grav.} {\bf 3} (1986) 733.

\bibitem{Buchel:2006gb}
A.~Buchel and J.~T. Liu, {\it {Gauged supergravity from type IIB string theory
  on Y**p,q manifolds}},  {\em Nucl.Phys.} {\bf B771} (2007) 93--112,
  [\href{http://arxiv.org/abs/hep-th/0608002}{{\tt hep-th/0608002}}].

\bibitem{Gauntlett:2006ai}
J.~P. Gauntlett, E.~O~Colgain, and O.~Varela, {\it {Properties of some
  conformal field theories with M-theory duals}},  {\em JHEP} {\bf 02} (2007)
  049, [\href{http://arxiv.org/abs/hep-th/0611219}{{\tt hep-th/0611219}}].

\bibitem{Gauntlett:2007sm}
J.~P. Gauntlett and O.~Varela, {\it {D=5 SU(2) x U(1) Gauged Supergravity from
  D=11 Supergravity}},  {\em JHEP} {\bf 02} (2008) 083,
  [\href{http://arxiv.org/abs/0712.3560}{{\tt arXiv:0712.3560}}].

\bibitem{Passias:2015gya}
A.~Passias, A.~Rota, and A.~Tomasiello, {\it {Universal consistent truncation
  for 6d/7d gauge/gravity duals}},  {\em JHEP} {\bf 10} (2015) 187,
  [\href{http://arxiv.org/abs/1506.05462}{{\tt arXiv:1506.05462}}].

\bibitem{Malek:2017njj}
E.~Malek, {\it {Half-Maximal Supersymmetry from Exceptional Field Theory}},
  {\em Fortsch. Phys.} {\bf 65} (2017), no.~10-11 1700061,
  [\href{http://arxiv.org/abs/1707.00714}{{\tt arXiv:1707.00714}}].

\bibitem{Hong:2018amk}
J.~Hong, J.~T. Liu, and D.~R. Mayerson, {\it {Gauged Six-Dimensional
  Supergravity from Warped IIB Reductions}},  {\em JHEP} {\bf 09} (2018) 140,
  [\href{http://arxiv.org/abs/1808.04301}{{\tt arXiv:1808.04301}}].

\bibitem{Malek:2018zcz}
E.~Malek, H.~Samtleben, and V.~Vall~Camell, {\it {Supersymmetric AdS$_{7}$ and
  AdS$_6$ vacua and their minimal consistent truncations from exceptional field
  theory}},  {\em Phys. Lett.} {\bf B786} (2018) 171--179,
  [\href{http://arxiv.org/abs/1808.05597}{{\tt arXiv:1808.05597}}].

\bibitem{Liu:2019cea}
J.~T. Liu and B.~McPeak, {\it {Gauged Supergravity from the Lunin-Maldacena
  background}},  \href{http://arxiv.org/abs/1905.06861}{{\tt
  arXiv:1905.06861}}.

\bibitem{Larios:2019lxq}
G.~Larios and O.~Varela, {\it {Minimal $D=4 ${\cal N}=2 supergravity from
  $D=11$: an M-theory free lunch}},  {\em JHEP} {\bf 10} (2019) 251,
  [\href{http://arxiv.org/abs/1907.11027}{{\tt arXiv:1907.11027}}].

\bibitem{Petrini:2009ur}
M.~Petrini and A.~Zaffaroni, {\it {N=2 solutions of massive type IIA and their
  Chern-Simons duals}},  {\em JHEP} {\bf 09} (2009) 107,
  [\href{http://arxiv.org/abs/0904.4915}{{\tt arXiv:0904.4915}}].

\bibitem{Lust:2009mb}
D.~Lust and D.~Tsimpis, {\it {New supersymmetric AdS(4) type II vacua}},  {\em
  JHEP} {\bf 0909} (2009) 098, [\href{http://arxiv.org/abs/0906.2561}{{\tt
  arXiv:0906.2561}}].

\bibitem{Passias:2018zlm}
A.~Passias, D.~Prins, and A.~Tomasiello, {\it {A massive class of $\mathcal{N}
  = 2$ AdS$_4$ IIA solutions}},  {\em JHEP} {\bf 10} (2018) 071,
  [\href{http://arxiv.org/abs/1805.03661}{{\tt arXiv:1805.03661}}].

\bibitem{Guarino:2015vca}
A.~Guarino and O.~Varela, {\it {Consistent $ \mathcal{N}=8 $ truncation of
  massive IIA on S$^{6}$}},  {\em JHEP} {\bf 12} (2015) 020,
  [\href{http://arxiv.org/abs/1509.02526}{{\tt arXiv:1509.02526}}].

\bibitem{Ciceri:2016dmd}
F.~Ciceri, A.~Guarino, and G.~Inverso, {\it {The exceptional story of massive
  IIA supergravity}},  {\em JHEP} {\bf 08} (2016) 154,
  [\href{http://arxiv.org/abs/1604.08602}{{\tt arXiv:1604.08602}}].

\bibitem{Cassani:2016ncu}
D.~Cassani, O.~de~Felice, M.~Petrini, C.~Strickland-Constable, and D.~Waldram,
  {\it {Exceptional generalised geometry for massive IIA and consistent
  reductions}},  {\em JHEP} {\bf 08} (2016) 074,
  [\href{http://arxiv.org/abs/1605.00563}{{\tt arXiv:1605.00563}}].

\bibitem{Inverso:2016eet}
G.~Inverso, H.~Samtleben, and M.~Trigiante, {\it {Type II supergravity origin
  of dyonic gaugings}},  {\em Phys. Rev.} {\bf D95} (2017), no.~6 066020,
  [\href{http://arxiv.org/abs/1612.05123}{{\tt arXiv:1612.05123}}].

\bibitem{Dall'Agata:2012bb}
G.~Dall'Agata, G.~Inverso, and M.~Trigiante, {\it {Evidence for a family of
  SO(8) gauged supergravity theories}},  {\em Phys.Rev.Lett.} {\bf 109} (2012)
  201301, [\href{http://arxiv.org/abs/1209.0760}{{\tt arXiv:1209.0760}}].

\bibitem{Dall'Agata:2014ita}
G.~Dall'Agata, G.~Inverso, and A.~Marrani, {\it {Symplectic Deformations of
  Gauged Maximal Supergravity}},  {\em JHEP} {\bf 1407} (2014) 133,
  [\href{http://arxiv.org/abs/1405.2437}{{\tt arXiv:1405.2437}}].

\bibitem{Inverso:2015viq}
G.~Inverso, {\it {Electric-magnetic deformations of D = 4 gauged
  supergravities}},  {\em JHEP} {\bf 03} (2016) 138,
  [\href{http://arxiv.org/abs/1512.04500}{{\tt arXiv:1512.04500}}].

\bibitem{Guarino:2015qaa}
A.~Guarino and O.~Varela, {\it {Dyonic ISO(7) supergravity and the duality
  hierarchy}},  {\em JHEP} {\bf 02} (2016) 079,
  [\href{http://arxiv.org/abs/1508.04432}{{\tt arXiv:1508.04432}}].

\bibitem{Gallerati:2014xra}
A.~Gallerati, H.~Samtleben, and M.~Trigiante, {\it {The $ \mathcal{N}>2 $
  supersymmetric AdS vacua in maximal supergravity}},  {\em JHEP} {\bf 12}
  (2014) 174, [\href{http://arxiv.org/abs/1410.0711}{{\tt arXiv:1410.0711}}].

\bibitem{Guarino:2019jef}
A.~Guarino, J.~Tarrio, and O.~Varela, {\it {Halving ISO(7) supergravity}},
  \href{http://arxiv.org/abs/1907.11681}{{\tt arXiv:1907.11681}}.

\bibitem{deWit:2008ta}
B.~de~Wit, H.~Nicolai, and H.~Samtleben, {\it {Gauged Supergravities, Tensor
  Hierarchies, and M-Theory}},  {\em JHEP} {\bf 02} (2008) 044,
  [\href{http://arxiv.org/abs/0801.1294}{{\tt arXiv:0801.1294}}].

\bibitem{Varela:2015uca}
O.~Varela, {\it {AdS$_{4}$ solutions of massive IIA from dyonic ISO(7)
  supergravity}},  {\em JHEP} {\bf 03} (2016) 071,
  [\href{http://arxiv.org/abs/1509.07117}{{\tt arXiv:1509.07117}}].

\bibitem{Bergshoeff:2009ph}
E.~A. Bergshoeff, J.~Hartong, O.~Hohm, M.~Huebscher, and T.~Ortin, {\it {Gauge
  Theories, Duality Relations and the Tensor Hierarchy}},  {\em JHEP} {\bf
  0904} (2009) 123, [\href{http://arxiv.org/abs/0901.2054}{{\tt
  arXiv:0901.2054}}].

\bibitem{Pang:2017omp}
Y.~Pang, J.~Rong, and O.~Varela, {\it {Spectrum universality properties of
  holographic Chern-Simons theories}},  {\em JHEP} {\bf 01} (2018) 061,
  [\href{http://arxiv.org/abs/1711.07781}{{\tt arXiv:1711.07781}}].

\bibitem{Guarino:2016ynd}
A.~Guarino, J.~Tarrio, and O.~Varela, {\it {Romans-mass-driven flows on the
  D2-brane}},  {\em JHEP} {\bf 08} (2016) 168,
  [\href{http://arxiv.org/abs/1605.09254}{{\tt arXiv:1605.09254}}].

\bibitem{Gauntlett:2009zw}
J.~P. Gauntlett, S.~Kim, O.~Varela, and D.~Waldram, {\it {Consistent
  supersymmetric Kaluza-Klein truncations with massive modes}},  {\em JHEP}
  {\bf 0904} (2009) 102, [\href{http://arxiv.org/abs/0901.0676}{{\tt
  arXiv:0901.0676}}].

\bibitem{Cassani:2010uw}
D.~Cassani, G.~Dall'Agata, and A.~F. Faedo, {\it {Type IIB supergravity on
  squashed Sasaki-Einstein manifolds}},  {\em JHEP} {\bf 1005} (2010) 094,
  [\href{http://arxiv.org/abs/1003.4283}{{\tt arXiv:1003.4283}}].

\bibitem{Gauntlett:2010vu}
J.~P. Gauntlett and O.~Varela, {\it {Universal Kaluza-Klein reductions of type
  IIB to N=4 supergravity in five dimensions}},  {\em JHEP} {\bf 1006} (2010)
  081, [\href{http://arxiv.org/abs/1003.5642}{{\tt arXiv:1003.5642}}].

\bibitem{Liu:2010sa}
J.~T. Liu, P.~Szepietowski, and Z.~Zhao, {\it {Consistent massive truncations
  of IIB supergravity on Sasaki-Einstein manifolds}},  {\em Phys. Rev.} {\bf
  D81} (2010) 124028, [\href{http://arxiv.org/abs/1003.5374}{{\tt
  arXiv:1003.5374}}].

\bibitem{Cassani:2011fu}
D.~Cassani and P.~Koerber, {\it {Tri-Sasakian consistent reduction}},  {\em
  JHEP} {\bf 1201} (2012) 086, [\href{http://arxiv.org/abs/1110.5327}{{\tt
  arXiv:1110.5327}}].

\bibitem{Cassani:2012pj}
D.~Cassani, P.~Koerber, and O.~Varela, {\it {All homogeneous N=2 M-theory
  truncations with supersymmetric AdS4 vacua}},  {\em JHEP} {\bf 1211} (2012)
  173, [\href{http://arxiv.org/abs/1208.1262}{{\tt arXiv:1208.1262}}].

\bibitem{Donos:2010ax}
A.~Donos, J.~P. Gauntlett, N.~Kim, and O.~Varela, {\it {Wrapped M5-branes,
  consistent truncations and AdS/CMT}},  {\em JHEP} {\bf 12} (2010) 003,
  [\href{http://arxiv.org/abs/1009.3805}{{\tt arXiv:1009.3805}}].

\bibitem{Malek:2019ucd}
E.~Malek, H.~Samtleben, and V.~Vall~Camell, {\it {Supersymmetric AdS$_7$ and
  AdS$_6$ vacua and their consistent truncations with vector multiplets}},
  {\em JHEP} {\bf 04} (2019) 088, [\href{http://arxiv.org/abs/1901.11039}{{\tt
  arXiv:1901.11039}}].

\bibitem{Cheung:2019pge}
K.~C.~M. Cheung, J.~P. Gauntlett, and C.~Rosen, {\it {Consistent KK truncations
  for M5-branes wrapped on Riemann surfaces}},  {\em Class. Quant. Grav.} {\bf
  36} (2019), no.~22 225003, [\href{http://arxiv.org/abs/1906.08900}{{\tt
  arXiv:1906.08900}}].

\bibitem{Larios:2019kbw}
G.~Larios, P.~Ntokos, and O.~Varela, {\it {Embedding the SU(3) sector of SO(8)
  supergravity in $D=11$}},  {\em Phys. Rev.} {\bf D100} (2019), no.~8 086021,
  [\href{http://arxiv.org/abs/1907.02087}{{\tt arXiv:1907.02087}}].

\bibitem{deWit:1982ig}
B.~de~Wit and H.~Nicolai, {\it {N=8 Supergravity}},  {\em Nucl.Phys.} {\bf
  B208} (1982) 323.

\bibitem{Warner:1983vz}
N.~Warner, {\it {Some New Extrema of the Scalar Potential of Gauged $N=8$
  Supergravity}},  {\em Phys.Lett.} {\bf B128} (1983) 169.

\bibitem{deWit:1986iy}
B.~de~Wit and H.~Nicolai, {\it {The Consistency of the $S^7$ Truncation in
  $D=11$ Supergravity}},  {\em Nucl.Phys.} {\bf B281} (1987) 211.

\bibitem{Varela:2015ywx}
O.~Varela, {\it {Complete $D=11$ embedding of SO(8) supergravity}},  {\em Phys.
  Rev.} {\bf D97} (2018), no.~4 045010,
  [\href{http://arxiv.org/abs/1512.04943}{{\tt arXiv:1512.04943}}].

\bibitem{Cremmer:1978km}
E.~Cremmer, B.~Julia, and J.~Scherk, {\it {Supergravity Theory in
  Eleven-Dimensions}},  {\em Phys.Lett.} {\bf B76} (1978) 409--412.

\bibitem{Corrado:2001nv}
R.~Corrado, K.~Pilch, and N.~P. Warner, {\it {An N=2 supersymmetric membrane
  flow}},  {\em Nucl. Phys.} {\bf B629} (2002) 74--96,
  [\href{http://arxiv.org/abs/hep-th/0107220}{{\tt hep-th/0107220}}].

\bibitem{Guarino:2017eag}
A.~Guarino and J.~Tarrío, {\it {BPS black holes from massive IIA on S$^{6}$}},
  {\em JHEP} {\bf 09} (2017) 141, [\href{http://arxiv.org/abs/1703.10833}{{\tt
  arXiv:1703.10833}}].

\bibitem{Hosseini:2017fjo}
S.~M. Hosseini, K.~Hristov, and A.~Passias, {\it {Holographic microstate
  counting for AdS$_{4}$ black holes in massive IIA supergravity}},  {\em JHEP}
  {\bf 10} (2017) 190, [\href{http://arxiv.org/abs/1707.06884}{{\tt
  arXiv:1707.06884}}].

\bibitem{Caldarelli:2003pb}
M.~M. Caldarelli and D.~Klemm, {\it {All supersymmetric solutions of N=2, D = 4
  gauged supergravity}},  {\em JHEP} {\bf 09} (2003) 019,
  [\href{http://arxiv.org/abs/hep-th/0307022}{{\tt hep-th/0307022}}].

\bibitem{Fluder:2015eoa}
M.~Fluder and J.~Sparks, {\it {D2-brane Chern-Simons theories: F-maximization =
  a-maximization}},  {\em JHEP} {\bf 01} (2016) 048,
  [\href{http://arxiv.org/abs/1507.05817}{{\tt arXiv:1507.05817}}].

\end{thebibliography}\endgroup

\end{document}